\documentclass{aa} 

\usepackage{graphicx}
\usepackage{txfonts}
\usepackage{natbib}
\bibpunct{(}{)}{;}{a}{}{,}

\newcommand{\be}{\begin{equation}}
\newcommand{\ee}{\end{equation}}
\newcommand{\angstrom}{\mbox{\normalfont\AA}}

\def\apjl{ApJL}
\def\apjs{ApJS}

\newcommand{\Mpc}{$h^{-1}$\thinspace Mpc}

\newcommand{\vmh}{h^{-1}\mathrm{Mpc} }

\begin{document}  

\title{
Death at watersheds:\\
galaxy quenching in low-density environments
} 


\author {Maret~Einasto\inst{1} 
\and Rain~Kipper\inst{1} 
\and Peeter~Tenjes\inst{1} 
\and Jaan~Einasto\inst{1,2,3}
\and Elmo~Tempel\inst{1,2} 
\and Lauri Juhan~Liivam\"agi\inst{1} 
}
\institute{Tartu Observatory, University of Tartu, Observatooriumi 1, 61602 T\~oravere, Estonia
\and
Estonian Academy of Sciences, Kohtu 6, 10130 Tallinn, Estonia
\and
ICRANet, Piazza della Repubblica 10, 65122 Pescara, Italy
}

\authorrunning{Einasto, M. et al. }

\offprints{Einasto, M.}

\date{ Received   / Accepted   }

\titlerunning{QW}

\abstract
{
The evolution of galaxies is influenced by their local and 
global environment in the cosmic web. 
Galaxies with very old stellar populations (VO galaxies
with $D_n(4000)$ index $\geq 1.75$) 
mostly 
lie in the  centres of galaxy clusters,
where they evolve under the influence of processes characteristic of high-density
cluster environments. However, VO galaxies have also been found
in poor groups in global low-density environments between superclusters, 
which we call watershed regions.
}
{
Our aim is to analyse the properties of galaxies in various cosmic 
  environments with a focus on VO galaxies 
in the watershed regions
to understand their evolution, 
and the origin of the large-scale morphology--density relation.
}
{
We employ the Sloan Digital Sky Survey DR10 MAIN spectroscopic galaxy sample in the redshift 
range $0.009 \leq z \leq 0.200$ to calculate the luminosity--density
field of galaxies, to determine groups and filaments in the galaxy distribution,
and to obtain data on galaxy properties.
The luminosity--density field with smoothing length $8$~\Mpc, $D8$,
characterises the global environment of galaxies.
We analyse the group and galaxy contents of regions with various $D8$
thresholds. We divide groups into low- and high-luminosity groups
based on the highest luminosity of groups in the watershed region, 
$L_{gr} \leq 15 \times10^{10} h^{-2} L_{\sun}$. We
compare the stellar masses, the concentration index, and the stellar velocity
dispersions of quenched and star-forming galaxies
among single galaxies, 
satellite galaxies, and the brightest group galaxies (BGGs) in various environments.
}
{
We show that 
the global density is most strongly related to the richness of galaxy groups.  Its influence on the
overall star formation quenching in galaxies is less strong. Correlations between the morphological properties of galaxies
and the global density field are the weakest.
The watershed regions  with $D8 < 1$ 
are populated mostly by single galaxies, constituting $70$~\%
of all galaxies there, and by low-luminosity groups. 
Still, approximately one-third of all galaxies in 
the watershed regions are VO galaxies. They
have lower stellar masses, smaller stellar velocity dispersions,
and stellar populations that are  up to 2~Gyr  younger than those of
VO galaxies 
in other global environments. 
In higher density global environments ($D8 > 1$), the morphological properties of galaxies are very similar.
Differences in galaxy properties are the largest between satellites and BGGs in groups.
}
{
Our results suggest that 
galaxy evolution is determined by the birthplace of galaxies in the cosmic web, 
and mainly 
by internal processes which lead to the present-day properties of galaxies. 
This may explain the similarity of (VO) galaxies in extremely different environments.
}


\maketitle

\section{Introduction} 
\label{sect:intro}

Galaxies are systems of stars, gas, and dust held together by gravity within a dark matter halo. Broadly speaking, galaxies 
can be divided into two major classes based on their ability to form stars
and their morphological and kinematic properties: actively star forming, blue, late-type galaxies, 
and passive, 
red, early-type galaxies
\citep{2001AJ....122.1861S, Bell:2017vy, Bluck:2020vt, Bluck:2022vu, Brownson:2022uf, 2022MNRAS.513..439D}.
The ability of a galaxy to sustain star formation
is determined  by how well it can provide and cool gas.
If  the gas supply in a galaxy, which is governed by accretion, outflows, and mixing, 
is not sufficient to continue star formation, the star formation rate (SFR) drops and
the galaxy becomes quenched or passive.

The processes that lead to star-formation quenching in galaxies can be divided 
into  `internal' and `external'  \citep{2006PASP..118..517B, 2020A&A...633A..70P}.
Internal processes, also referred to as mass quenching, depend first of all on the mass 
of the dark halo of a galaxy. 
Processes that blow out galactic gas include stellar winds, 
supernova explosions, nuclear activity, active galactic nucleus (AGN) feedback, and so on
 \citep{2006MNRAS.372..265M, 2006MNRAS.365...11C,
2019MNRAS.485.3446H}. These processes are more effective in massive galaxies 
 \citep{2020ApJ...889..156C} and at higher redshifts.
 
External processes that cause quenching of galaxies depend on  the local environment
of galaxies within clusters and/or groups 
\citep[environmental quenching; see][for definitions and references]{2019MNRAS.484.1702P, 2021gcf2.confE..20W}.  
External processes involve the stripping away of galactic gas by the ram pressure of the hot gas in a cluster or group 
\citep{1972ApJ...176....1G, 2019MNRAS.483.1042Y},
cold gas removal by viscous stripping 
\citep{1982MNRAS.198.1007N}, 
starvation due to prevention of the arrival of fresh gas 
by removal of their feeding primordial filaments 
\citep{2019OJAp....2E...7A, 2019A&A...621A.131M},
or harassment due to multiple high-speed mergers 
\citep{1996Natur.379..613M}. 
External processes depend on environmental density, 
the orbital properties of galaxies, and also galaxy mass. 

The formation and/or evolution of galaxies and 
the environment of galaxies are related 
(referred to as the nature-versus-nurture problem). 
While rich galaxy clusters and especially their central parts are
mostly populated by early-type, 
red, quenched galaxies, late-type, blue,
 star-forming galaxies 
 are preferentially found in poor groups, in the outskirts of clusters, or in the low-global-density 
environment.
This is known as the  morphological segregation of galaxies or the morphology--density relation, where
`morphology' denotes various properties of galaxies:
their morphological type, star formation activity, colour, and so on.
\citet{1974Natur.252..111E}, \citet{1980ApJ...236..351D}, \citet{1984ApJ...281...95P},
\citet{1987MNRAS.226..543E}, and \citet{1991MNRAS.252..261E} 
demonstrated that the 
morphology--density relation extends from rich clusters to isolated galaxies in low-density environments.
\citet{1978MNRAS.185..357J} and \citet{1980MNRAS.193..353E} showed that the central parts of the
Perseus-Pisces supercluster are populated by early-type galaxies, while the fraction of
late-type galaxies increases in the supercluster outskirts. 
The dependence of galaxy properties on environment 
was studied in detail by, for example, \citet{2003MNRAS.346..601G}, \citet{2003ApJ...584..210G},
\citet{2011A&A...529A..53T}, \citet{2007ApJ...658..898P}, \citet{2009ApJ...691.1828P},
\citet{2009ARA&A..47..159B}, \citet{2012A&A...545A.104L}, and \citet{2022A&A...665A..44A}.
If galaxies stop forming stars  in groups or filaments before falling into
clusters, this is called preprocessing \citep{2004PASJ...56...29F, 2009MNRAS.400..937M, 2014MNRAS.442L.105M,
2019OJAp....2E...7A, 2019MNRAS.490L...6D, 2022A&A...657A...9C, 2022ApJS..259...43C}.
Even galaxies that do not belong to any group are more likely
to be red and passive if they are located in superclusters 
\citep{2007A&A...464..815E, 2012A&A...545A.104L,2014A&A...562A..87E}. 
This shows that global environment, which refers to scales larger than groups and clusters,
may also be important in shaping galaxy properties.

In galaxy evolution, internal and external quenching processes often act together 
and it is not easy to decipher which one is dominating in different local and/or global environments.  
Understanding the formation of stars within galaxies and the
end of star formation via the quenching 
processes
is one of the most important
unresolved problems in extragalactic astrophysics and cosmology.
In this context, detailed studies of galaxies in various environments lead us to a better understanding
of the formation and evolution of galaxies.

Most galaxies lie in groups of various richness 
and luminosity. In groups, galaxies can be divided into 
the brightest group galaxies (BGGs; often also called the central galaxies in groups)
and satellite galaxies. 
There are also galaxies that do not belong to any group 
according to the criteria used to determine  groups in galaxy distribution; 
we call these single galaxies.
Single galaxies may represent various galaxy populations. They may be
outer members of groups
having closest group member galaxies that are too faint to be included
in the given sample, in the Sloan Digital Sky Survey 
(SDSS) spectroscopic sample in our study. 
Single galaxies can also be isolated galaxies, that is,  
galaxies that do not have close bright galaxies according to certain criteria
\citep{2006A&A...449..937S, 2016A&A...588A..79L}.

The large-scale distribution of galaxies can be described as the cosmic web: 
a network of galaxies 
and their groups and clusters, connected by galaxy filaments and separated by 
low-density regions with almost no visible galaxies.
Systems of galaxies of various richness in the cosmic web represent global and local
environments in which galaxies form and evolve.
The present study focuses on the global environment of galaxies. 
We  quantify global environment using the  luminosity--density field. 
This field is calculated using a
smoothing length $8$~\Mpc\ (in units of
the mean luminosity--density, and denote as $D8$). 
In many studies, a threshold density of $D8 = 5$ is used to separate
the largest overdensity regions, namely superclusters, from the underdense regions between them.
\citet{2018A&A...616A.141E, 2019A&A...623A..97E} showed that superclusters occupy $\approx 1$\%
of the total SDSS sample volume, approximately $65$\% of the SDSS volume
has very low global luminosity--density with $D8 < 1$, and
in $90$\% of the SDSS volume,  $D8 < 2$. 
Superclusters contain rich galaxy clusters and groups of galaxies connected
by filaments. In underdense regions, galaxy groups and clusters are poor, and are typically connected by longer 
filaments than in superclusters
\citep[see ][for details and  references]{2020A&A...641A.172E}.

Simulations show that galaxies and their systems (groups and clusters)
can form  in the cosmic density field
where large-scale density perturbations in combination
with small-scale overdensities are sufficiently high 
\citep{2021arXiv210602672P, 2011A&A...534A.128E, 2011A&A...531A.149S}. 
In the underdense regions, galaxy groups are poor, and are too
far apart to merge and form richer groups and clusters.
Density fields and velocity fields are related; 
the velocities of galaxies are the lowest in the regions of 
lowest global density,
where the formation of groups is also the slowest 
\citep{2009A&A...495...37T, 2021A&A...652A..94E}.
\citet{2020A&A...641A.172E} found that while in the underdense region around the 
massive rich supercluster SCl~A2142 ---at a distance of $260$~\Mpc\  
(with mass of $M \approx 4.3\times~10^{15}h^{-1}M_\odot$ and more than 1000 member galaxies)--- 
all groups are poor
and do not show signatures of dynamical activity,
even the poor groups in the supercluster have possibly started merging.
Simulations show that 
in the regions  of lowest global density where the global luminosity--density value
is $D < 1.5$ (in units of mean luminosity--density), 
the sizes of haloes remain the same in a wide redshift interval,
$\approx 1$~\Mpc, while in the regions of  highest global luminosity--density 
(supercluster cores with $D8 > 7$),
the sizes of haloes decrease by about five times during their evolution 
\citep{2009A&A...495...37T}.
In underdense regions, the peculiar velocities of 
group-sized haloes  have values below $500$~$km/s$,
and may be close to zero at the minima of the density field, while these velocities may have 
values higher than $500$~$km/s$ in superclusters
\citep{1991ApJ...379....6N, 1994ApJ...436...23B, 1998aums.conf...78D}. 

In regions with increasingly high global density, 
groups are located closer together, and they can grow by merging 
and infall of galaxies or other groups
\citep{2009MNRAS.400..937M, 2015MNRAS.452.2528M, 2016A&A...586A.112R}. 
The richest clusters form at the locations of the
highest density peaks (deep potential wells) where they attract matter from the surroundings more strongly
than than  poor clusters found elsewhere.
In the densest regions of the cosmic web (seeds of present-day rich 
clusters in supercluster cores), galaxies may have started to form earlier than elsewhere
\citep{2022ApJ...937...15P}.
 Groups that are found near rich clusters are also richer 
and more luminous than groups far from rich clusters,
a phenomenon called environmental enhancement of poor groups
\citep{2003A&A...401..851E, 2005A&A...436...17E}.


In accordance with earlier studies,
\citet{2018A&A...610A..82E, 2021A&A...649A..51E} found that the inner, virialised parts of the 
richest clusters in superclusters are populated by galaxies with
a $D_n(4000)$ index  (ratio of the average flux densities
in the bands $4000 - 4100 \angstrom$ and $3850 - 3950 \angstrom$)
$D_n(4000) \geq 1.75$.
\citet{2003MNRAS.341...33K} showed that such galaxies may  have stellar populations 
with ages of at least $4$~Gyr.
Therefore, following \citet{2020A&A...641A.172E}, we refer to these 
galaxies with very old stellar populations as VO galaxies in order to distinguish them from 
galaxies that were quenched recently or are still forming stars. 
Somewhat unexpectedly, \citet{2020A&A...641A.172E} found that VO galaxies
 are also common in poor groups and even among single galaxies in the underdense region around SCl~A2142.
There,  $40$\% of galaxies in groups and one-third of single galaxies are VO galaxies.
Isolated elliptical galaxies have also been found 
in filaments around the Coma cluster in the Coma supercluster \citep{2016A&A...588A..79L}.
Moreover, \citet{2020A&A...641A.172E} showed that the properties of VO galaxies
(morphology, concentration index, and other properties) are very similar 
across galaxy systems that are very different in terms of  mass, from rich clusters in the supercluster
to the poorest groups and single galaxies in global low-density environments.

The analyses in \citet{2020A&A...641A.172E} and \citet{2016A&A...588A..79L} focused on the properties of a small sample of galaxies in the SCl~A2142 and Coma 
superclusters and their environments.
In the present study we aim  to analyse the properties and environment of  galaxies
at various stages of star formation
in a larger volume covered by the SDSS.
One particular
aim of our study is to search for VO galaxies in environments with the lowest global
density, and to understand their  properties. 

To avoid mixing the influence of local (group) and global environment 
on galaxy properties, we analyse the properties of single galaxies, satellites,
and brightest group galaxies in various global environments separately.
Our study helps us to identify the main types of systems where galaxies 
are likely to be quenched. 
We aim to explore whether galaxy quenching occurs predominantly in rich clusters, which typically 
reside in global high-density environments, or
also in poor groups and among single galaxies in the global low-density environment where 
conditions are very different from those in rich clusters.

We refer to the regions  of lowest global luminosity--density as watersheds; that is, watersheds are 
the lowest-density parts of underdense regions between superclusters.
We may expect the evolution
of galaxies and groups in watershed regions to differ from that of galaxies 
in the high-density cores of rich clusters. 
Therefore, this study helps us to better understand the processes involved in galaxy
evolution in various environments.

We compare the group and galaxy content, as well as the properties of galaxies
in the watersheds with those in the higher global luminosity-density regions
(higher-density parts of underdensity regions and superclusters).
Our first aim is to clarify whether or not VO galaxies 
also populate the watershed regions. 
Secondly, if we  find such a population then we want to understand
whether or not the properties of VO galaxies differ 
  depending on whether they are  located in watersheds 
or environments with higher global
density. We also analyse whether there is a smooth change in galaxy properties with density or
a sharp change in terms of galaxy and group properties at a certain threshold density that separates watersheds
from  regions of
global higher density.
In another series of comparisons, we investigate environmental trends in the properties of
galaxies with young stellar populations with $D_n(4000) < 1.75$.

As in \citet{2020A&A...641A.172E}, we use 
data from the SDSS data release 10
(SDSS DR10) to analyse properties of galaxies and galaxy
groups, with the following cosmological parameters: the Hubble parameter $H_0=100~ 
h$ km~s$^{-1}$ Mpc$^{-1}$, matter density $\Omega_{\rm m} = 0.27$, 
parameter $h = 0.7$,  and 
dark energy density $\Omega_{\Lambda} = 0.73$ 
\citep{2011ApJS..192...18K}.

\section{Data on galaxies, groups, and their environment} 
\label{sect:data} 

We first describe  how we determine the global environments of galaxies.
For this task, we employ the luminosity--density field.  In the
luminosity--density field, the lowest density regions define the watersheds
between superclusters, and the highest density regions correspond to
superclusters.
Global environment properties may be mixed with those caused by local environment. 
Examples of local environment are galaxy groups and single galaxies. 
Therefore  we describe our group catalogue next. 
In groups, it is well known that the BGGs are more luminous and massive than satellite
galaxies. Therefore, in order to avoid mixing large-scale effects and differences
in galaxy properties within groups, we analyse single galaxies,
satellites, and BGGs  separately. 
Finally, we describe morphological properties of the galaxies analysed in this study,
 such as their stellar masses and other properties.

\subsection{Global environment of galaxies: luminosity--density field} 
\label{sect:env} 

\begin{figure}
\centering
\resizebox{0.44\textwidth}{!}{\includegraphics[angle=0]{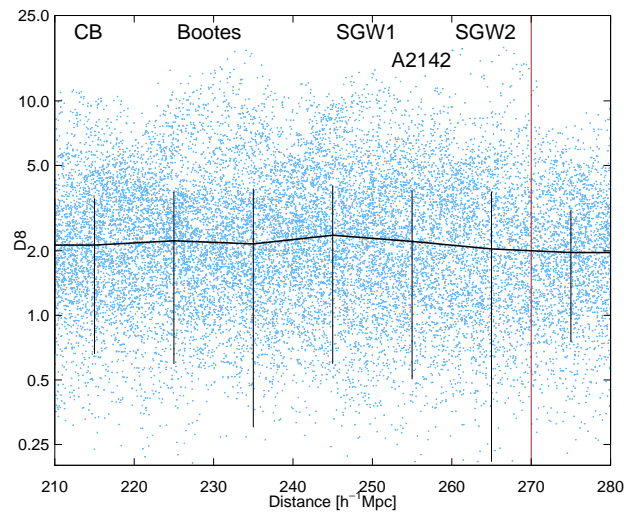}}
\caption{
Distribution of global luminosity density $D8$ around galaxies versus distance. 
The black line with error bars shows the median value of global luminosity density $D8$
at a given distance. We note that the logarithmic scale enhances low values of density and
the associated errors. We mark several rich superclusters where $D8$ values are
high. CB denotes the Corona Borealis supercluster, A2142 marks the supercluster
SCL~A2142, and SGW1 and SGW2 mark the locations 
of the two richest superclusters in the Sloan Great Wall.
The dark red line 
shows the distance limit $270$~\Mpc\ used to define our final sample of galaxies.
}
\label{fig:d8dist}
\end{figure}

We use galaxy data 
from the SDSS DR10 MAIN spectroscopic galaxy sample  with
apparent Galactic extinction corrected $r$ magnitudes $r \leq 
17.77$ and redshifts $0.009 \leq z \leq 0.200$
\citep{2011ApJS..193...29A, 2014ApJS..211...17A}.
We calculated the absolute magnitudes of galaxies as
\begin{equation}
M_r = m_r - 25 -5\log_{10}(d_L)-K,
\end{equation} 
where $d_L$ is the luminosity distance in units of $h^{-1}$Mpc and
$K$ is the $k$+$e$-correction calculated as in 
\citet{2007AJ....133..734B} and  \citet{2003ApJ...592..819B}
\citep[see][for details]{2014A&A...566A...1T}.
As an order of magnitude reference, the limiting magnitude $r = 
17.77$ at $z = 0.1$ corresponds to the stellar mass 
$M^\star \simeq 1.8\times 10^{10}~\mathrm{M}_{\sun}$, 
which is twice the stellar mass of large satellites in the Local Group,
the  Large Magellanic Cloud, or the Triangulum Nebula.

We characterise the global environment of galaxies using 
the luminosity--density field \citep{2012A&A...539A..80L}. This field is calculated using a
smoothing kernel based on the $B_3$ spline function:
\begin{equation}
    B_3(x) = \frac{|x-2|^3 - 4|x-1|^3 + 6|x|^3 - 4|x+1|^3 + |x+2|^3}{12}.
\end{equation}
The details of the calculation of density field using a $B_3$ spline kernel
can be found in \citet{2007A&A...476..697E,  2014A&A...566A...1T}.
We use a smoothing length of $8$~\Mpc\ to define the global luminosity--density
field,
and denote global luminosity--density as $D8$. 
Luminosity--density values are expressed in units of mean 
luminosity density, $\ell_{\mathrm{mean}}$ = 
1.65$\cdot10^{-2}$ $\frac{10^{10} h^{-2} L_\odot}{(\vmh)^3}$. 
In the luminosity--density field, superclusters can be defined as the connected 
volumes above a threshold density of $D8 = 5.0$ \citep[as in e.g. ][]{2012A&A...539A..80L,
2014A&A...562A..87E,2020A&A...641A.172E}.
In the following analysis, we use a series of global luminosity--density $D8$ intervals
to study the trends in galaxy and group properties with global density.
The lowest global density interval corresponds to low-density watershed regions between superclusters,
and the highest global densities ($D8 \geq 5$) correspond to superclusters. The global density
regions with intermediate densities represent medium-density regions around superclusters.

We also employ luminosity--density field with 
smoothing length 1~\Mpc\ to define local luminosity--density field around
single galaxies, denoted as $D1$. 
For galaxies in groups, local environment is defined by group membership, 
which we define in the following subsection.

In studies of environmental trends of galaxy properties, 
density is typically defined by Nth nearest neighbours \citep[see e.g. ][for a review and references]
{2009ARA&A..47..159B}. For our purposes, the use of the luminosity--density
field is preferable because with this approach we can distinguish  between
various global environments.

\begin{figure*}
\centering
\resizebox{0.44\textwidth}{!}{\includegraphics[angle=0]{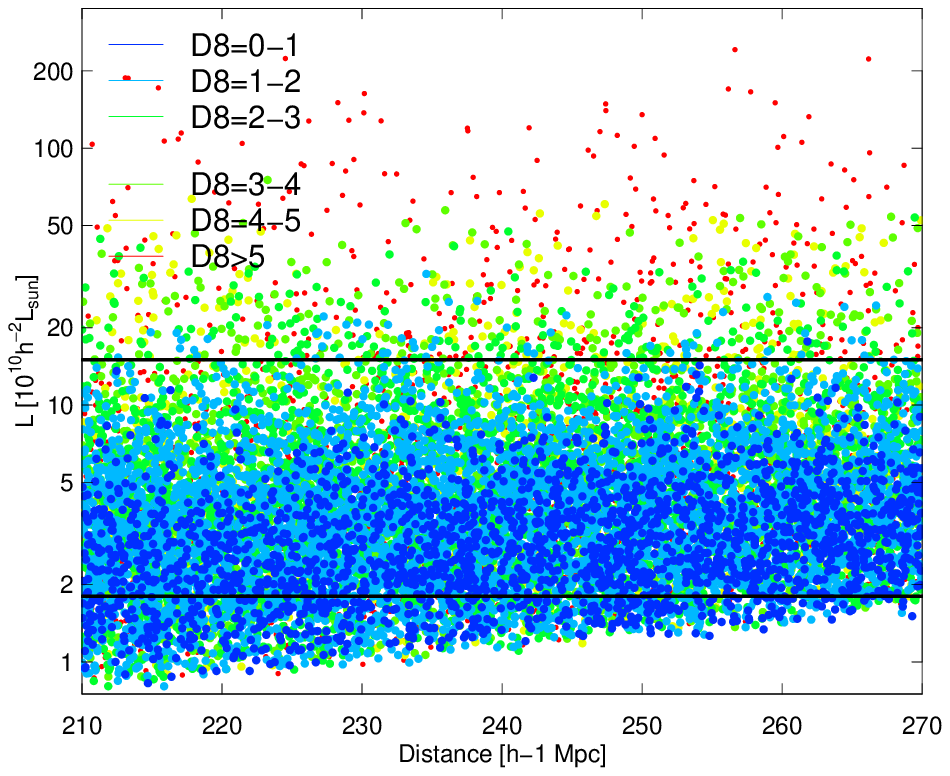}} 
\resizebox{0.43\textwidth}{!}{\includegraphics[angle=0]{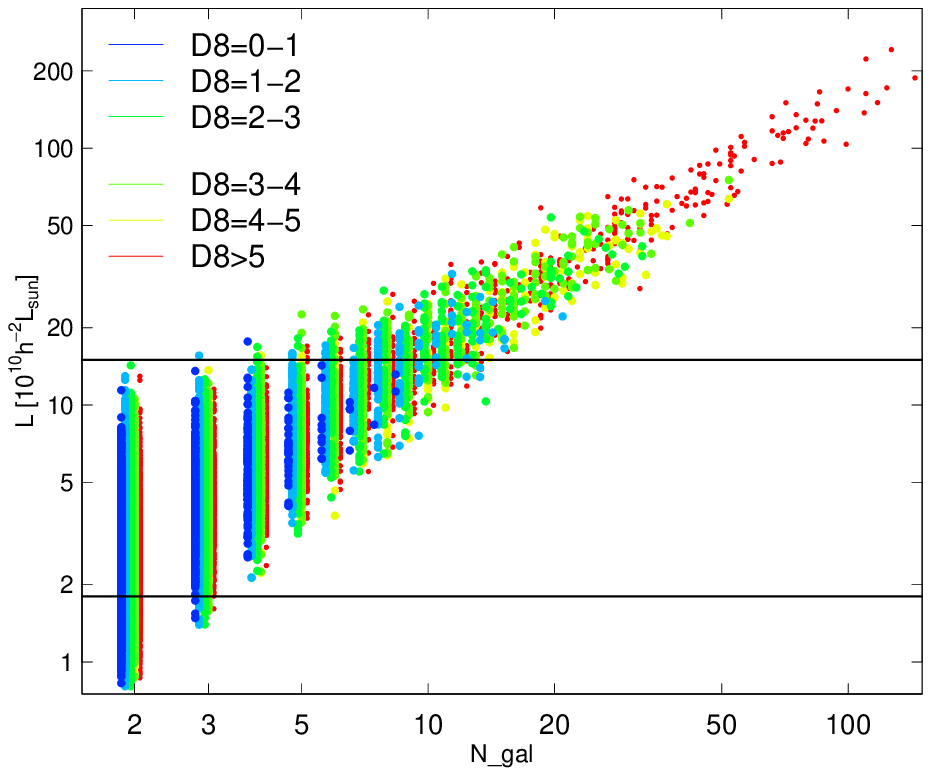}} 
\caption{
Total luminosity of groups versus group distances (left panel) and 
group richness (right panel).
In the right panel, group richness is shifted to avoid overlapping, which is especially 
strong for groups with up to five galaxies.
}
\label{fig:grlumrich}
\end{figure*}

\subsection{Group data}
\label{sect:gr}

The majority of galaxies belong to groups of  different richness, from
galaxy pairs to the richest clusters. To take this into account
we obtained data on group membership for galaxies from group catalogues by 
\citet{2012A&A...540A.106T, 2014A&A...566A...1T,
2014MNRAS.438.3465T}. 
These catalogues are based on the SDSS DR10 MAIN spectroscopic galaxy sample
described in the previous section.
Catalogues of galaxy groups are
available at  the CDS
\footnote{cdsarc.u-strasbg.fr}.

Galaxy groups in these catalogues were determined using 
the friends-of-friends cluster analysis 
method \citep{1982Natur.300..407Z, 1982ApJ...257..423H}, where
a galaxy is considered a member of a group 
if it lies less than  linking length from at least one group member galaxy. In a flux-limited sample, the density of galaxies slowly 
decreases with distance. To properly  take this selection effect into account 
when a group catalogue is constructed, the 
linking length is re-scaled with distance, calibrating the scaling relation by observed 
groups. As a result, the 
maximum sizes in the sky projection and the velocity dispersions of the groups 
are similar at all distances. The  redshift-space distortions (also known as Fingers of God) 
for groups were suppressed,
as described in detail in \citet{2014A&A...566A...1T}.

In the \citet{2014A&A...566A...1T} group catalogue, the luminosities of groups $L_{gr}$
are calculated
using $r$-band luminosities of group member galaxies, corrected for the missing (unobserved)
galaxies at a distance of a given group. In groups, it is well known that the BGGs differ in their properties from satellite
galaxies. 
Therefore, in order to avoid mixing large-scale effects and differences
in galaxy properties within groups, we distinguish between satellite galaxies
and BGGs in our analysis. 

There are galaxies in our sample that are not members of any group.
We call these single galaxies and analyse them separately. 
Single galaxies may also be the brightest galaxies of groups of
which the other members are too faint to be included in the SDSS MAIN spectroscopic
sample. They may also be outer satellites of groups, 
connected to the other group member galaxies by faint galaxies that are not in the sample.
Single galaxies may belong to samples of isolated galaxies if  the magnitude difference between 
the galaxy and its brightest satellites and the distance to the nearest
bright galaxy are large enough to satisfy the conditions
to be included in such a sample; as described, for example, in \citet{2006A&A...449..937S}
and in \citet{2016A&A...588A..79L}. 
Isolated galaxies may be hypergalaxy-type systems
in which giant galaxies are surrounded by dwarf satellites
\citep{1974TarOT..48....3E}.

\begin{figure}
\centering
\resizebox{0.44\textwidth}{!}{\includegraphics[angle=0]{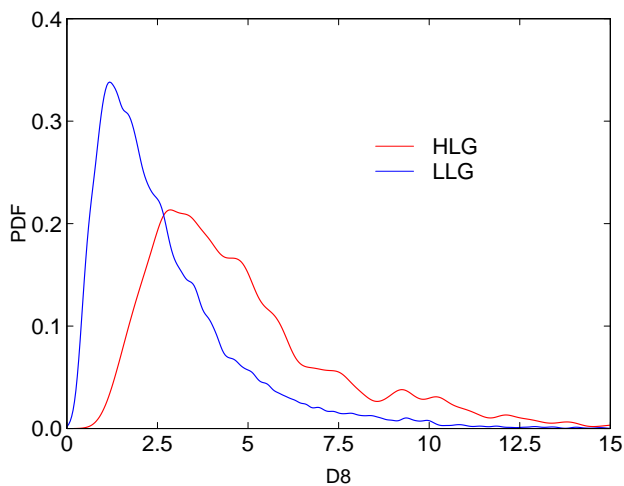}}
\caption{
Distribution of global densities $D8$ at the location of 
HLGs (red line) and LLGs (blue line). 
}
\label{fig:d8grlum}
\end{figure}

\subsection{Selection effects and final sample}

The SDSS MAIN sample covers various global density environments from voids to
superclusters. At distances shorter than  $\approx 200$~\Mpc\ the SDSS sample 
is narrow, and it crosses an underdense region between the Hercules supercluster
and the Bootes and the Corona Borealis superclusters.
At distances greater than $270$~\Mpc\ there is a large underdense region without
rich superclusters. 
The distance interval
$210 - 270$~\Mpc\ covers the region with rich and very rich superclusters,
such as the Sloan Great Wall, the Corona Borealis supercluster, the Bootes supercluster, and
SCl~A2142, and the underdense regions between them. More details about 
these structures can be found, for example, in 
\citet{2012A&A...542A..36E} and in \citet{2014A&A...562A..87E}.
To minimise distance-dependent selection effects
in our study,  we need to have  a variety of global environments at the same distances.
Therefore, we use the distance range $210 - 270$~\Mpc\ which satisfies this condition. 
Figure~\ref{fig:d8dist}  shows the luminosity--density $D8$ around galaxies
versus distance. Several peaks in the density distribution can be seen
at the location of rich  superclusters. 
The median value of global luminosity--density in this distance interval is 
$D8_{\mathrm{med}} \approx 2.1$.
Figure~\ref{fig:d8dist} shows that the median value of $D8$ is almost unchanged
with distance, but the scatter of densities at a given distance interval is large.
This is related to the presence of both over- and underdensity structures at a given
distance interval.

As we use a flux-limited (apparent magnitude limited) 
galaxy sample to construct our group catalogue, the
group luminosities and richness are affected by distance-dependent selection
effects. This is shown in
Fig.~\ref{fig:grlumrich}, where we plot group luminosities versus their distance and 
group richness versus group luminosity in a 
distance interval of $210 - 270$~\Mpc\ (redshift range $0.07 \leq z \leq 0.100$). 
We highlight groups in the lowest global density environment with $D8 \leq 1$.
The  flux-limited nature of our galaxy sample also affects group luminosities; low-luminosity groups 
are absent among more distant groups. The right panel of Fig.~\ref{fig:grlumrich}
shows that this selection affects mostly galaxy pairs in all global density environments.
To minimise this effect, we  exclude groups with luminosity limit 
$L_{gr} < 1.8\times10^{10} h^{-2} L_{\sun}$ from a complete sample of groups
(mostly galaxy pairs).

\begin{figure*}
\centering
\resizebox{0.90\textwidth}{!}{\includegraphics[angle=0]{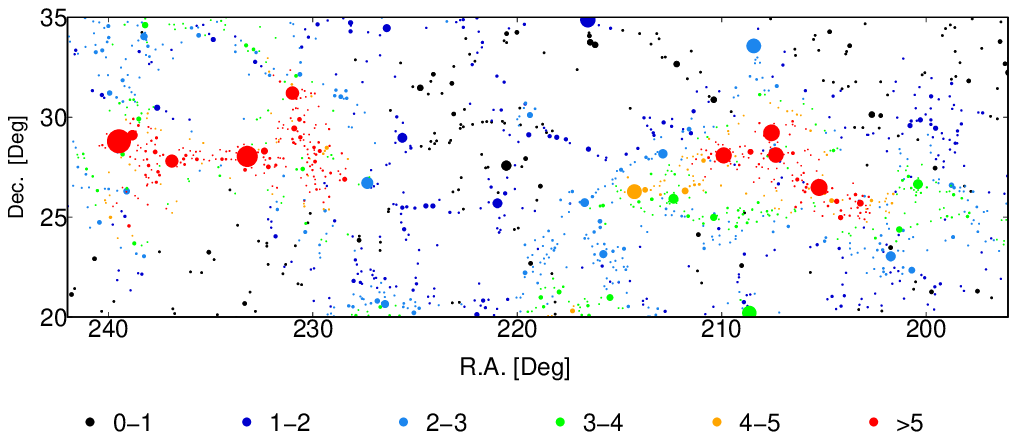}}
\caption{
Distribution of groups in the sky plane in a slice of thickness of $20$~\Mpc\ at a distance
interval $210 - 230$~\Mpc\ which partly covers the Corona Borealis supercluster at $R.A. = 230$\degr and $Dec. = +27$\degr,
and the Bootes supercluster at $R.A. = 207$\degr and $Dec. = +28$\degr, and the low-density region between and around them.   
The sizes of the circles are proportional to the richness of the groups.
We only plot the location of the brightest galaxy of any given group.
Colours code the limits of global density $D8$ in the environment of galaxies.
}
\label{fig:radecwide}
\end{figure*}

Figure~\ref{fig:grlumrich} shows that in the lowest global density region, $D8 \leq 1$,
all groups are poor, with $N_{gal} \leq 9$ and of low luminosity,
$L_{gr} \leq 15\times10^{10} h^{-2} L_{\sun}$ (only one group with four galaxies in this 
global density interval has luminosity
$L_{gr} > 15\times10^{10} h^{-2} L_{\sun}$). 
As the group richness is more strongly affected by selection effects, in what follows
we use group luminosities to define samples of low-luminosity groups (LLGs) and 
high-luminosity groups (HLGs). 
Low-luminosity groups have group luminosity limits
$1.8\times10^{10} h^{-2} L_{\sun} \leq L_{gr} \leq 15\times10^{10} h^{-2} L_{\sun}$, while
HLGs  
have $L_{gr} > 15\times10^{10} h^{-2} L_{\sun}$. 
These limits are shows in Fig.~\ref{fig:grlumrich}.
For a statistical comparison
of galaxy properties, we apply an absolute magnitude limit of $M_r = -19.6$~mag. 
This limit corresponds to a volume-limited sample in the distance interval used in our study.
In total, within the distance interval  $210 - 270$~\Mpc\ and given luminosity limits, we
have data for 81527 galaxies.

\begin{figure}
\centering
\resizebox{0.44\textwidth}{!}{\includegraphics[angle=0]{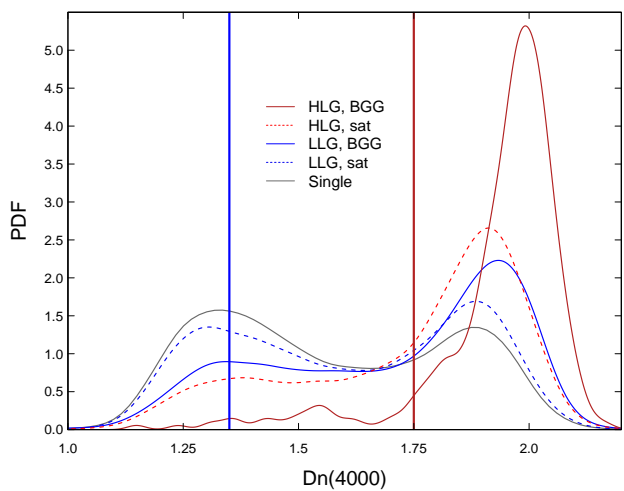}}
\caption{
Distribution of $D_n(4000)$ index for single galaxies (solid grey line),
and for satellite galaxies (dashed lines) and BGGs (solid lines)
in LLGs (blue lines) and HLGs (red lines).  Vertical lines show the values
of $D_n(4000)$ index $D_n(4000) = 1.35$ ($D_n(4000)$ limit for galaxies with very 
young stellar populations), and  $D_n(4000) = 1.75$ (limit for VO galaxies
with $D_n(4000) \geq 1.75$, see text).
}
\label{fig:dn4gr}
\end{figure}

Figure~\ref{fig:d8grlum} shows the distribution of global densities $D8$ at the locations
of groups. From this figure, we see that although LLGs are
present in all global density environments, and HLGs are present everywhere except
the lowest global density region with $D8 \leq 1$, the distribution of global densities
at their location is different. LLGs  preferentially occupy 
regions outside superclusters ($D8 < 5$), while over 50\%  of HLGs lie in superclusters.
High-density cores of superclusters with $D8 > 8$ are populated mostly by HLGs (mostly rich clusters).
The number of such groups in our sample is small, and we do not analyse them separately.

We show the sky distribution of groups from our sample in Fig.~\ref{fig:radecwide} 
where we show groups in a $20$~\Mpc\ thin slice in the distance range of
$210 - 230$~\Mpc\ which covers the regions of two superclusters
(part of the Corona Borealis supercluster at $R.A. = 207$\degr and $Dec. = +28$\degr,
and the Bootes supercluster at $R.A. = 230$\degr and $Dec. = +27$\degr), and a low-density region between and around them. 
The locations of these superclusters are also marked in Fig.~\ref{fig:d8dist}.
Figure~\ref{fig:radecwide} shows how 
groups in the low-global-density region 
separate superclusters, which occupy only $1$\% of the total volume. 
Groups appear to form
filaments, and richer groups lie at the crossings of these filaments.

\subsection{Morphological data of galaxies}
\label{sect:gal}

To compare galaxy samples of various classes we use galaxy
morphological data: $D_n(4000)$ index, $\mathrm{SFR}$,
stellar masses $M^\star$, stellar velocity
dispersion $\sigma^\star$, and concentration index $C$. 
Data on galaxy properties for our study are taken from the SDSS DR10 web page  
\footnote{\url{http://skyserver.sdss3.org/dr10/en/help/browser/browser.aspx}}.
Below, we list  the data for the galaxies used in this paper.

The data on $D_n(4000)$ index  are from 
the MPA-JHU spectroscopic catalogue \citep{2004ApJ...613..898T, 2004MNRAS.351.1151B}.  
$\text{The }D_n(4000)$ index is the ratio of the average flux densities
in the band $4000 - 4100 \angstrom$ to those in the band $3850 - 3950 \angstrom$, 
and is correlated with the time passed since 
the most recent star formation event in a galaxy \citep{2003MNRAS.341...33K}.
This index can be used as proxy for the ages of the stellar populations of galaxies,
and their SFRs.
We use the $D_n(4000)$ index of galaxies as calculated in \citet{1999ApJ...527...54B}.

Data on stellar masses $M^\star$ and $\mathrm{SFRs}$ 
are taken from
the MPA-JHU spectroscopic catalogue \citep{2004ApJ...613..898T, 2004MNRAS.351.1151B}.  
In this catalogue, the properties of galaxies are calculated using 
the stellar population synthesis models and fitting SDSS photometry and spectra 
with \citet{2003MNRAS.344.1000B} models.
The stellar masses of galaxies are calculated 
as described in \citet{2003MNRAS.341...33K}. 
The $\mathrm{SFR}$s 
were computed using the photometry and emission lines as described 
by \citet{2004MNRAS.351.1151B} and \citet{2007ApJS..173..267S}. 
Stellar velocity dispersions of galaxies, $\sigma^{\mathrm{*}}$, are  
calculated by fitting galaxy spectra 
using publicly available codes, namely the Penalized PiXel Fitting code
\citep[pPXF,][]{2004PASP..116..138C} 
and the Gas and Absorption Line Fitting code 
\citep[GANDALF,][]{2006MNRAS.366.1151S}.
The concentration index of galaxies $C$ is calculated as the ratio of the 
Petrosian radii $R_{50}$ and  $R_{90}$: $C = R_{50}/R_{90}$, which themselves are defined as the radii containing
50\% and 90\% of the Petrosian flux of a galaxy, respectively \citep{2001AJ....121.2358B,
2001AJ....122.1104Y}.


We divide galaxies into those with old and young stellar populations using the $D_n(4000)$ index.
Earlier studies showed that the central, virialised parts of galaxy clusters
(ancient infall regions in the projected phase space diagrams) 
are populated mostly by galaxies with  $D_n(4000) \geq 1.75$
 \citep{2018A&A...610A..82E, 2020A&A...641A.172E, 2021A&A...649A..51E}.
According to \citet{2003MNRAS.341...33K}, the value $D_n(4000) = 1.75$ corresponds to a mean age of about $4$~Gyr (for Solar metallicity) or older 
(for lower metallicities). 
Values $D_n(4000) > 2.0$ approximately correspond to a stellar age of $10$~Gyr
\citep{2003MNRAS.341...33K}, that is, 
such galaxies stopped forming stars at redshifts $z \approx 1.5 - 2$.
In the present study, we apply the limit $D_n(4000) = 1.75,$ above which we define galaxies 
as having very old stellar populations (VO galaxies).

\begin{figure*}
\centering
\resizebox{0.23\textwidth}{!}{\includegraphics[angle=0]{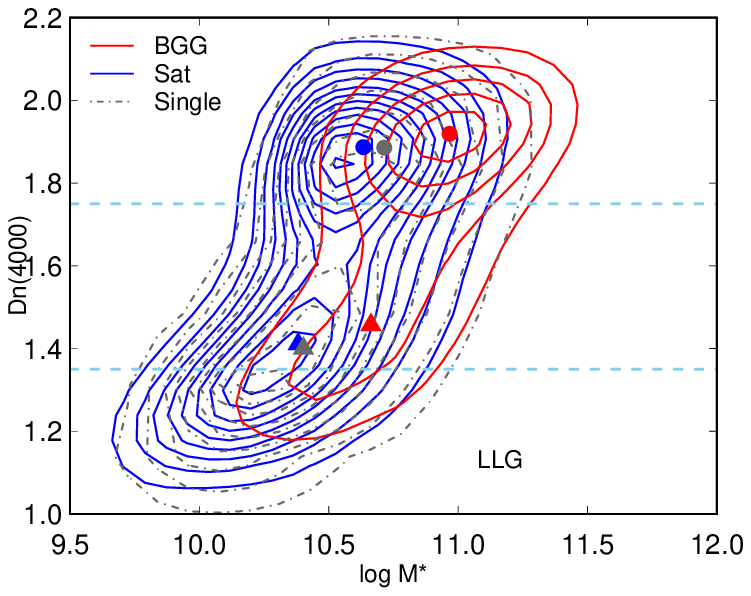}}
\resizebox{0.23\textwidth}{!}{\includegraphics[angle=0]{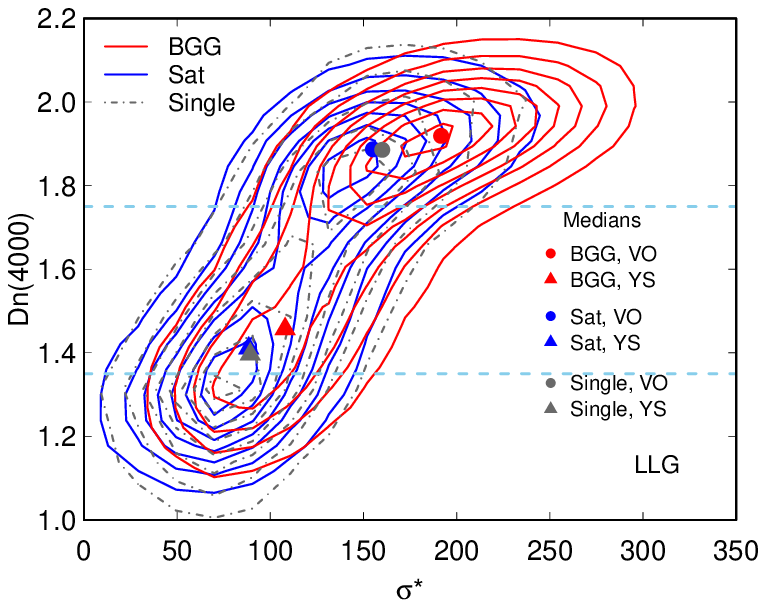}}
\resizebox{0.23\textwidth}{!}{\includegraphics[angle=0]{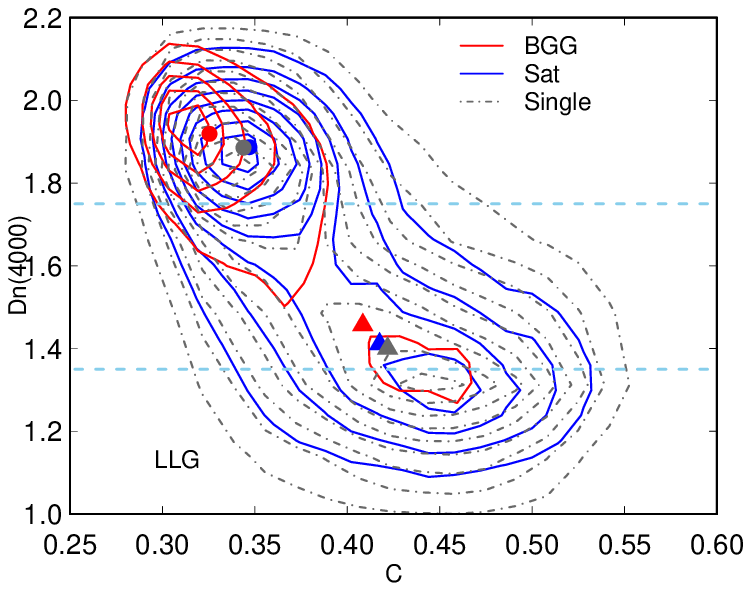}}
\resizebox{0.23\textwidth}{!}{\includegraphics[angle=0]{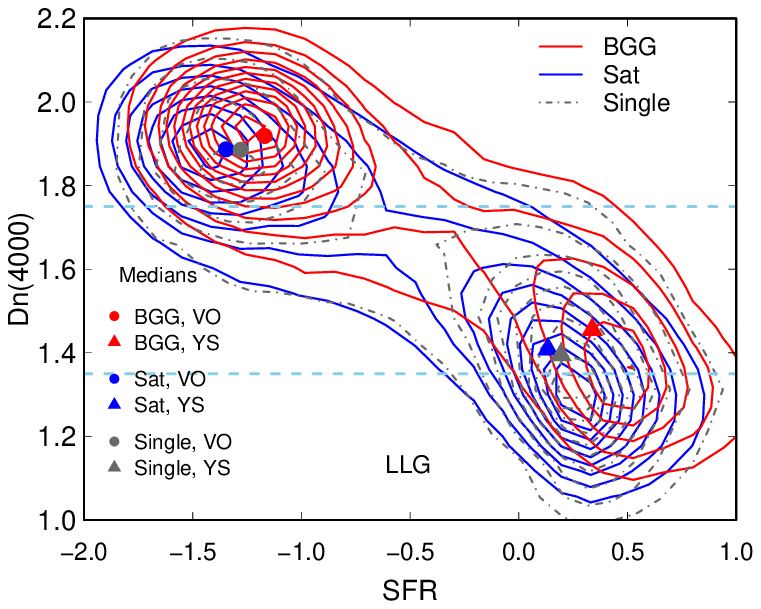}}

\resizebox{0.23\textwidth}{!}{\includegraphics[angle=0]{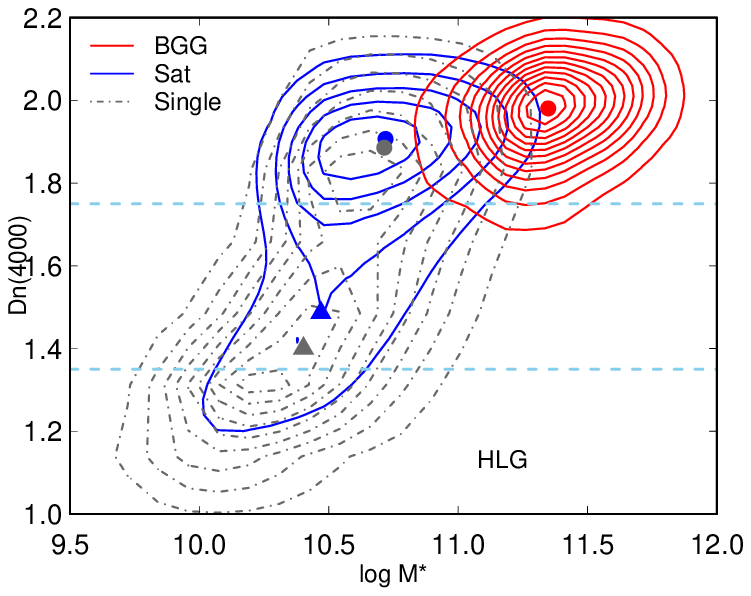}}
\resizebox{0.23\textwidth}{!}{\includegraphics[angle=0]{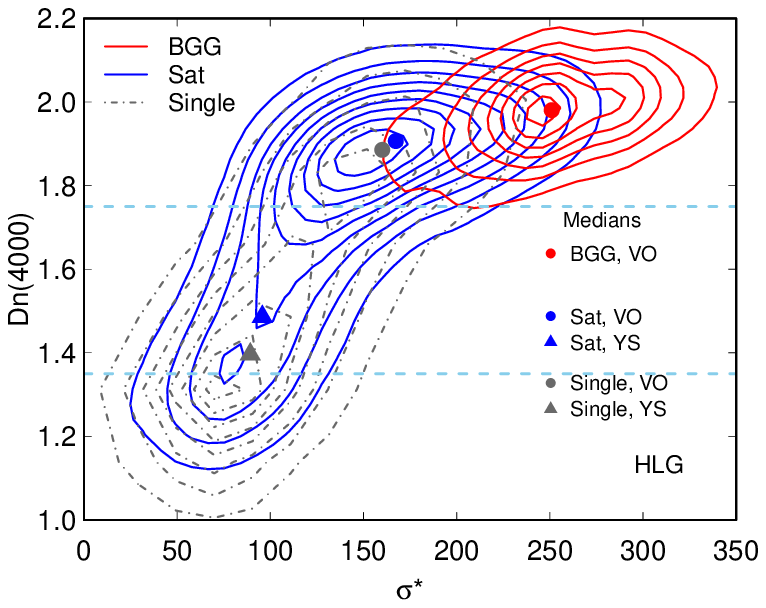}}
\resizebox{0.23\textwidth}{!}{\includegraphics[angle=0]{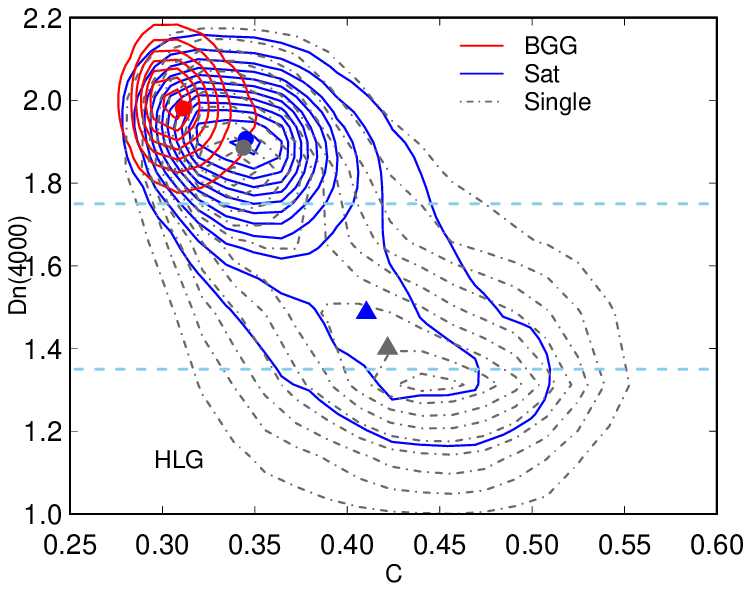}}
\resizebox{0.23\textwidth}{!}{\includegraphics[angle=0]{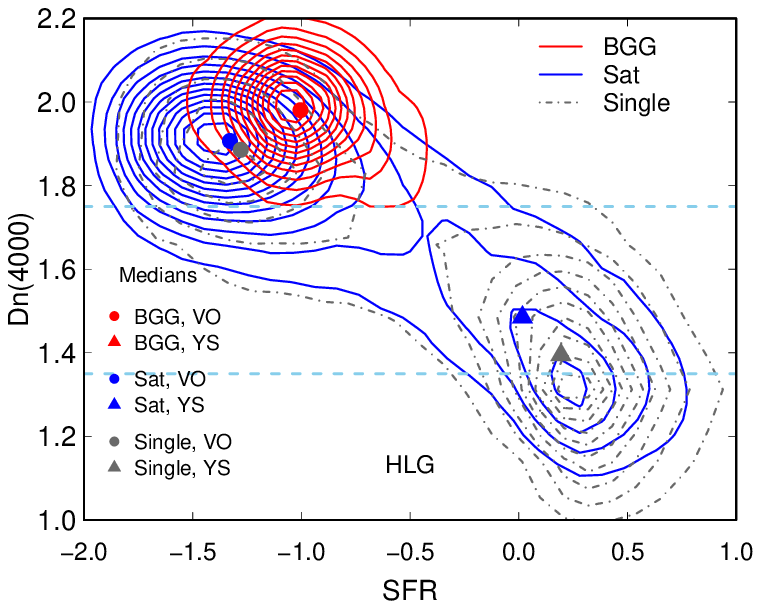}}
\caption{
Stellar mass $\log M^\star$ (leftmost panels), stellar velocity dispersions  $\sigma^{\mathrm{*}}$
(left central panels), concentration index of galaxies $C$ (right central panels),
and $\log \mathrm{SFR}$ (rightmost panels)
versus $D_n(4000)$ index 
for galaxies in LLGs (upper row), and 
in HLGs (lower row).
Filled circles and triangles show median values of galaxy properties 
for VO (circles) and YS (triangles) galaxies, as described in the legend in $\sigma^{\mathrm{*}}$
and in  $log \mathrm{SFR}$ panels. 
Horizontal lines show the values
of $D_n(4000)$ index $D_n(4000) = 1.35$ and  $D_n(4000) = 1.75$.
}
\label{fig:dn4galpara}
\end{figure*}

To separate galaxies with old and young
stellar populations  the limit $D_n(4000) = 1.55$ can be used; this limit approximately corresponds to 
a mean age of  stellar populations of a galaxy  about $1.5$~Gyr
\citep{2003MNRAS.341...33K}. 
Galaxies with $D_n(4000) > 1.55$ are often referred to as passive or red and dead.  
We use the term VO to emphasise the use of a higher $D_n(4000)$ index value,
$D_n(4000) = 1.75$.  
Below the value $D_n(4000) = 1.35,$  galaxies are considered to have very young stellar populations
with a mean age of approximately  $0.6$~Gyr only \citep{2003MNRAS.341...33K}.
\citet{2020A&A...641A.172E} and \citet{2021A&A...649A..51E} showed that in
the $D_n(4000)$--stellar mass plot, galaxies with $D_n(4000)$ index
in the interval of $1.35 < D_n(4000) \leq 1.75$ belong to mixed populations, 
containing quenched galaxies with no active star formation,
red and blue
star forming galaxies, and recently quenched galaxies (those with young stellar
populations but with no active star formation). We analysed the trends with global
density for each of these populations, and because these were similar 
 for these populations, we show the results
for them taken together. Therefore, in what follows, we
analyse two populations of galaxies: VO galaxies with
$D_n(4000) \geq 1.75$, and all other galaxies with $D_n(4000) < 1.75$ taken together
as galaxies with young stellar populations (hereafter YS galaxies).  

Figure~\ref{fig:dn4gr} shows the distribution of the $D_n(4000)$ index for single galaxies,
and for satellites galaxies and BGGs in low- and high-luminosity groups.
This figure already shows important difference between BGGs of HLGs, and all other galaxy classes:
almost all of them are VO galaxies, and there are very few  YS galaxies among them. 
The percentage of VO galaxies among other galaxy classes 
increases from 31\%  of single galaxies to 61\% of satellites of HLGs; 
see Table~\ref{tab:ngalgr}.

The 
$D_n(4000)$ index can be used as an indicator of the  SFRs in galaxies.
We also analyse the $\mathrm{SFR}$ of galaxies directly. 
Galaxies with $\log \mathrm{SFR} \geq -0.5$ are considered as star forming,
and those with $\log \mathrm{SFR} < -0.5$ are passive, quenched galaxies.

Earlier studies showed that the concentration index of galaxies $C$ 
is related to the galaxy structure parameters.
Early-type elliptical galaxies have lower values of  $C$ 
than 
disk-dominated late-type galaxies \citep{2005AJ....130.1535G, 2020A&A...641A.172E}.
\citet{2001AJ....122.1861S} showed that the value $C \approx 0.38$ separates
early- from late-type galaxies.

The stellar velocity dispersion of galaxies, $\sigma^{\mathrm{*}}$, 
also correlates with other galaxy properties.
Typically, star forming galaxies (i.e. the YS galaxies in the present study) 
have a lower stellar velocity dispersion
$\sigma^{\mathrm{*}}$ than quiescent 
galaxies (i.e. the VO galaxies here).
The stellar velocity dispersion of a galaxy 
correlates with the mass of its supermassive black hole 
\citep[see][for details and references]{2012ApJ...760...62B, 2020A&A...641A.172E}.
Observations show that 
at the high end of the stellar velocity dispersion, $\sigma^{\mathrm{*}} > 300~\mathrm{km/s}$,
 $\sigma^{\mathrm{*}}$ has changed 
very little since $z \approx 1.5,$ suggesting that
the galaxies with the most massive black holes have almost not changed over the last $10$~Gyr
\citep{2012ApJ...760...62B}.

\subsection{Abbreviations}
To make the paper easier to follow, below we list some of the abbreviations 
introduced above. They are used in our paper to characterise group membership
and the star formation properties of galaxies.

\begin{itemize}
\item VO = galaxy with very old stellar populations with $D_n(4000) \geq 1.75$;
\item YS = young, star-forming galaxies with $D_n(4000) < 1.75$ (this is mixed populations,
and includes quenched galaxies with no active star formation, red and blue
star forming galaxies, and recently quenched galaxies); 

\item BGG = brightest group galaxies;
\item Sat = satellite galaxies in groups;
\item Single = galaxies which do not belong to any group;

\item LLGs = low-luminosity groups with luminosity limits
$1.8\times10^{10} h^{-2} \mathrm{L}_{\sun} \leq L_{gr} \leq 15\times10^{10} h^{-2} \mathrm{L}_{\sun}$;
\item HLGs = high-luminosity groups with $L_{gr} > 15\times10^{10} h^{-2} \mathrm{L}_{\sun}$;
\item CWD = cosmic web detachment. 
\end{itemize}

\section{Results}
\label{sect:results}  

In this section, we present our main results. We start with our 
analysis of  the morphological data of single galaxies, satellites,
and BGGs.
Thereafter, we
analyse galaxy and group content in various global density environments
using data on the group membership of galaxies and the $D_n(4000)$ index.
The section ends with a comparison of other morphological properties
of galaxies with old and young stellar populations 
in various global density regions.

\subsection{Morphological properties of single galaxies, satellites,
and BGGs}
\label{sect:galprop}  

We plot stellar mass $\log M^{\mathrm{*}}$, 
stellar velocity dispersion  $\sigma^{\mathrm{*}}$, $SFR$, 
and  $C$ of galaxies versus
$D_n(4000)$ index in Fig.~\ref{fig:dn4galpara}.  
In each panel, we plot satellite galaxies, BGGs, and single galaxies 
separately. Single galaxies in both rows are the same, they are plotted
for comparison.
Figure~\ref{fig:dn4galpara} also shows  $D_n(4000)$ index limits $D_n(4000) = 1.75$, 
and  $D_n(4000) = 1.35$. 
Table~\ref{tab:galpop} presents the median values of galaxy properties in each population.
Errors are very small, typically less than $0.1$~\%, and are 
not given so as to avoid overcrowding the table.

A bimodality in galaxy morphological properties is clearly seen in Fig. 6
as a `red cloud' above  $D_n(4000)\approx  1.55$ and $\log \mathrm{SFR} < -0.5$, 
and `blue cloud' below this limit, around $\log \mathrm{SFR} \geq -0.5$.
The
$D_n(4000)$ index interval $1.35 - 1.75$ (`green valley') covers galaxies from mixed populations,
including galaxies in transformation (red star forming galaxies at the high-stellar-mass end
in this $D_n(4000)$ interval, and recently
quenched galaxies at the low-stellar-mass end). 
These galaxies are mostly of late type, although  there are also early-type galaxies 
among them, as suggested by their concentration index in the right
panels. This agrees with the recent results of \citet{2022arXiv220604707Q}, who show that
galaxies in the green valley are mostly of late type.
We also note that a small fraction of galaxies have
$D_n(4000)\geq  1.75$ and $\log \mathrm{SFR} \geq -0.5$. These are VO galaxies that are
still forming stars.
Such galaxies form 2\% 
of all galaxy populations (single galaxies, and galaxies in LLGs and HLGs).
This very small percentage tells us that the high value of $D_n(4000)$ index, $D_n(4000)= 1.75,$
can be used to separate quenched and star forming galaxies.

Let us now analyse what Fig.~\ref{fig:dn4galpara} tells us about galaxy populations in LLGs (upper row) 
and single galaxies. 
To begin with, we reiterate the fact that  BGGs were selected on the basis of galaxy luminosity, being the most luminous galaxies in a group. In what follows, we compare 
other properties of galaxies. Figure~\ref{fig:dn4galpara} shows that
in the case of 
both the red and blue cloud galaxies, 
BGGs occupy areas of higher stellar mass and velocity dispersion.
The concentration indices of BGGs are somewhat lower, meaning that their morphological types are slightly earlier compared to satellites and single galaxies.
BGGs avoid the area of lowest stellar mass in the left panel. 
BGGs in LLGs also avoid the lowest values of $D_n(4000)$ index
although some of them have higher values of $\mathrm{SFR}$s than LLG satellites or single galaxies.
The median
values for BGGs in all panels indicate the same trend.
At the same time, the properties of satellites and single galaxies are very similar.
As mentioned above, single galaxies can be the BGGs of faint groups 
of which the other member galaxies are too faint to be included in the SDSS spectroscopic sample.
They can also be outer satellites of groups, which could explain the similarity of their
properties. 
We discuss the possible connection between single galaxies and satellites further in 
Sect.~\ref{sect:waterq}.

In HLGs (lower row in  Fig.~\ref{fig:dn4galpara}), BGGs  clearly differ in their 
star formation properties from other galaxies. 
More than $90$\% of  these are
VO galaxies with 
$D_n(4000) \geq 1.75$, although even among  these  approximately
$2$~\% of galaxies have $\log \mathrm{SFR} \geq -0.5$. 
HLG BGGs have  much higher stellar masses than 
satellite galaxies, and also higher stellar velocity distributions and lower
concentration indexes than satellites galaxies. 
According to concentration index, they have clearly earlier morphological types than satellites or single galaxies. 
Galaxies with $\sigma^{\mathrm{*}} > 300~\mathrm{km/s}$ in our sample are all BGGs  of HLGs.
These galaxies have $D_n(4000) > 1.85$, 
and therefore both their $D_n(4000)$ index and $\sigma^{\mathrm{*}}$
values suggest that they stopped forming stars 
and have not changed  much over the last $10$~Gyr.

Figure~\ref{fig:dn4galpara} reveals that the properties of 
satellite galaxies in HLGs with $D_n(4000) \geq 1.75$ are close to those
of VO single galaxies, and the corresponding median values of galaxy properties almost overlap. At the  lower $D_n(4000)$ index end (YS population), satellites of HLGs
have, on average, higher stellar masses and stellar velocity dispersions
and lower concentration index than single galaxies. 
HLG satellites also avoid the lowest end of the $D_n(4000)$ index values,
meaning that their stellar populations are, on average, older than
the stellar populations of YS single galaxies. Nevertheless, it is interesting that
the differences between the properties of HLG satellites and single galaxies 
are rather small.

Our comparison of the galaxy parameters of single galaxies, satellites, and BGGs
of LLGs and HLGs in   Fig.~\ref{fig:dn4galpara} suggests that
galaxy population type is the major parameter determining the
quenching status of galaxies. The  difference between the fractions of BGGs
and satellite galaxies is the largest in HLGs, and is also
fairly large in LLGs, as seen in
Table~\ref{tab:ngalgr}. 

\begin{figure}
\centering
\resizebox{0.44\textwidth}{!}{\includegraphics[angle=0]{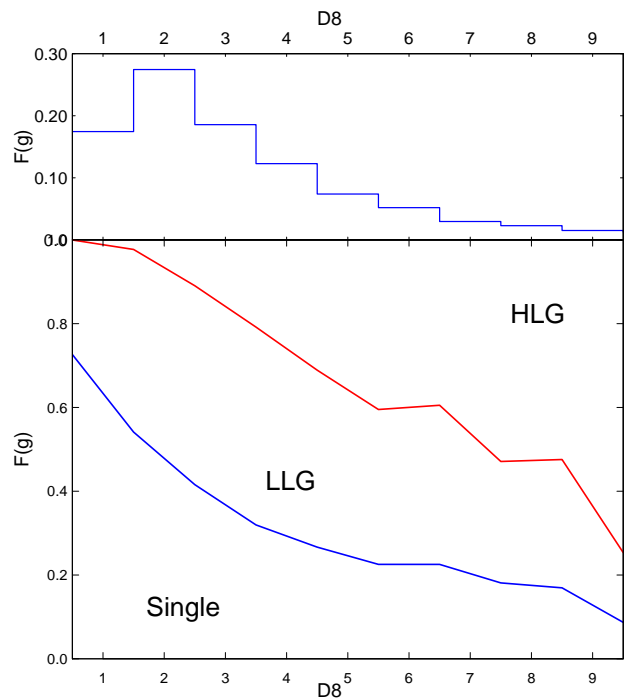}}
\caption{
Percentage of galaxies $F(g)$ from all galaxies in various global density 
$D8$ intervals (upper panel), and the 
percentage of single galaxies, LLG members, and HLG members 
in various global density  $D8$ intervals (lower panel).
}
\label{fig:frac}
\end{figure}

\begin{figure*}
\centering
\resizebox{0.32\textwidth}{!}{\includegraphics[angle=0]{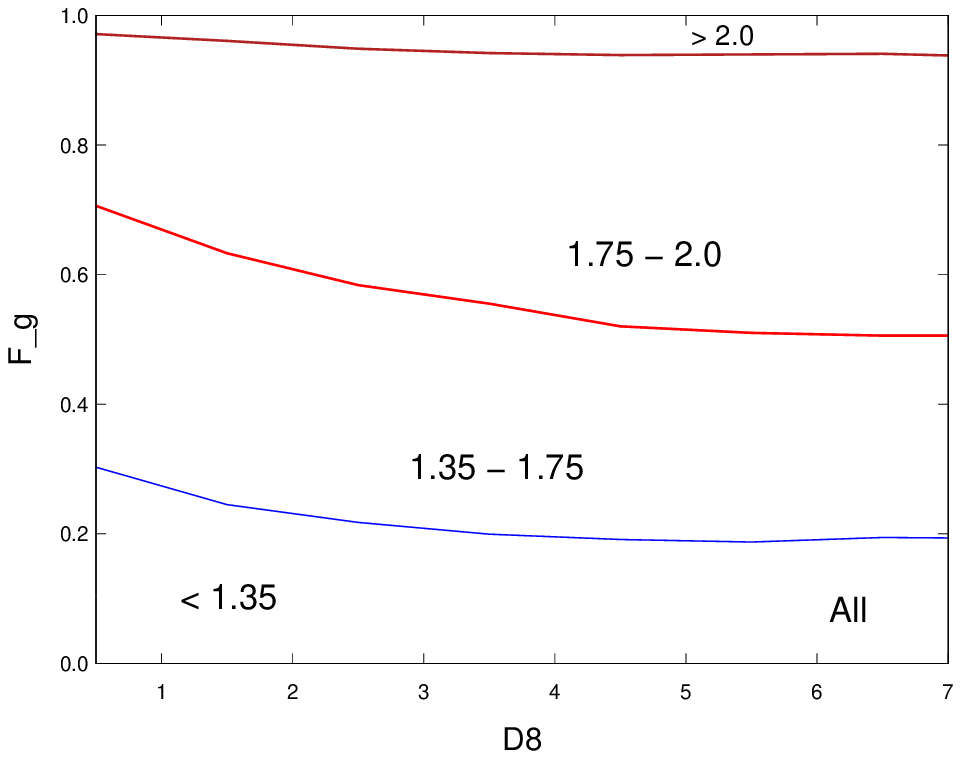}}
\resizebox{0.32\textwidth}{!}{\includegraphics[angle=0]{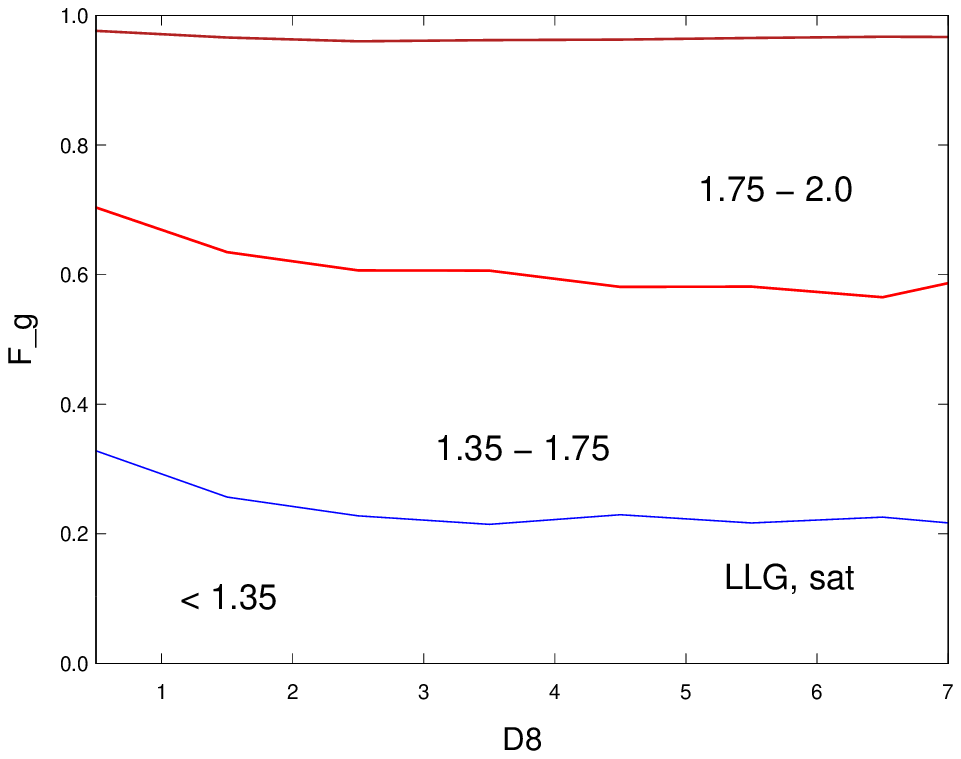}}
\resizebox{0.32\textwidth}{!}{\includegraphics[angle=0]{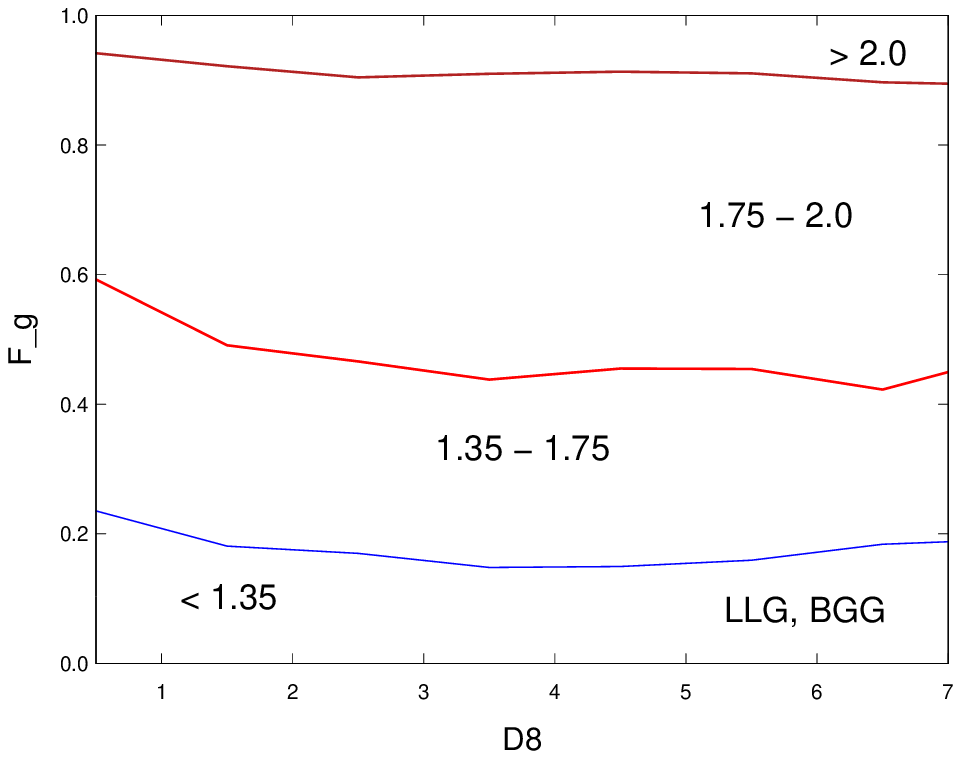}}

\resizebox{0.32\textwidth}{!}{\includegraphics[angle=0]{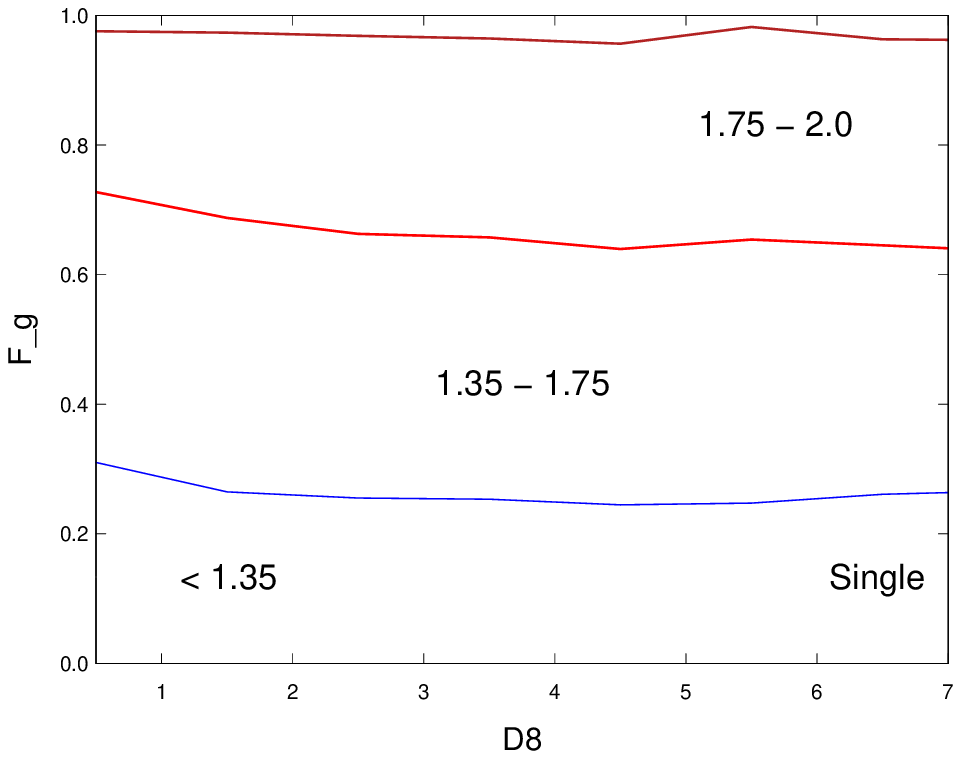}}
\resizebox{0.32\textwidth}{!}{\includegraphics[angle=0]{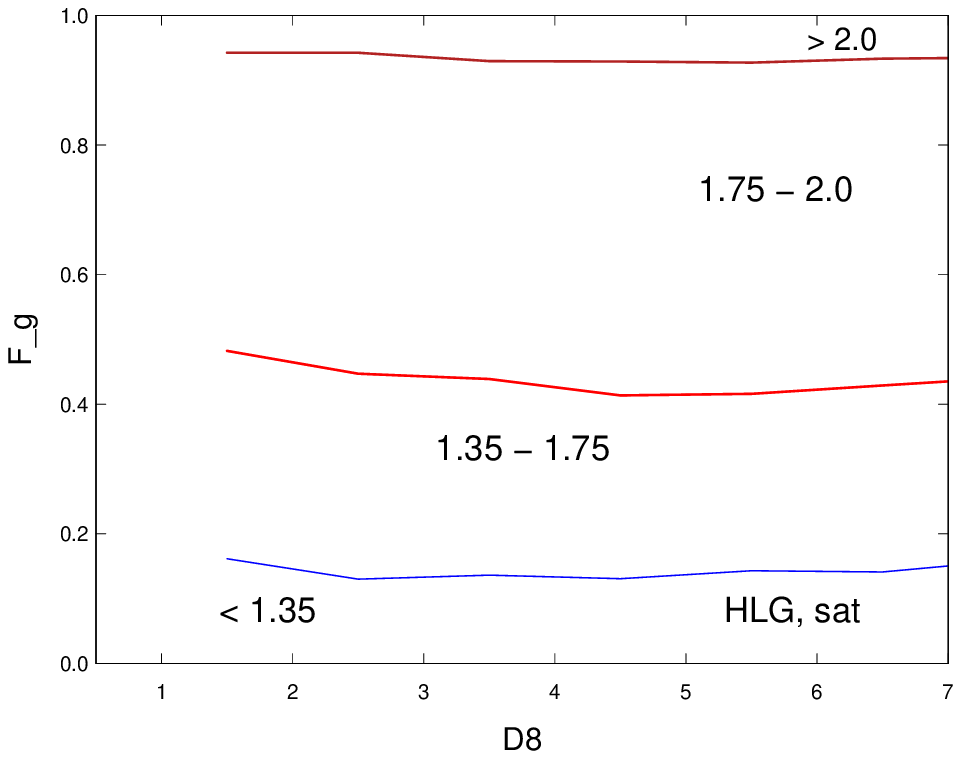}}
\resizebox{0.32\textwidth}{!}{\includegraphics[angle=0]{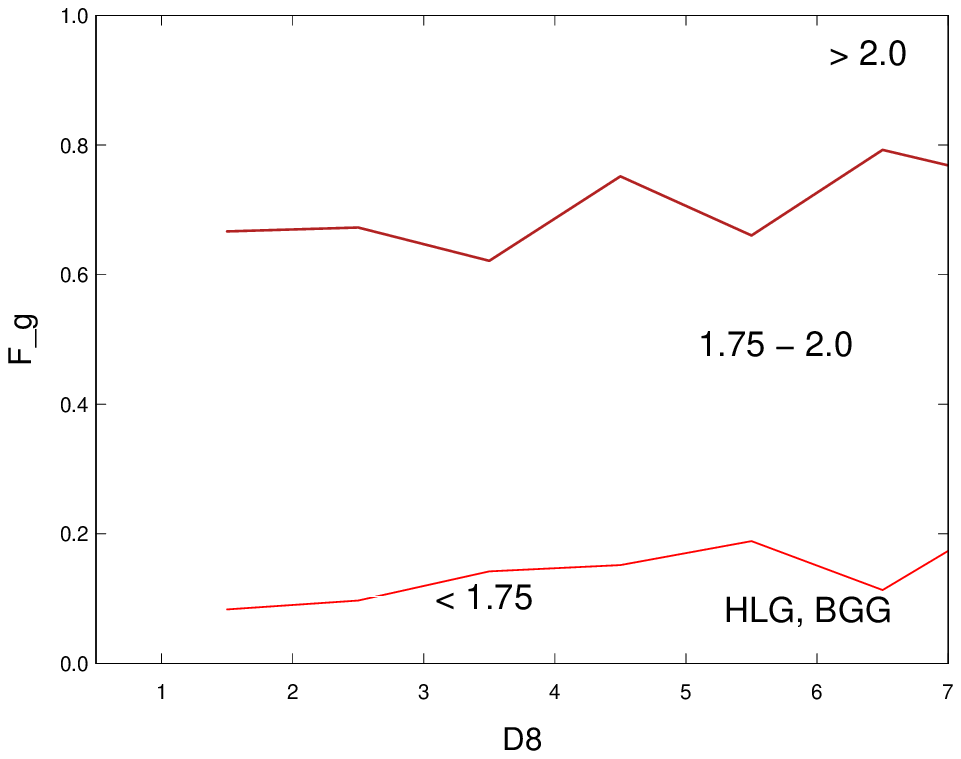}}
\caption{
Galaxy content of various global density $D8$ regions
according to the $D_n(4000)$ index in the following intervals:
$D_n(4000) \leq 1.35$, $1.35 < D_n(4000) \leq 1.75$, $1.75 < D_n(4000) \leq 2.0$,
and $D_n(4000) > 2.0$.
Upper left panel: All galaxies. Upper middle panel: LLG
satellite galaxies. Upper right panel: LLG BGGs.
Lower left panel: Single galaxies. Lower middle panel: HLG satellite galaxies. Lower right panel: HLG BBGs.
}
\label{fig:dn4frac}
\end{figure*}

\begin{figure*}
\centering
\resizebox{0.32\textwidth}{!}{\includegraphics[angle=0]{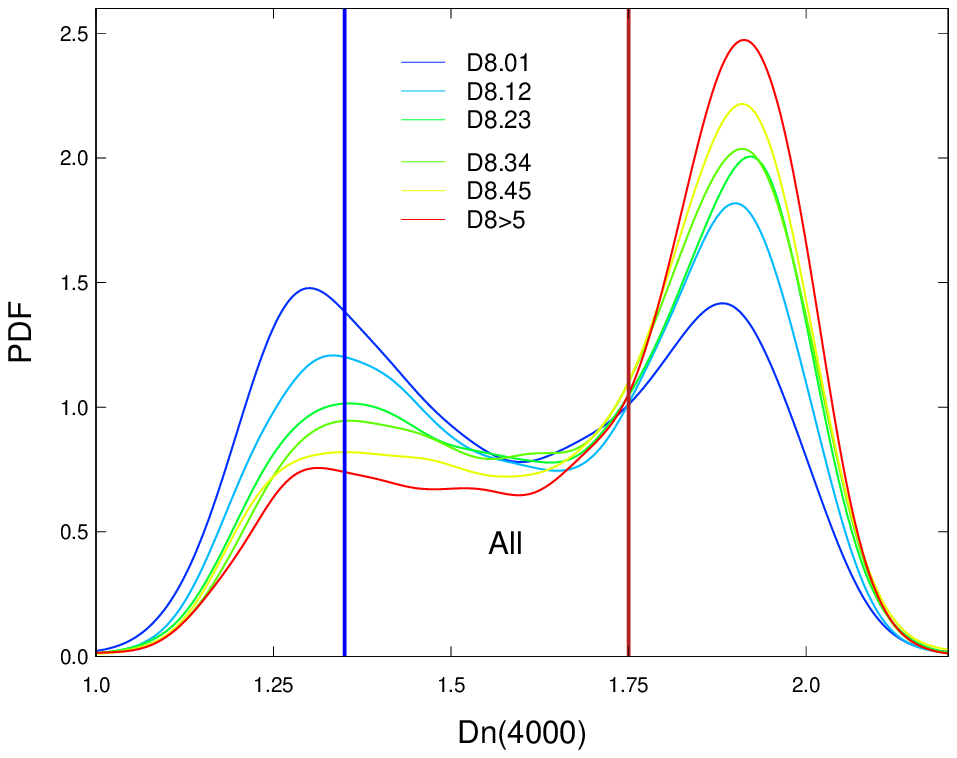}}
\resizebox{0.32\textwidth}{!}{\includegraphics[angle=0]{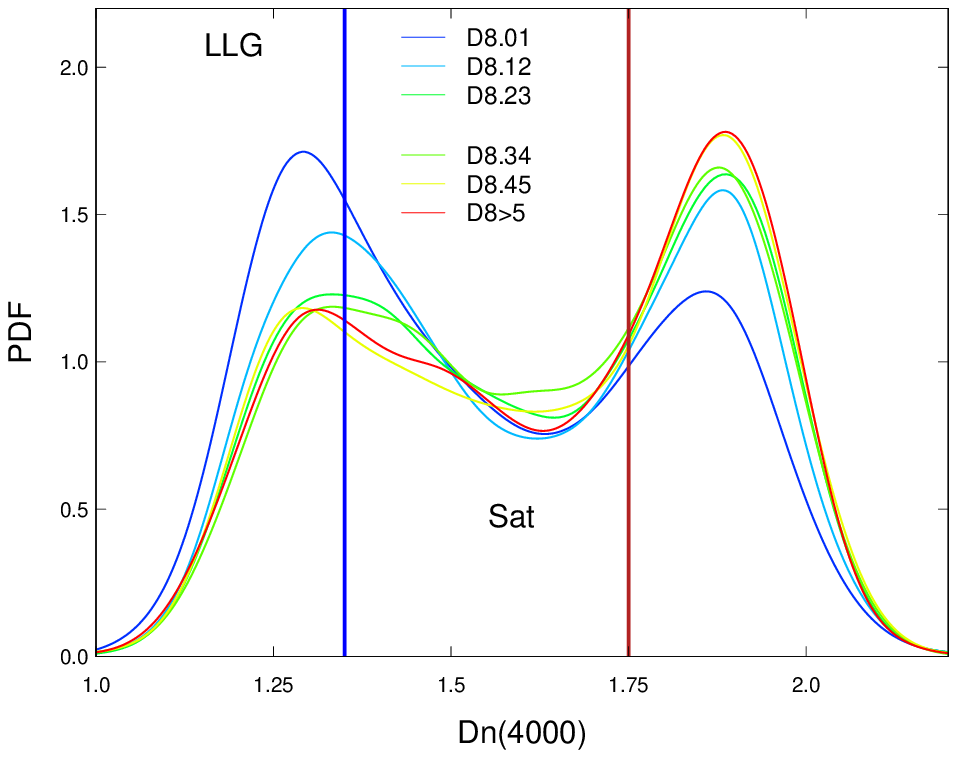}}
\resizebox{0.32\textwidth}{!}{\includegraphics[angle=0]{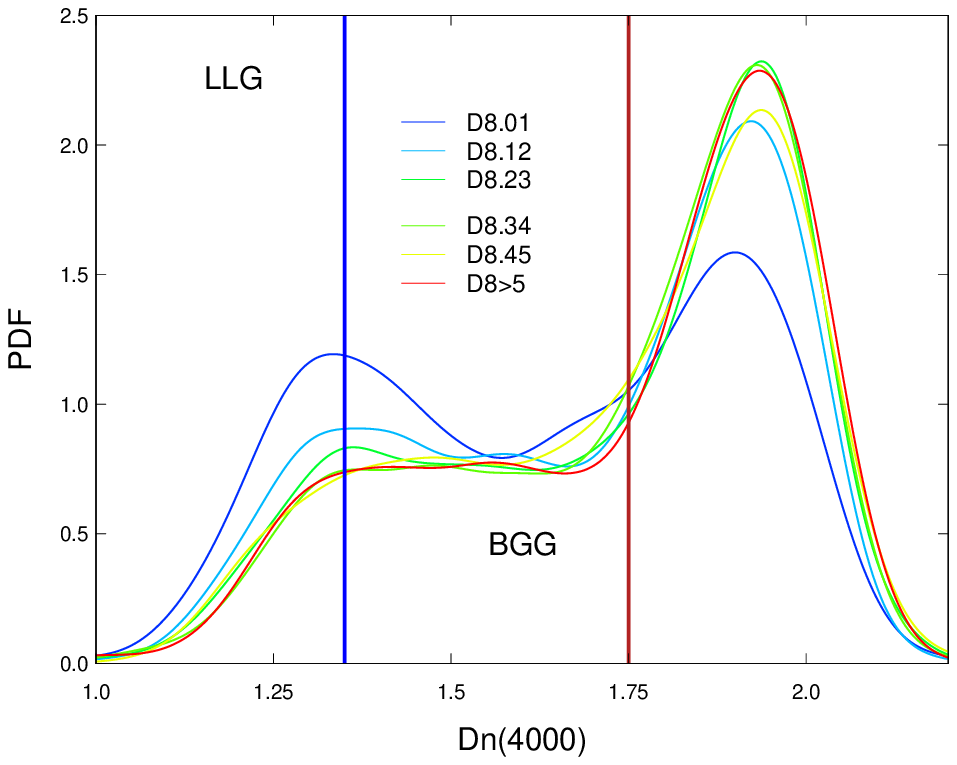}}

\resizebox{0.32\textwidth}{!}{\includegraphics[angle=0]{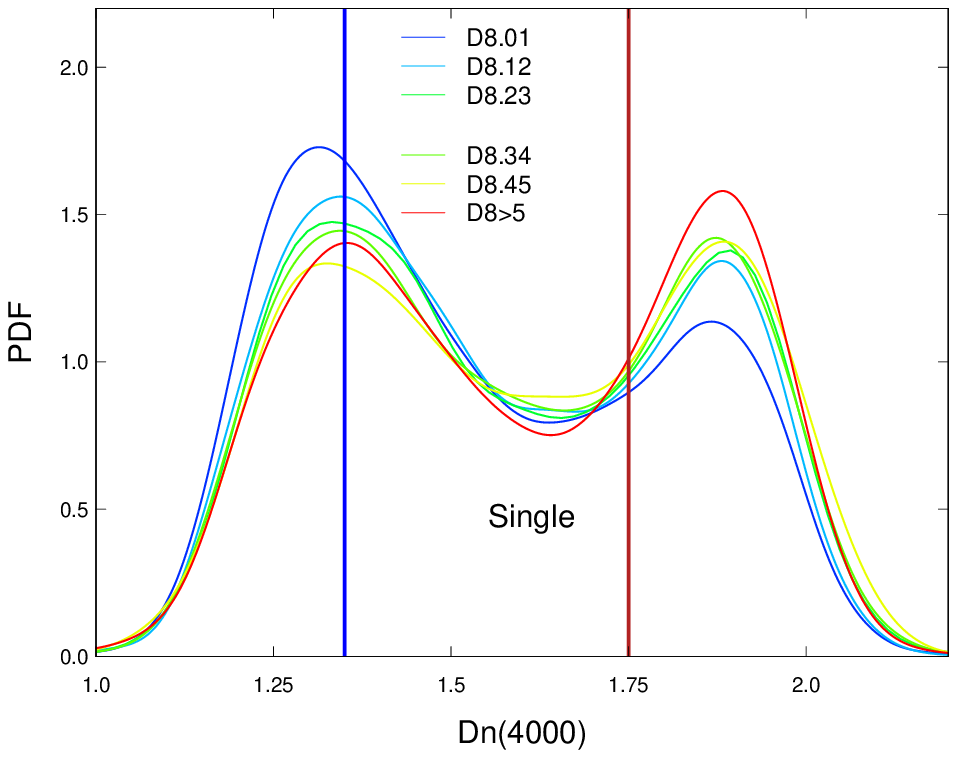}}
\resizebox{0.32\textwidth}{!}{\includegraphics[angle=0]{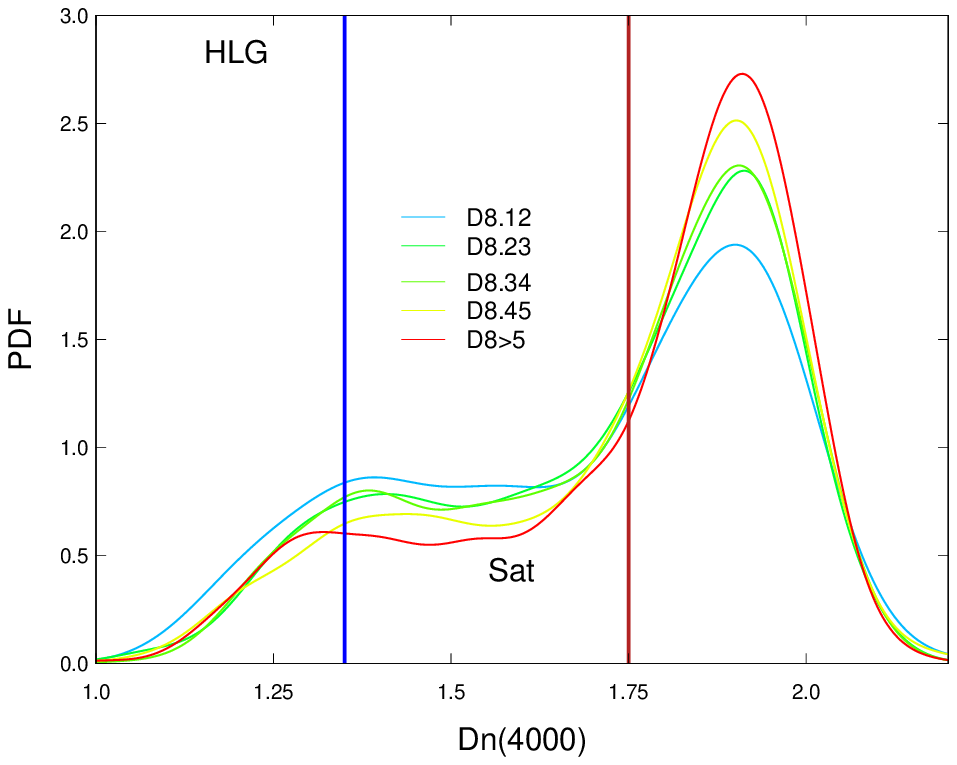}}
\resizebox{0.32\textwidth}{!}{\includegraphics[angle=0]{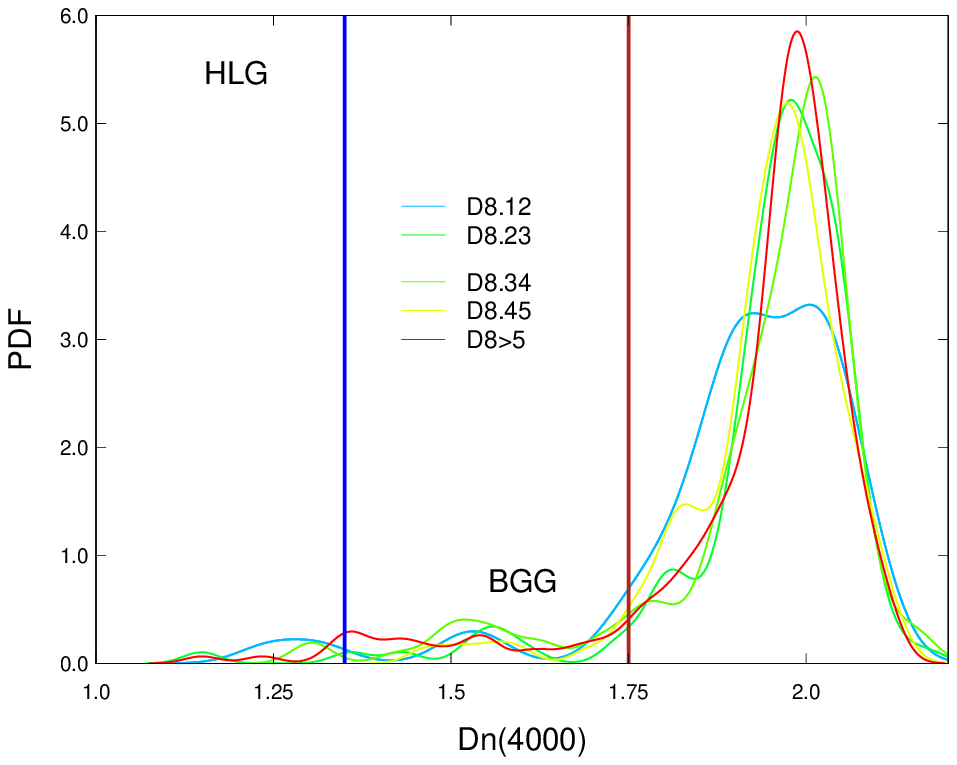}}
\caption{
Distribution of $D_n(4000)$ index of galaxies in various global density $D8$ regions.
Upper left panel: all galaxies, upper middle panel: LLG (low-luminosity groups)
satellite galaxies, upper right panel - LLG BGGs.
Lower left panel: single galaxies, lower middle panel: HLG 
(high-luminosity groups) satellite galaxies, and lower right panel: HLG BBGs.
Vertical lines show the values
of $D_n(4000)$ index $D_n(4000) = 1.35$, and  $D_n(4000) = 1.75$.
}
\label{fig:dn4d8}
\end{figure*}

\begin{table*}[ht]
\centering
\caption{Galaxy populations in various global density intervals.}
\begin{tabular}{rrrrrrrrrrrr} 
\hline\hline  
(1)&(2)&(3)&(4)&(5)&(6)&(7)&(8)&(9)&(10)&(11)&(12)\\      
\hline 
D8 limits &  $N_{\mathrm{gal}}$ &$F_{gal}$ & $F_{1}$ & $F_{LLG}$ & $F_{HLG}$ &
 $F^{\mathrm{VO}}_{all}$ &$F^{\mathrm{VO}}_{1}$ & \multicolumn{2}{c}{$F^{\mathrm{VO}}_{LLG}$}  
 & \multicolumn{2}{c}{$F^{\mathrm{VO}}_{HLG}$}   \\
&&&&&& &&Sat&BGG&Sat&BGG\\
\hline                                                    
 $  < 1$ & 14226 &0.17 & 0.73 &0.27&  0 &0.30 &0.27 & 0.30 & 0.42 &  0    & 0     \\
 $1 - 2$ & 22378 &0.27 & 0.54 &0.44&0.02&0.37 &0.31 & 0.37 & 0.53 &  0.53 & 0.92  \\
 $2 - 3$ & 15135 &0.19 & 0.42 &0.47&0.11&0.43 &0.33 & 0.40 & 0.56 &  0.57 & 0.93 \\
 $3 - 4$ & 10014 &0.12 & 0.32 &0.47&0.21&0.46 &0.34 & 0.39 & 0.59 &  0.58 & 0.91 \\
 $4 - 5$ &  6025 &0.07 & 0.27 &0.42&0.31&0.50 &0.35 & 0.41 & 0.57 &  0.60 & 0.95 \\
 $  > 5$ & 13749 &0.17 & 0.17 &0.30&0.53&0.55 &0.36 & 0.42 & 0.59 &  0.62 & 0.92 \\
\hline                                        
\label{tab:ngalgr}  
\end{tabular}\\
\tablefoot{                                                                                 
Columns are as follows:
(1): Interval of global luminosity density $D8$, in units of mean density (see text);
(2): Number of galaxies in the given density interval
(3): Percentage of galaxies in the given density interval among all galaxies;
(4--6): Percentage of single galaxies $F_{1}$, and percentages of galaxies in low- and high-
luminosity groups ($F_{LLG}$ and $F_{HLG}$) in the given density interval;
(7): percentage of VO galaxies with $D_n(4000) \geq 1.75$ among all galaxies 
in the given density interval;
(8): percentage of VO galaxies with $D_n(4000) \geq 1.75$ among single galaxies 
in the given density interval;
(9--10): percentage of VO galaxies among satellite galaxies and BGGs of 
low-luminosity groups (LLGs) in the given density interval; 
(11--12): percentage of VO galaxies among satellite galaxies and BGGs of 
low-luminosity groups (HLGs) in the given density interval. 
}
\end{table*}

\begin{table*}[ht]
\centering
\caption{Median values of the  galaxy properties
for single galaxies, and galaxies in LLGs and HLGs}
\begin{tabular}{lrrrrrrrrrrrr} 
\hline\hline  
(1)&(2)&(3)&(4)&(5)&(6)&(7)&(8)&(9)&(10)&(11)&(12)&(13)\\      
\hline 
 $ID$ & $N_{\mathrm{gal}}$ &$F_{VO}$ &\multicolumn{2}{c}{$M^{\mathrm{*}}_{\mathrm{med}}$} & 
 \multicolumn{2}{c}{$\sigma^{\mathrm{*}}_{\mathrm{med}}$} &
 \multicolumn{2}{c}{$C_{\mathrm{med}}$} & 
  \multicolumn{2}{c}{$D_n(4000)_{\mathrm{med}}$} & 
  \multicolumn{2}{c}{$\mathrm{SFR}_{\mathrm{med}}$} \\
  &   &             & $VO$ & $YS$ & $VO$ & $YS$&   $VO$ & $YS$ & $VO$ & $YS$ & $VO$ & $YS$    \\
\hline
 $Single$   &35530&0.31& 10.71 & 10.40 & 160  &  89 & 0.34 & 0.42 & 1.89 & 1.40 & -1.28 & 0.19\\
\hline                                                                          
 $Sat_{LLG}$&19106&0.39& 10.63 & 10.38 & 155  &  88 & 0.35 & 0.42 & 1.89 & 1.41 & -1.35 & 0.13\\
 $BGG_{LLG}$&12133&0.54& 10.97 & 10.66 & 192  & 108 & 0.33 & 0.41 & 1.92 & 1.46 & -1.17 & 0.34\\
\hline                                                                          
 $Sat_{HLG}$&12062&0.61& 10.72 & 10.47 & 167  & 96  & 0.35 &0.41  & 1.91 & 1.49 & -1.33 & 0.02\\
 $BGG_{HLG}$& 732 &0.92& 11.35 & 11.12 & 251  &149  & 0.31 &0.40  & 1.98 & 1.53 & -1.00 & 0.34\\
 \hline
\label{tab:galpop}  
\end{tabular}\\
\tablefoot{                                                                                 
Columns are as follows:
(1): Population ID;
(2): Number of galaxies in a population;
(3): Percentage of VO galaxies in a population;
(4--5): Median values of the stellar mass $M^{\mathrm{*}}_{\mathrm{med}}$
for VO and YS galaxies;
(6--7): Median values of the stellar velocity dispersion $\sigma^{\mathrm{*}}$ 
for VO and YS galaxies;
(8--9): Median values of the concentration index $C$  
for VO and YS galaxies;
(10--11): Median values of the $D_n(4000)$ index $D_n(4000)_{\mathrm{med}}$
for VO and YS galaxies. 
(12--13): Median values of SFR index $\mathrm{SFR}_{\mathrm{med}}$
for VO and YS galaxies. 
}
\end{table*}

According to \citet{2014MNRAS.438.3465T}, approximately 35\%--40\%\  
of all galaxies are members of filaments. To find the fraction
of the galaxies in our sample that belong to filaments, we used data on galaxy
filaments from filament catalogues 
by \citet{2014MNRAS.438.3465T} and \citet{2016A&C....16...17T}.
Our preliminary analysis shows that approximately 50\% of the galaxies in LLGs are  
filament members, compared to 40\% of those in HLGs.
A comparison of the properties of the galaxies found in and outside of filaments shows that 
approximately $25$\% of filament member galaxies
have $D_n(4000) \geq 1.75$.  Comparison with the percentages
given in Table~\ref{tab:galpop} shows that the percentages of VO galaxies 
in filaments are lower than the percentages of VO galaxies among
any population in Table~\ref{tab:galpop}.
This difference may be related to the fact that only
some groups lie in filaments, and the percentage
of galaxies in groups with $D_n(4000) \geq 1.75$ is higher than the same percentage of galaxies not in groups. 
 A detailed analysis of galaxy quenching in filaments 
is beyond the scope of this paper, and merits a separate study.

\subsection{Group and galaxy  content of 
various global environments}
\label{sect:d1d8}  

Now we analyse the overall group and galaxy content of various global 
density environments. 
We first show the percentage of galaxies 
from total sample, and the  percentages of single
galaxies and member galaxies of LLGs and HLGs in various global density environments
(Fig.~\ref{fig:frac}, upper and lower panels). 
These percentages are presented also in Table~\ref{tab:ngalgr}.

First, Table~\ref{tab:ngalgr} and Fig.~\ref{fig:frac} show that the galaxy and group populations 
change significantly with global density. 
Single galaxies and poor, low-luminosity groups 
can be found everywhere in the cosmic web, from the regions of lowest global
density to superclusters. 
Figure~\ref{fig:grlumrich} demonstrates that, in accordance with the general understanding 
of the formation of groups in the cosmic web, groups in the lowest global density regions  
are poor and of low luminosity.  Figure~\ref{fig:frac} reveals that in watersheds, even the formation of 
poor groups is suppressed; watersheds 
are mostly populated by single galaxies (73\% of all galaxies). LLGs host 27\% of all galaxies
here. At higher global densities, at first the percentage of galaxies in LLGs increases
and that of single galaxies decreases. 
In global density interval $1 < D8 < 2,$ the percentage of single galaxies drops to $54$\%,
and $44$\% of galaxies are members of LLGs. Only $2$\% of all galaxies in this density interval
reside in HLGs.   
Then, as the global density increases, the 
percentage of galaxies in HLGs starts to grow. In the environments of highest global density,
more than 50\% of all galaxies lie in HLGs, and only 17\% of galaxies are single.

In order to investigate the distribution of galaxies with different star formation properties
among galaxies in a given global density interval, we show in Fig.~\ref{fig:dn4frac}
the percentage of galaxies from different $D_n(4000)$ index ranges versus global density
$D8$. 
Here, we also show  the percentage of galaxies with $D_n(4000) > 2.0$,
which approximately corresponds to an age of $10$~Gyr for the stellar populations therein 
\citep{2003MNRAS.341...33K}. This suggest that 
these galaxies may have stopped  forming stars at redshifts $z \approx 1.5 - 2$.
We mention that less than $0.1$~\% of these galaxies may still be forming stars,
as they have $\log \mathrm{SFR} \geq -0.5$.
Observations have found  quenched galaxies even at redshifts as high as $z = 3.37$
\citep{2022ApJ...926...37M}, and therefore the high fraction of quenched BGGs and other galaxies at redshifts
 $z \approx 1.5 - 2$ does not come as a surprise.
Additionally, Fig.~\ref{fig:dn4d8} shows the distribution of $D_n(4000)$ index
for single galaxies, satellites, and BGGs in various global density environments.
The percentages of VO galaxies among galaxies according to their
group membership are given in Table~\ref{tab:ngalgr}.

Figures~\ref{fig:dn4frac} and \ref{fig:dn4d8} and Table~\ref{tab:ngalgr}  show
that VO galaxies are found in  environments of all global densities.
Even in the watershed region,
30\% of all galaxies, and 27\% of single galaxies have very old stellar populations.
In contrast, in the highest global density regions (superclusters), $55$~\%
of all galaxies are with  very old stellar populations.
In all global density intervals, the percentage of VO galaxies 
is lower in LLGs than in HLGs (both among satellites and BGGs, Table~\ref{tab:ngalgr}). 
Among galaxies in LLGs,
29\% of satellite galaxies and 40\% of BGGs are VO galaxies in the $D8 < 1$ region. 
These percentages increase up to 40\% and 60\% in higher global density
regions. 
Interestingly, the percentage of VO galaxies among BGGs of LLGs
is almost unchanged over a wide global density range, $D8 > 2$, while
among satellite galaxies of LLGs there is an increasing percentage of VO
galaxies with increasing global density. It is also interesting to 
see that among single galaxies, which are the least affected by the influence of
other galaxies, the percentage of VO galaxies in superclusters is higher than
in lower global density regions (36\% versus 27\%). 
 
The percentage of VO galaxies among 
galaxies in HLGs is higher, spanning from 
53\% to 62\% among satellite galaxies in the lowest and highest global
density regions, and being larger than   92\% among BGGs. 
We note that YS BGGs 
in HLGs lie in the lowest luminosity end of this group population, 
and only two of them have $D8 > 5$ at their location.
In total, eight galaxies among BGGs of HLGs have $D_n(4000) < 1.35$.

In general, these trends  reflect the well-known large-scale morphology--density relation
\citep{1987MNRAS.226..543E}. However, there is an important difference.
Typically, densities around galaxies have been calculated using the distance to Nth 
nearest neighbour, as in \citet{1987MNRAS.226..543E}  and \citet{Bluck:2020we}. 
In the present study, we distinguish
between group membership and global densities using the luminosity--density field.
This enables us to show that in groups of all luminosities
and for single galaxies, galaxy  populations in various 
global density regions are different. 

\begin{figure*}
\centering
\resizebox{0.19\textwidth}{!}{\includegraphics[angle=0]{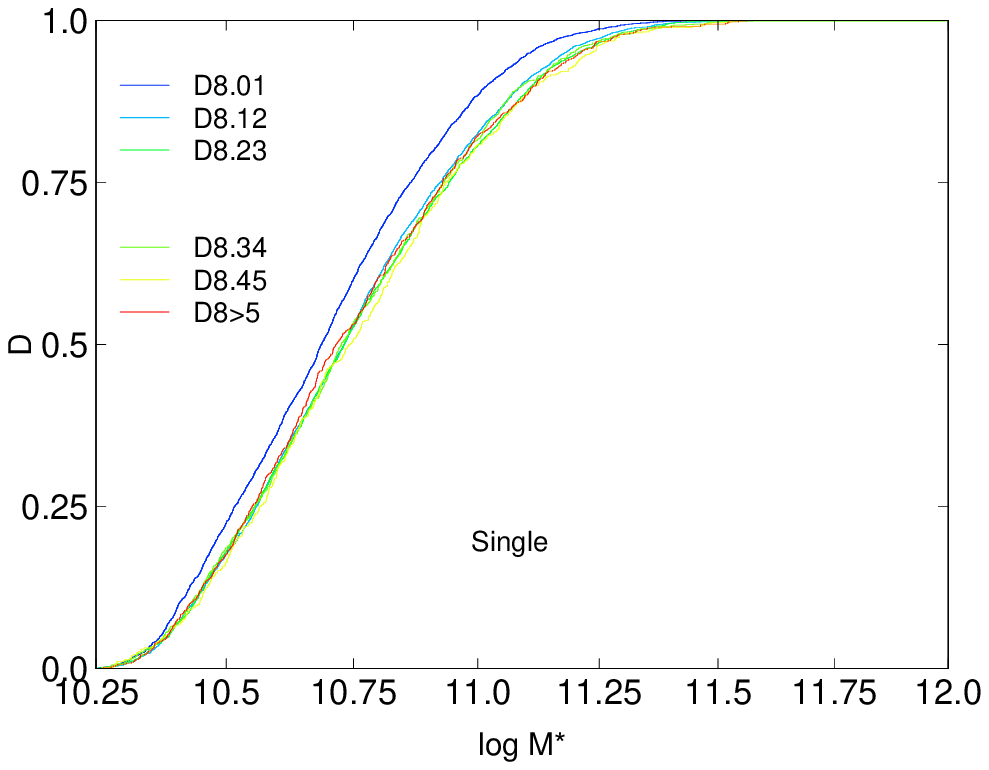}}
\resizebox{0.19\textwidth}{!}{\includegraphics[angle=0]{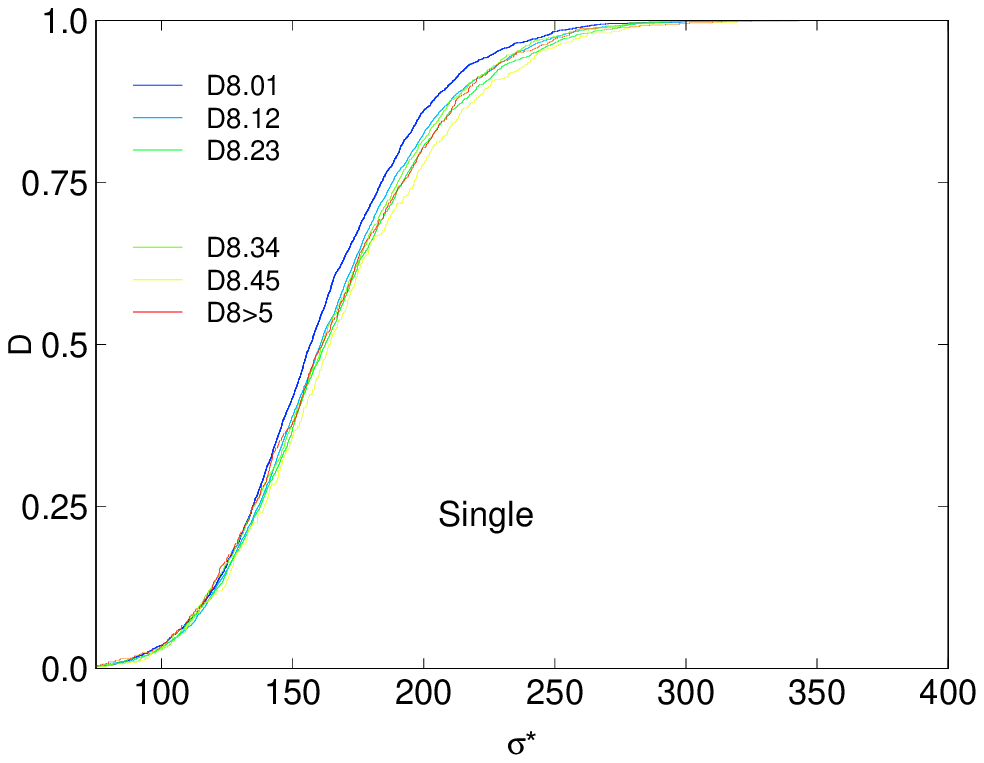}}
\resizebox{0.19\textwidth}{!}{\includegraphics[angle=0]{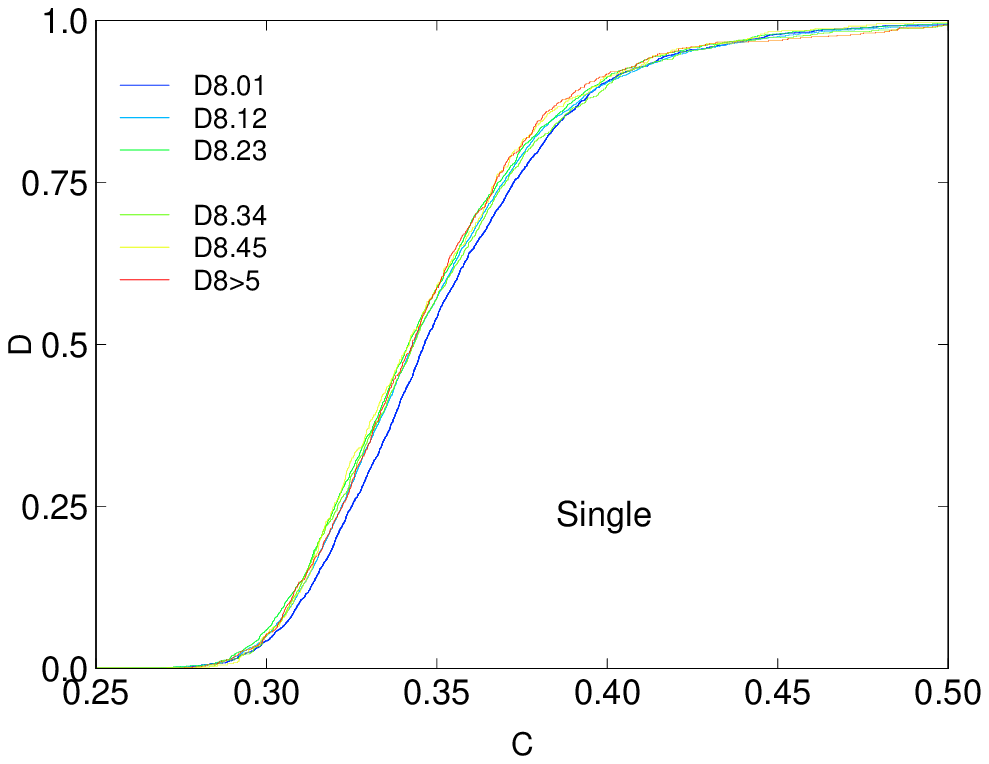}}
\resizebox{0.19\textwidth}{!}{\includegraphics[angle=0]{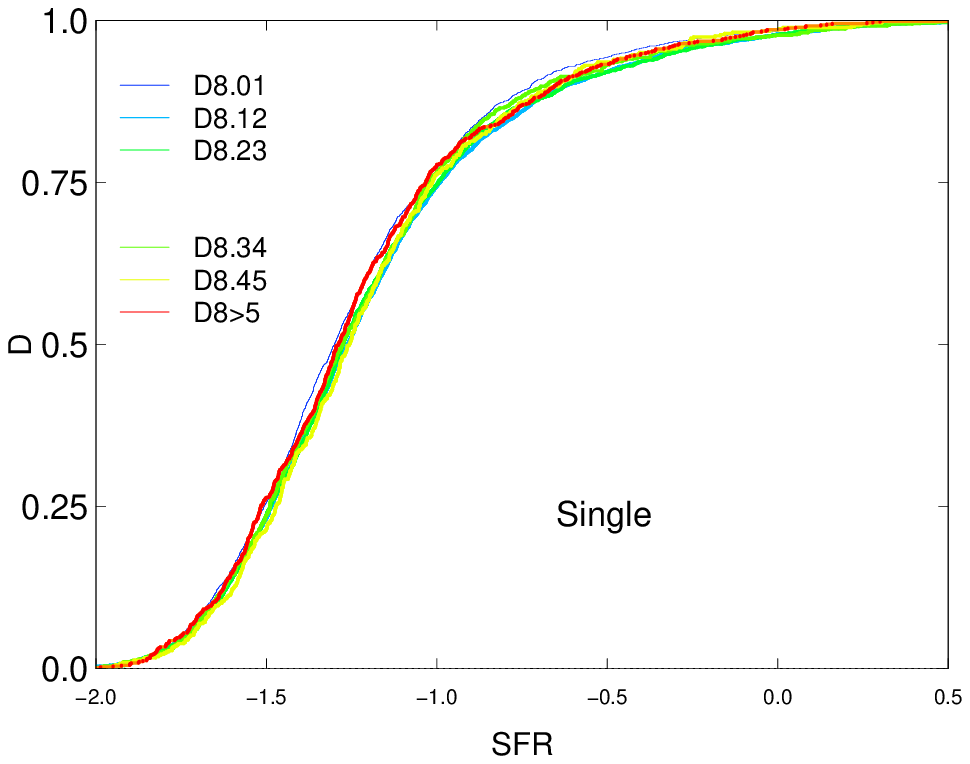}}
\resizebox{0.19\textwidth}{!}{\includegraphics[angle=0]{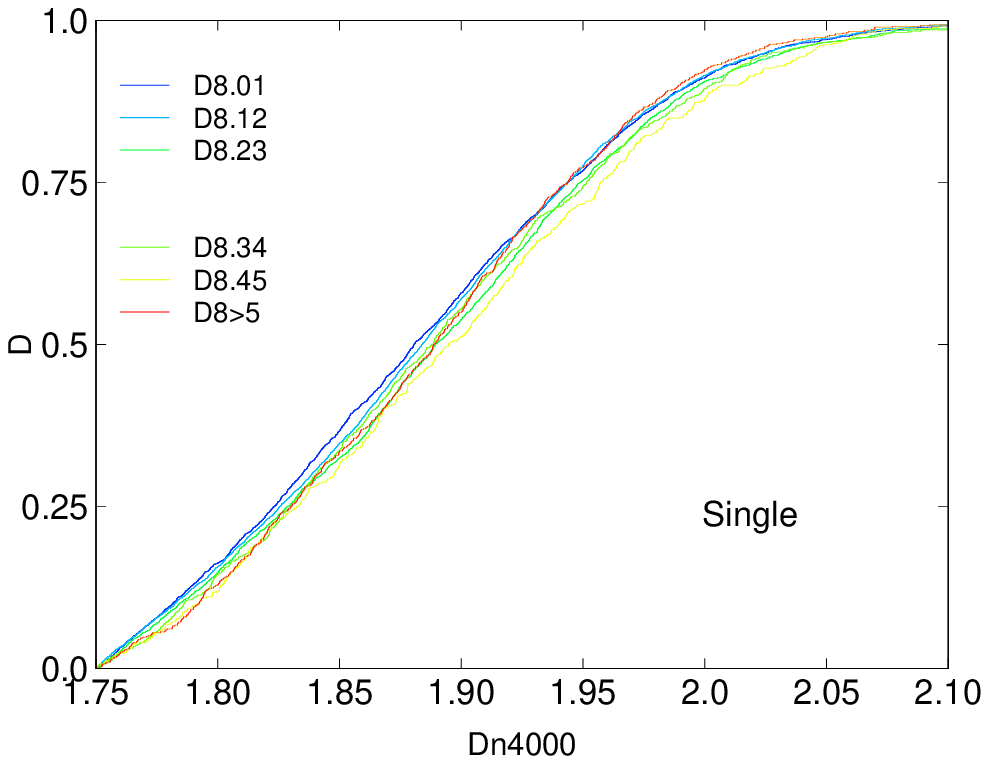}}

\resizebox{0.19\textwidth}{!}{\includegraphics[angle=0]{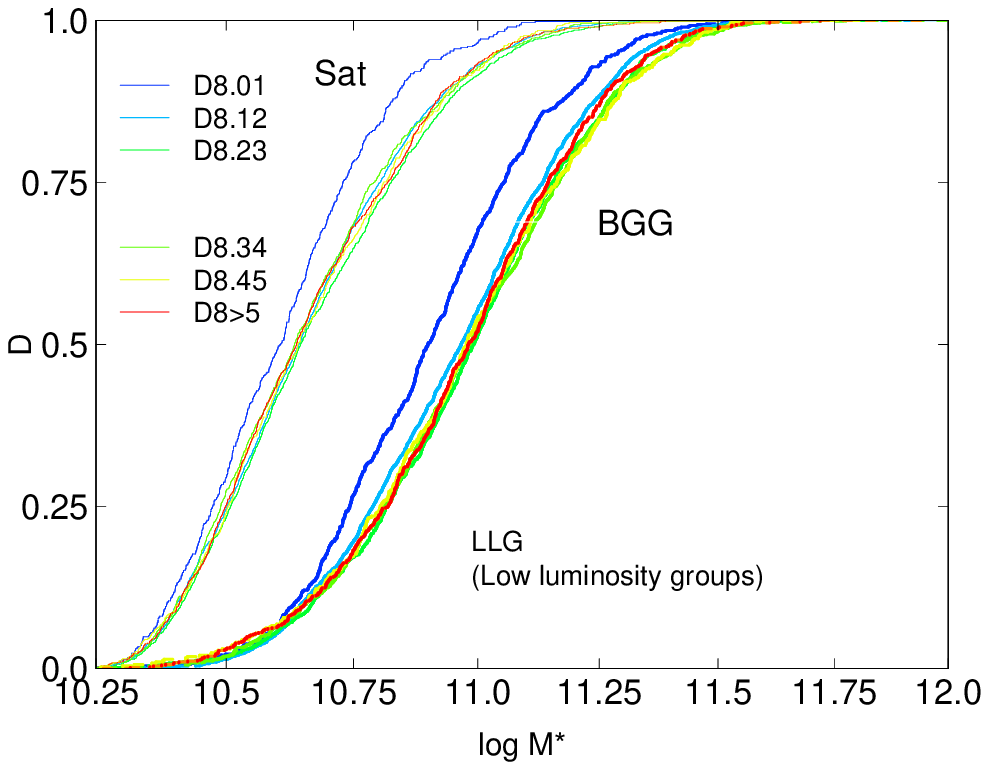}}
\resizebox{0.19\textwidth}{!}{\includegraphics[angle=0]{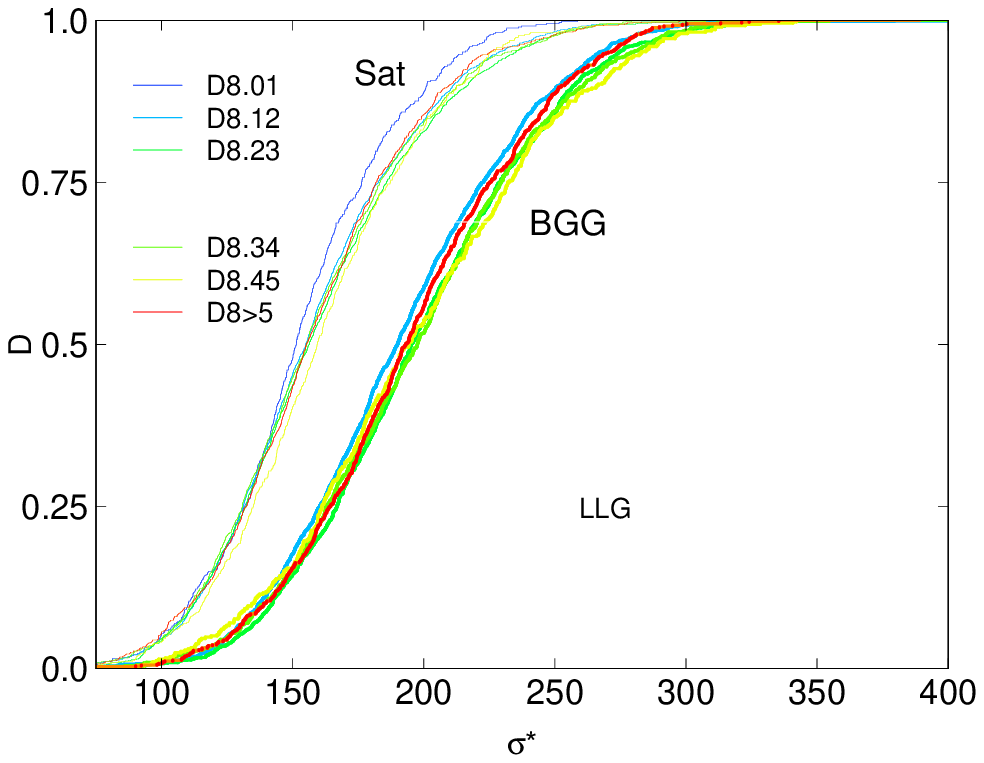}}
\resizebox{0.19\textwidth}{!}{\includegraphics[angle=0]{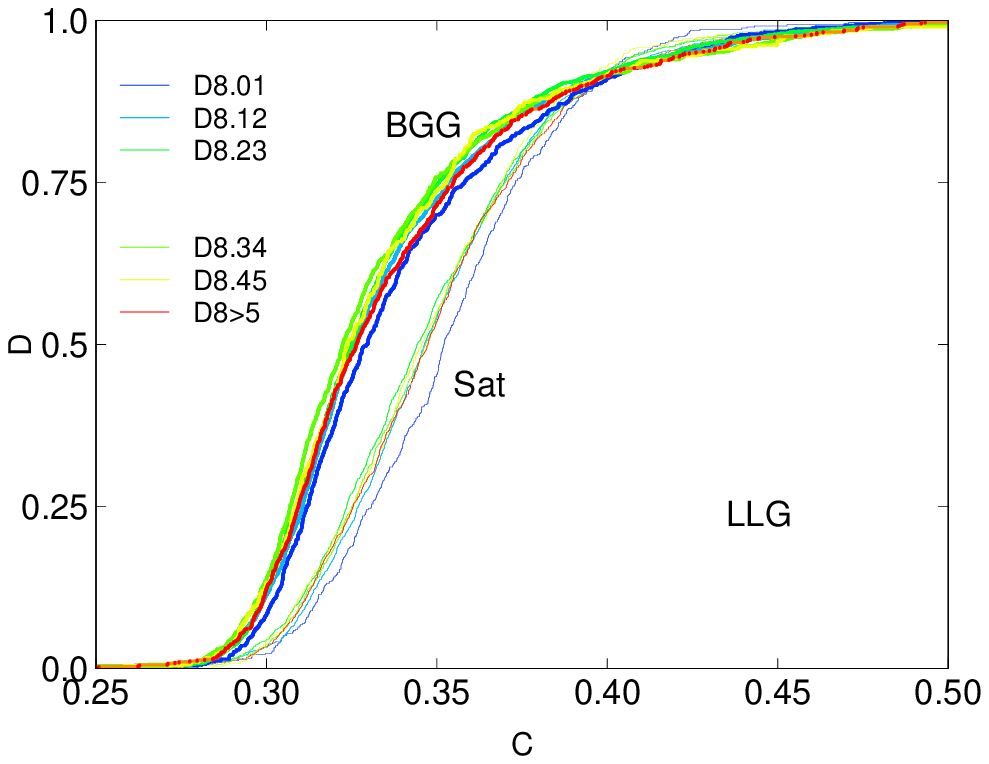}}
\resizebox{0.19\textwidth}{!}{\includegraphics[angle=0]{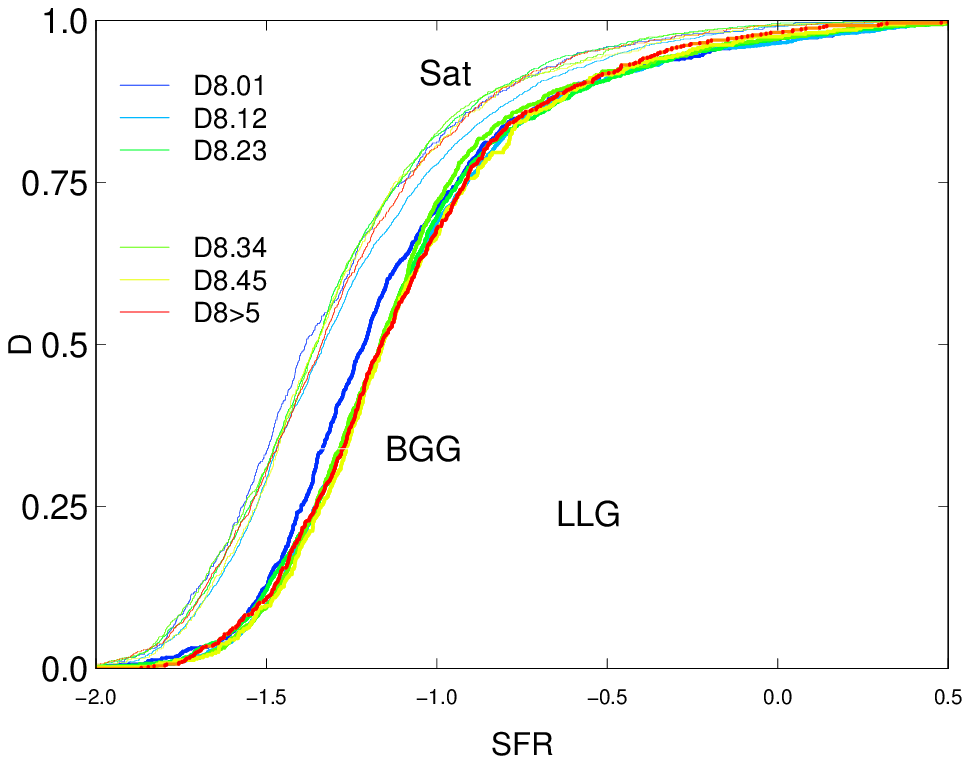}}
\resizebox{0.19\textwidth}{!}{\includegraphics[angle=0]{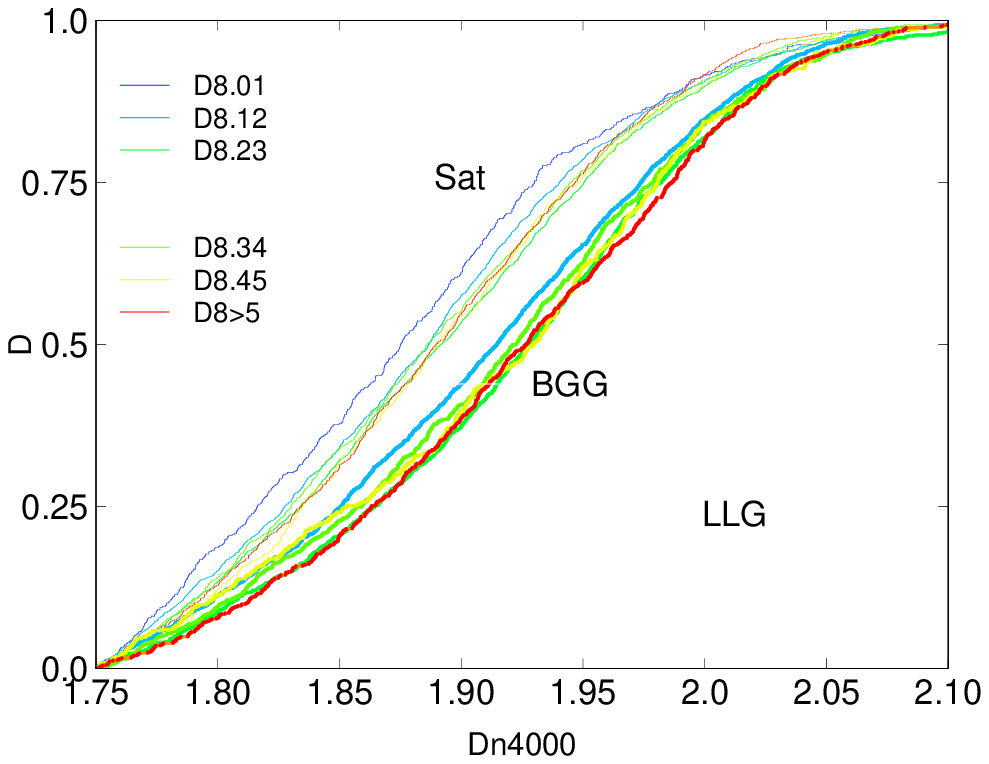}}

\resizebox{0.19\textwidth}{!}{\includegraphics[angle=0]{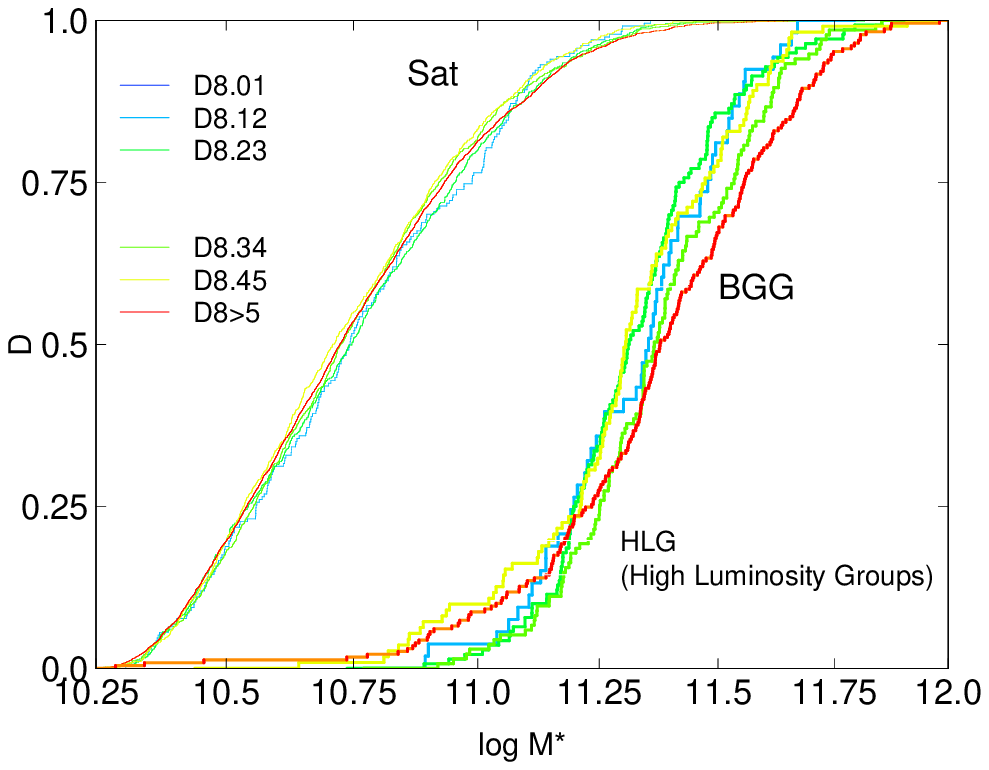}} 
\resizebox{0.19\textwidth}{!}{\includegraphics[angle=0]{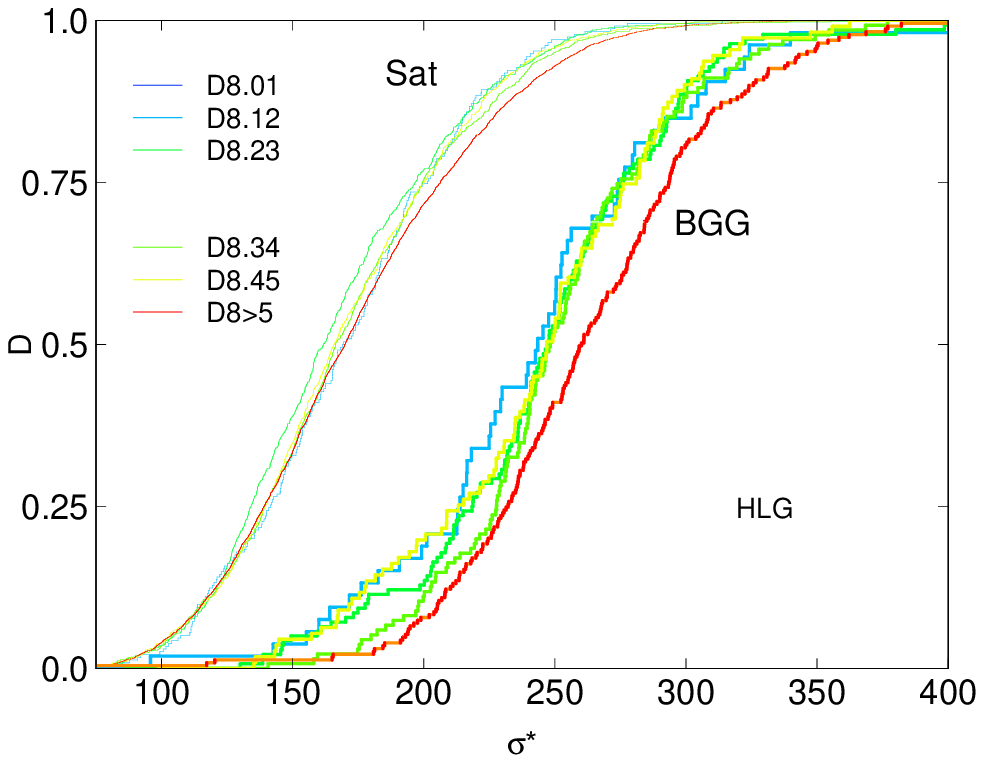}} 
\resizebox{0.19\textwidth}{!}{\includegraphics[angle=0]{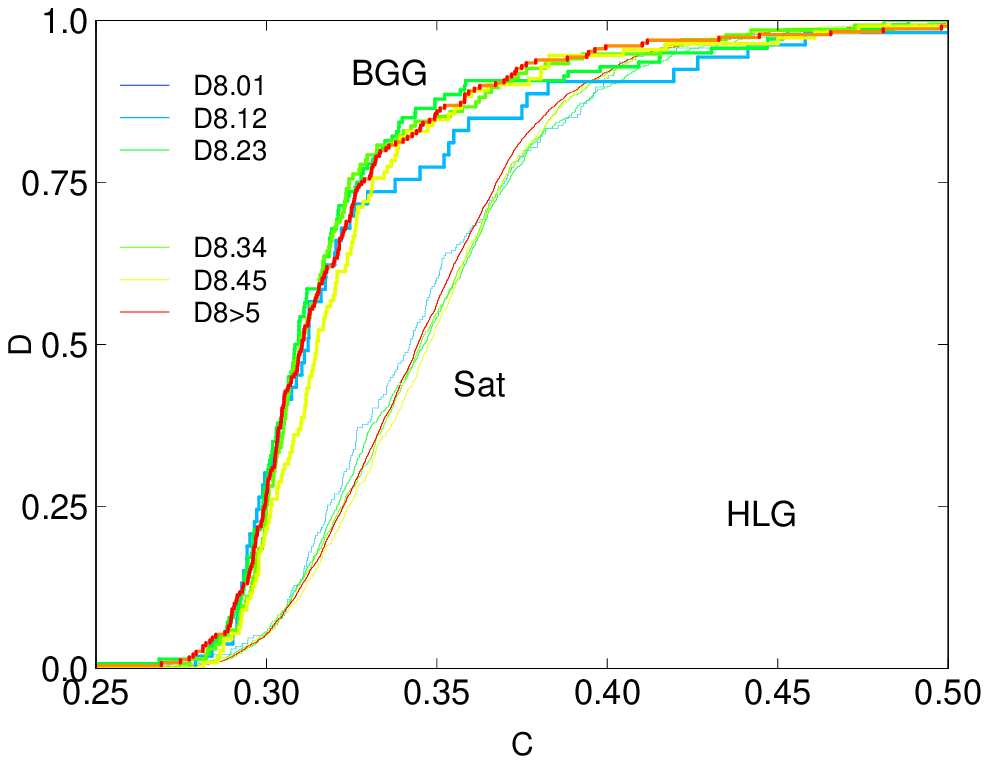}} 
\resizebox{0.19\textwidth}{!}{\includegraphics[angle=0]{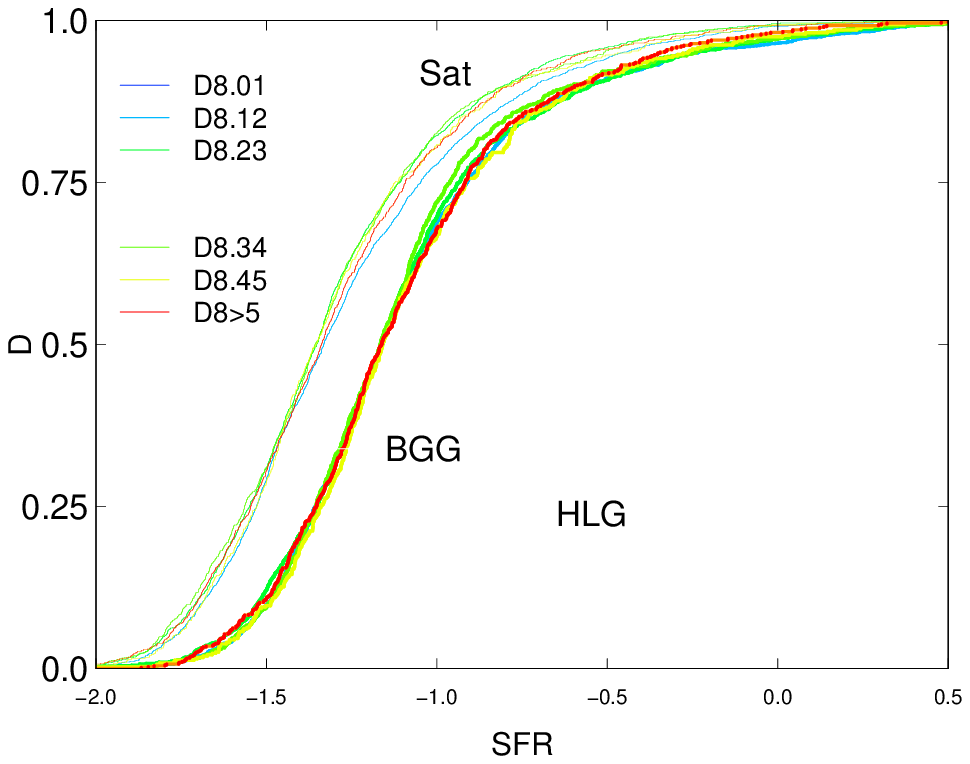}} 
\resizebox{0.19\textwidth}{!}{\includegraphics[angle=0]{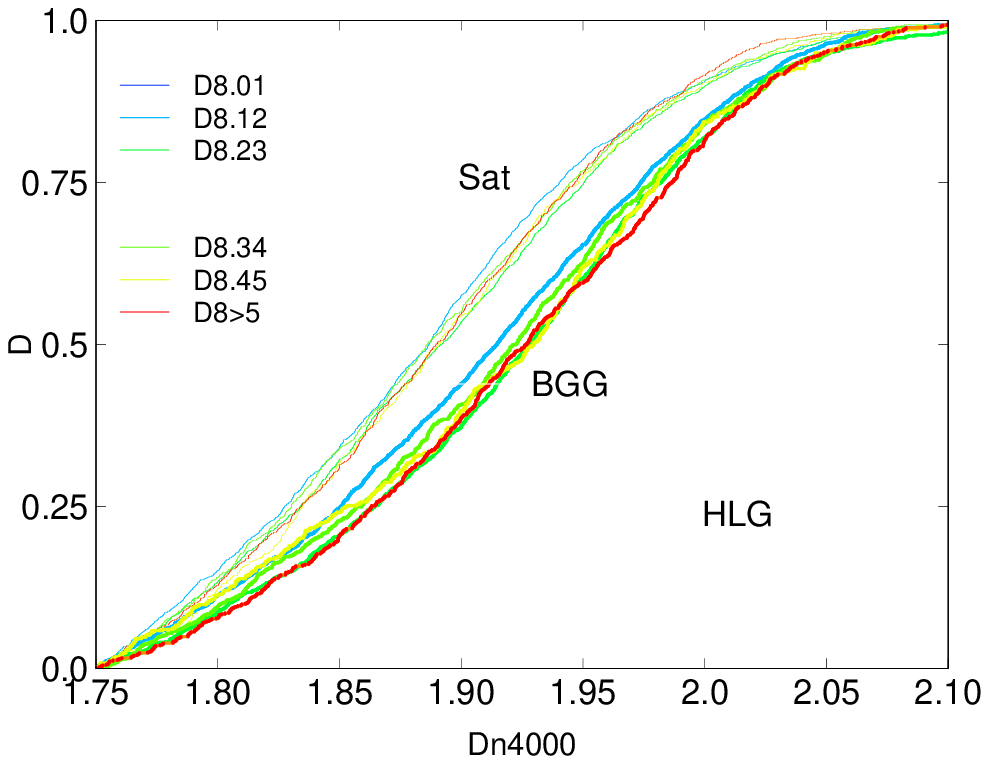}}
\caption{
VO galaxies:
distributions of stellar mass $M^{\mathrm{*}}$ (leftmost panels), 
stellar velocity dispersion $\sigma^{\mathrm{*}}$ (second panels),
concentration index $C$ (third panels), $\log \mathrm{SFR}$
(fourth panels), and $D_n(4000)$ index (rightmost panels)
for single galaxies (upper row), satellite galaxies and BGGs of 
LLGs (thin lines; middle row),
and satellite galaxies and BGGs of HLGs
(thick lines; lower row) in various global density $D8$ intervals, as shows in the legends.
}
\label{fig:g175d8int}
\end{figure*}

\begin{figure*}
\centering
\resizebox{0.19\textwidth}{!}{\includegraphics[angle=0]{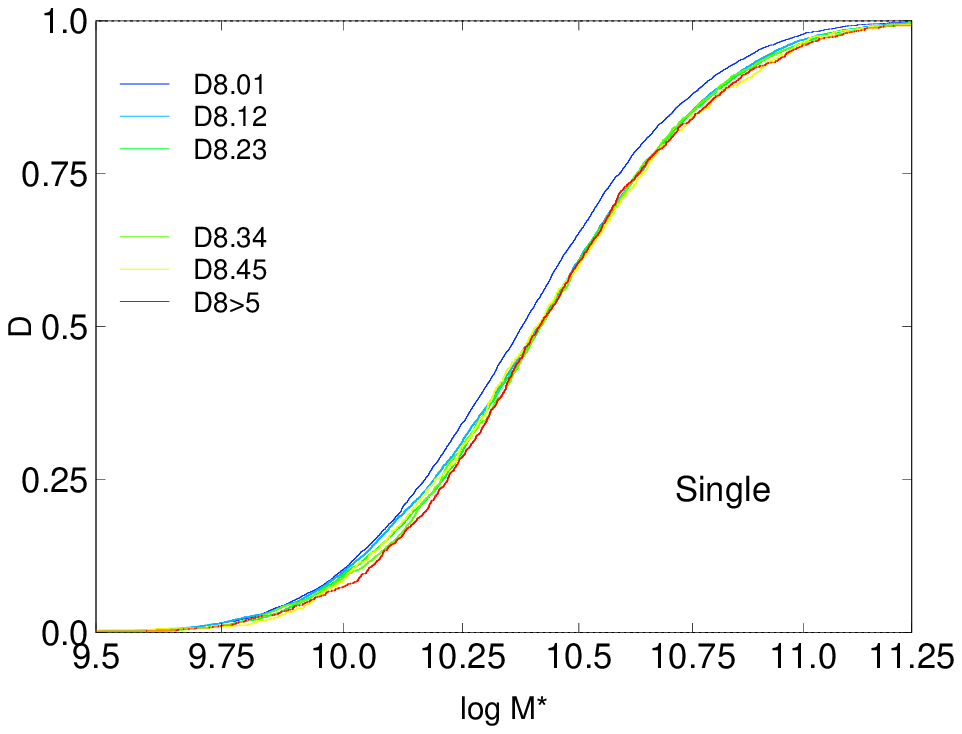}}
\resizebox{0.19\textwidth}{!}{\includegraphics[angle=0]{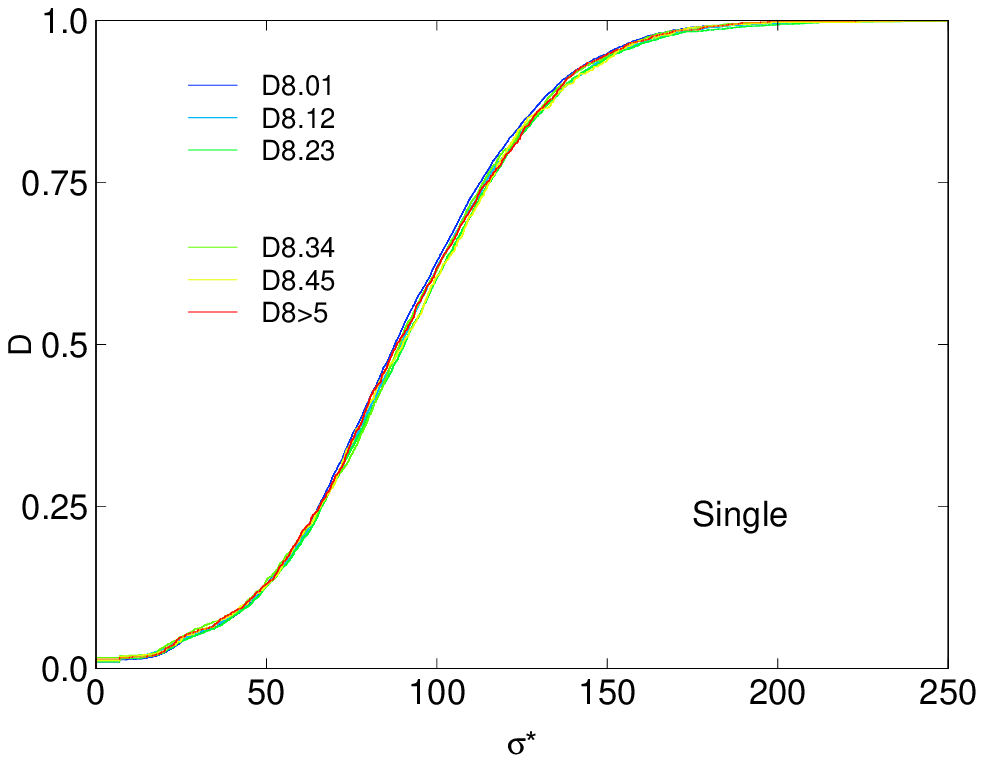}}
\resizebox{0.19\textwidth}{!}{\includegraphics[angle=0]{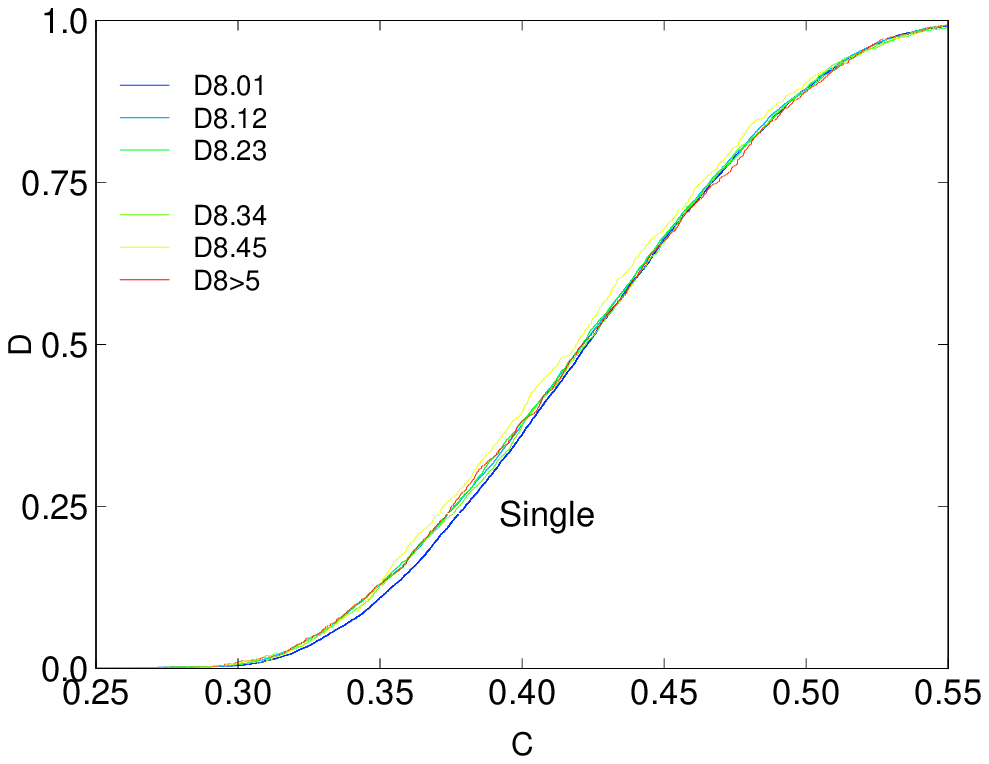}}
\resizebox{0.19\textwidth}{!}{\includegraphics[angle=0]{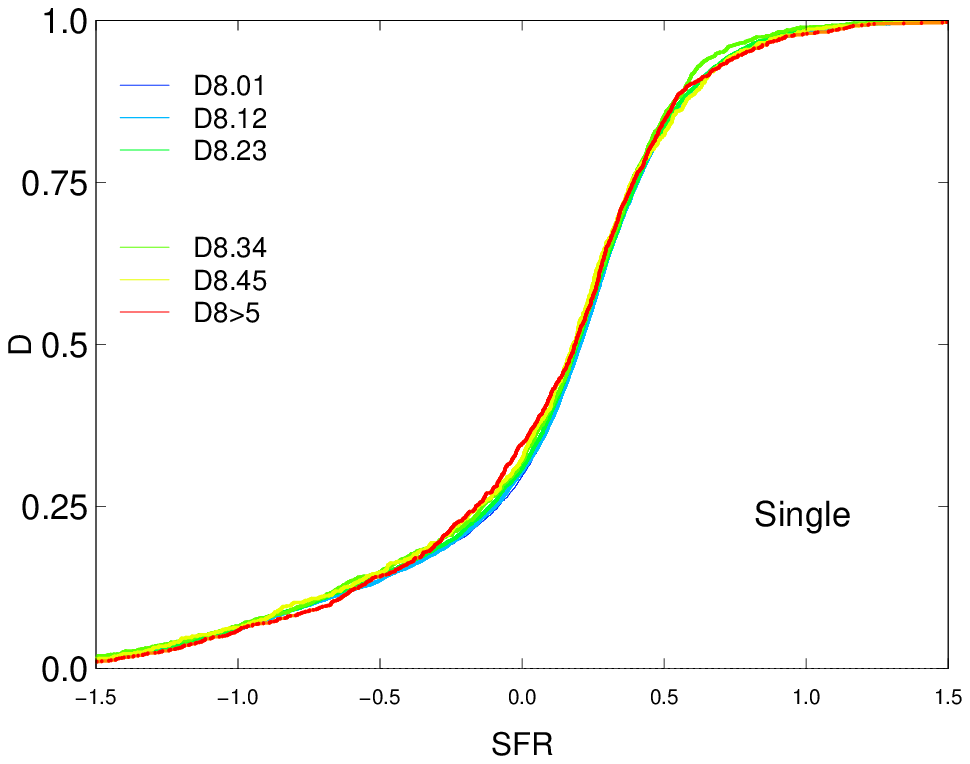}}
\resizebox{0.19\textwidth}{!}{\includegraphics[angle=0]{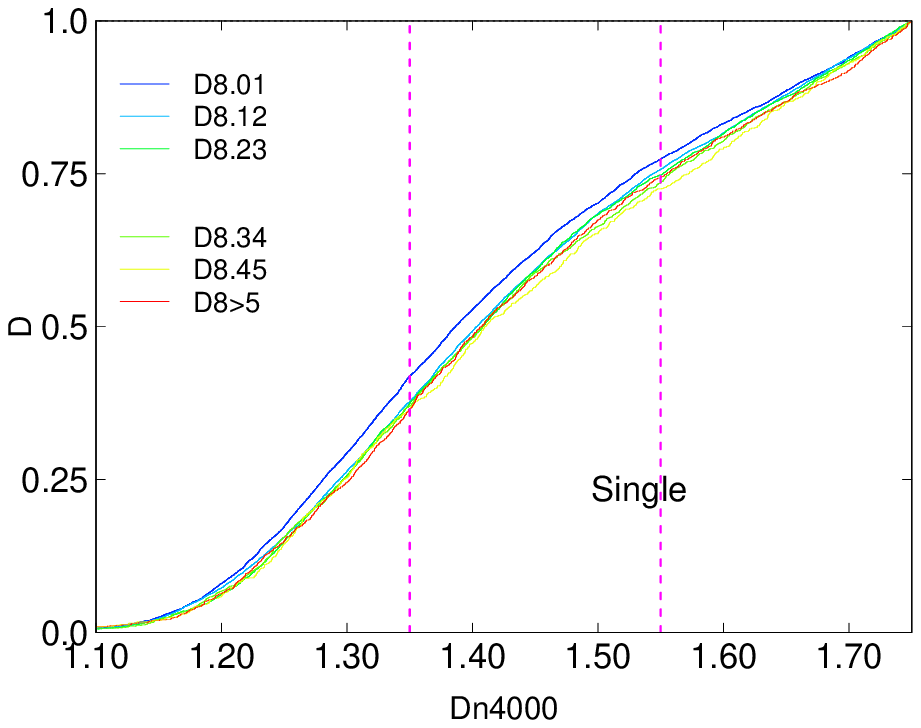}}

\resizebox{0.19\textwidth}{!}{\includegraphics[angle=0]{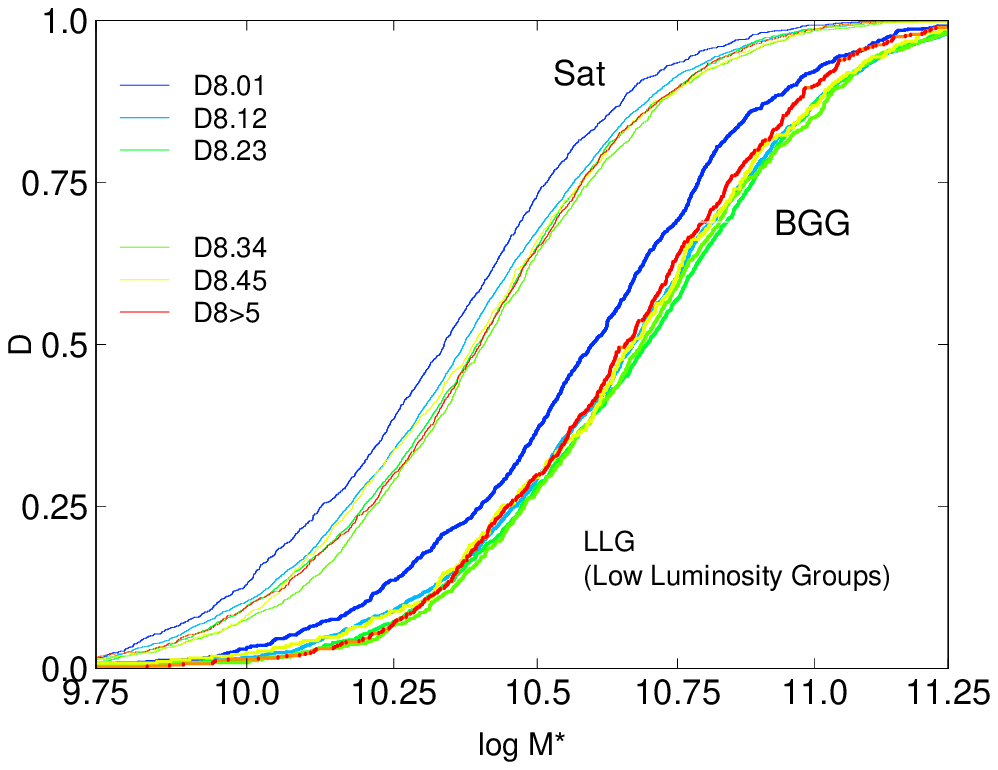}}
\resizebox{0.19\textwidth}{!}{\includegraphics[angle=0]{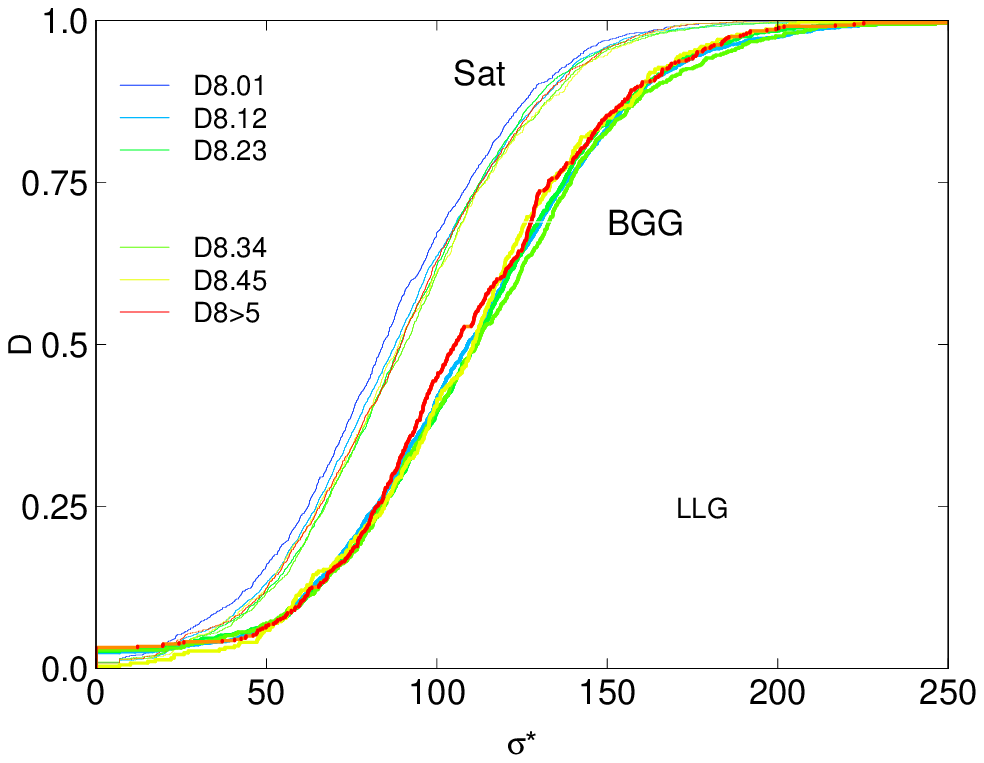}}
\resizebox{0.19\textwidth}{!}{\includegraphics[angle=0]{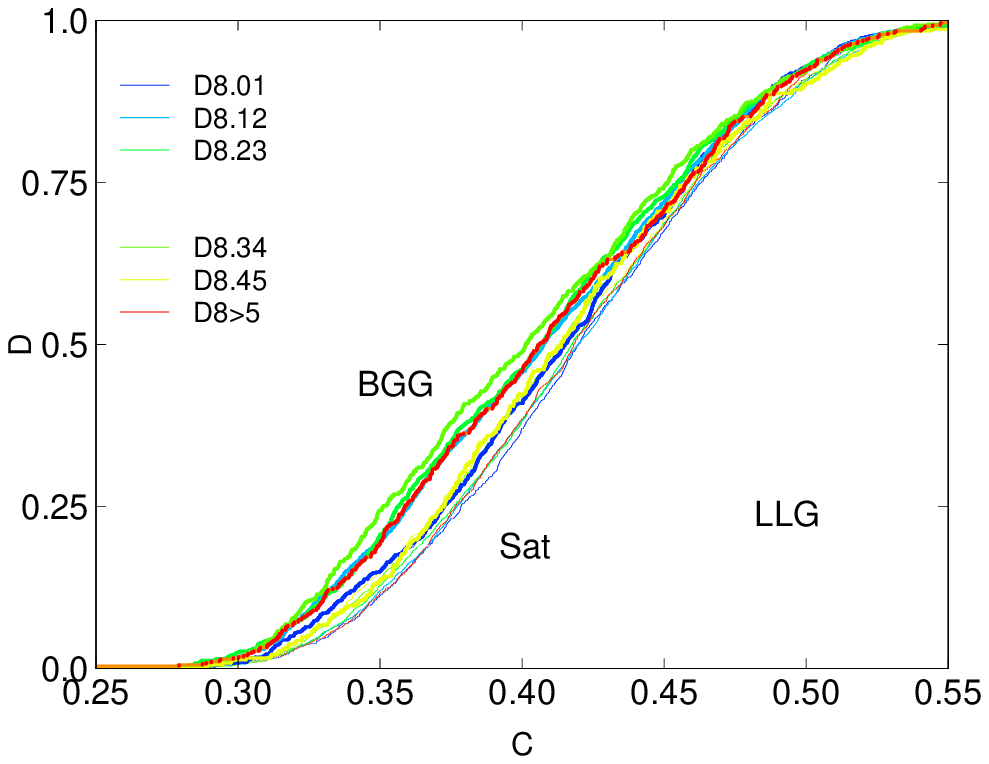}}
\resizebox{0.19\textwidth}{!}{\includegraphics[angle=0]{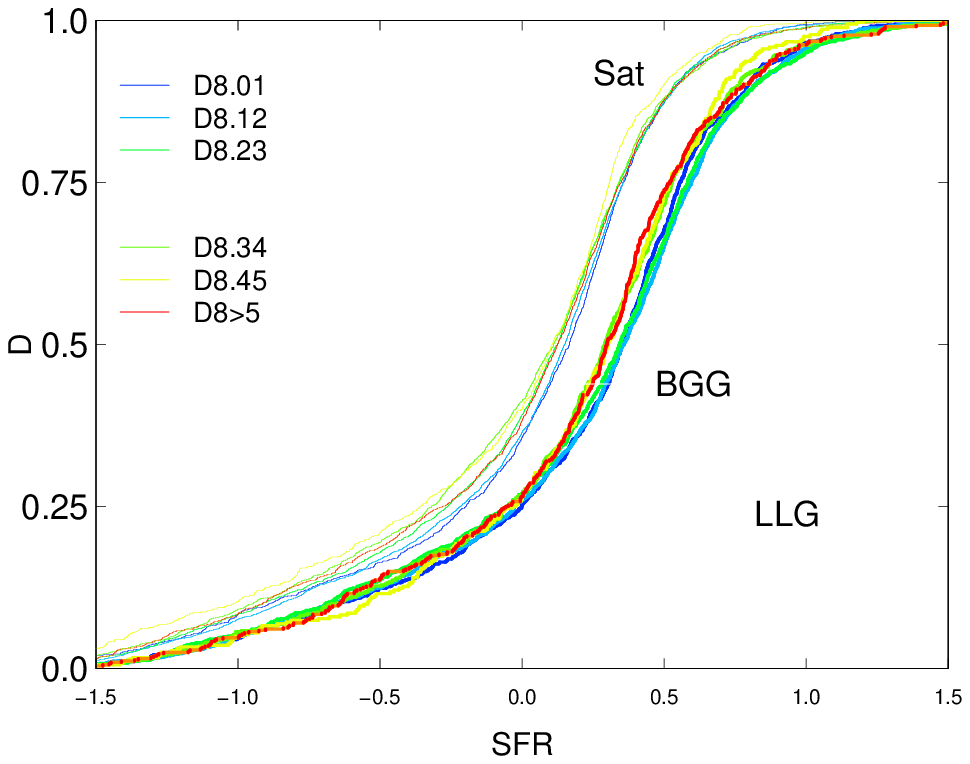}}
\resizebox{0.19\textwidth}{!}{\includegraphics[angle=0]{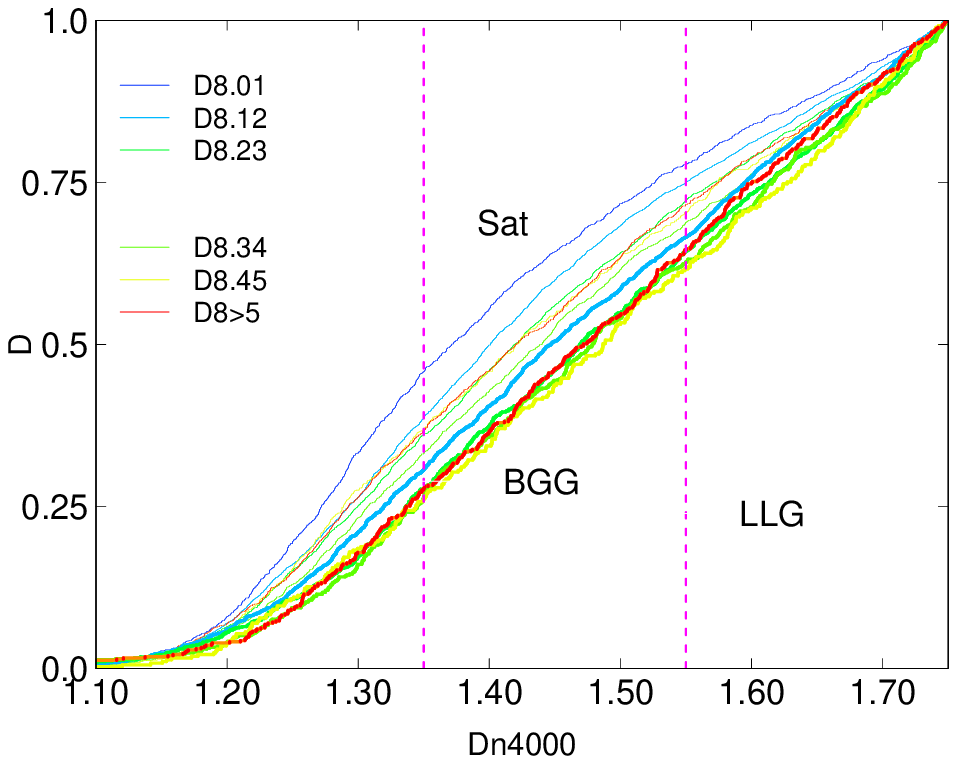}}

\resizebox{0.19\textwidth}{!}{\includegraphics[angle=0]{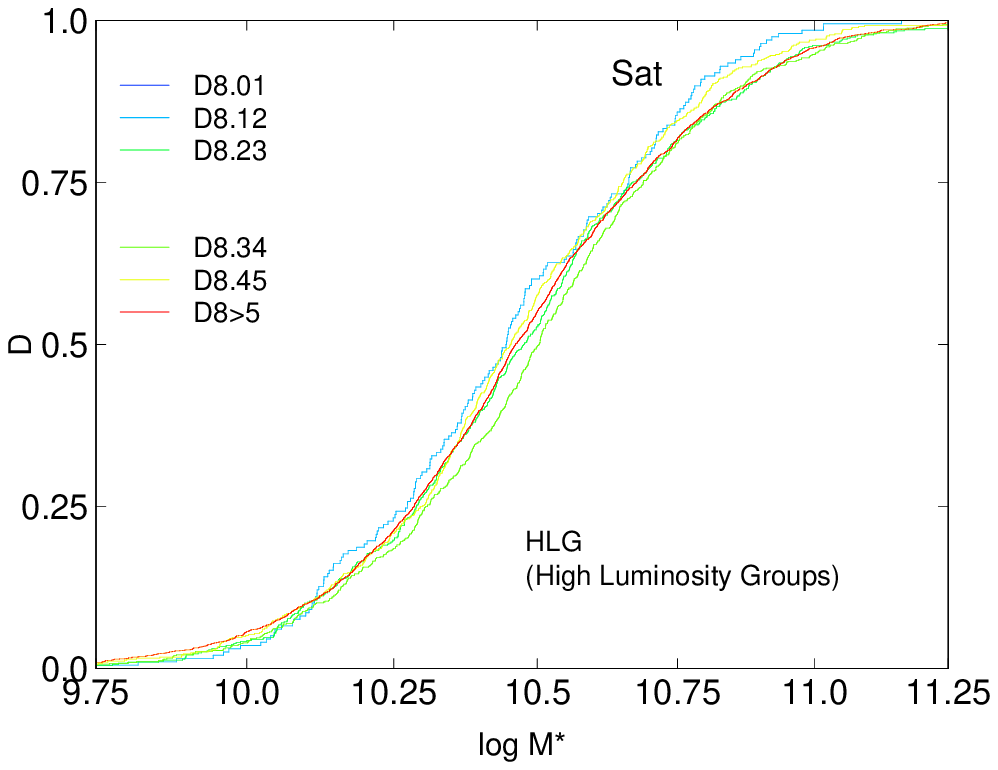}}
\resizebox{0.19\textwidth}{!}{\includegraphics[angle=0]{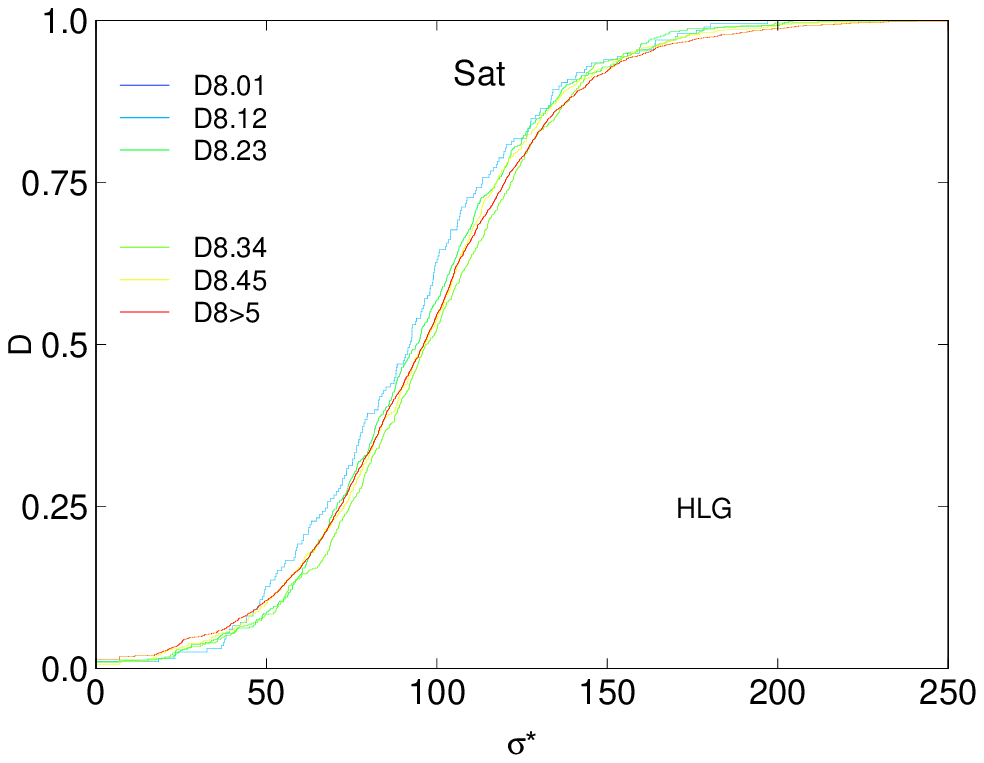}}
\resizebox{0.19\textwidth}{!}{\includegraphics[angle=0]{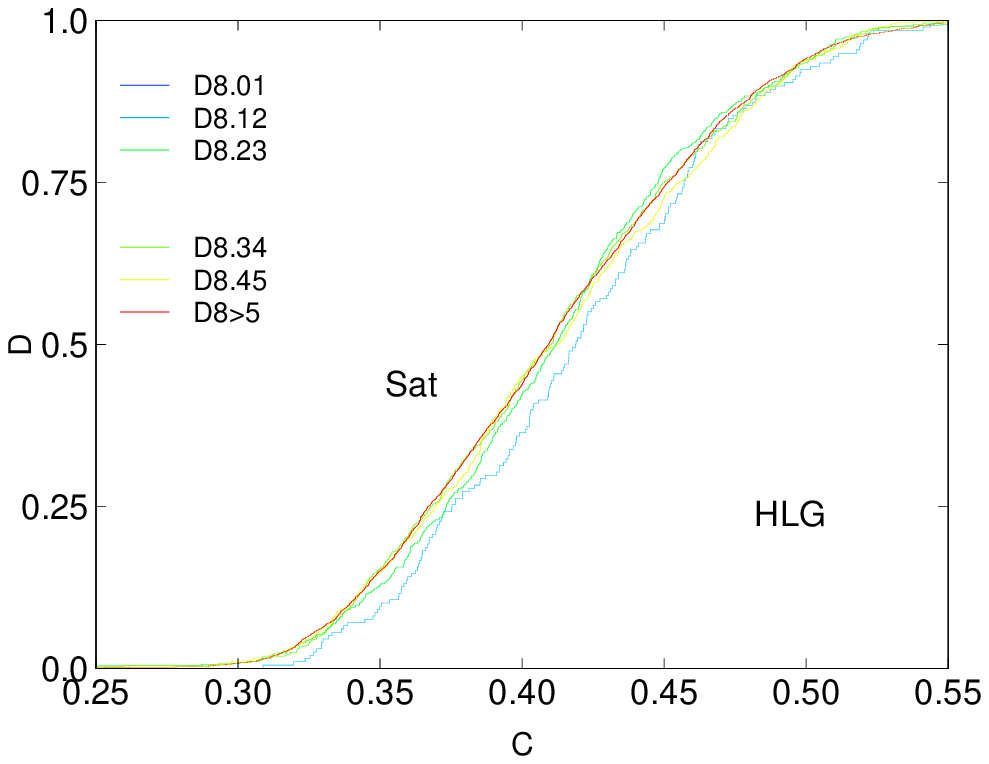}}
\resizebox{0.19\textwidth}{!}{\includegraphics[angle=0]{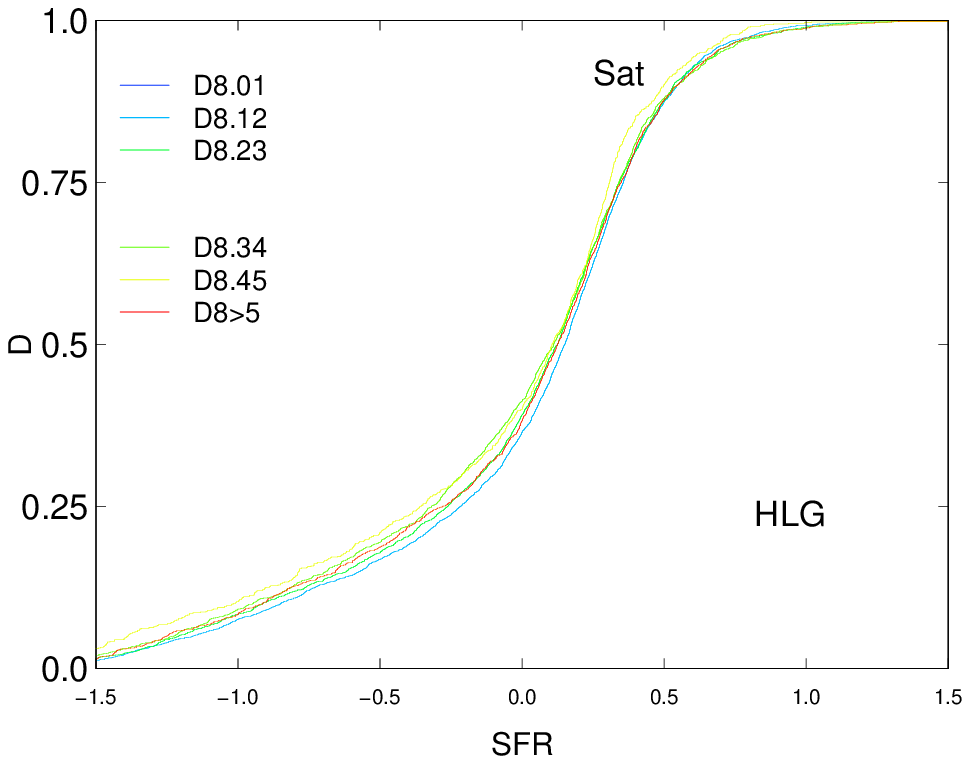}}
\resizebox{0.19\textwidth}{!}{\includegraphics[angle=0]{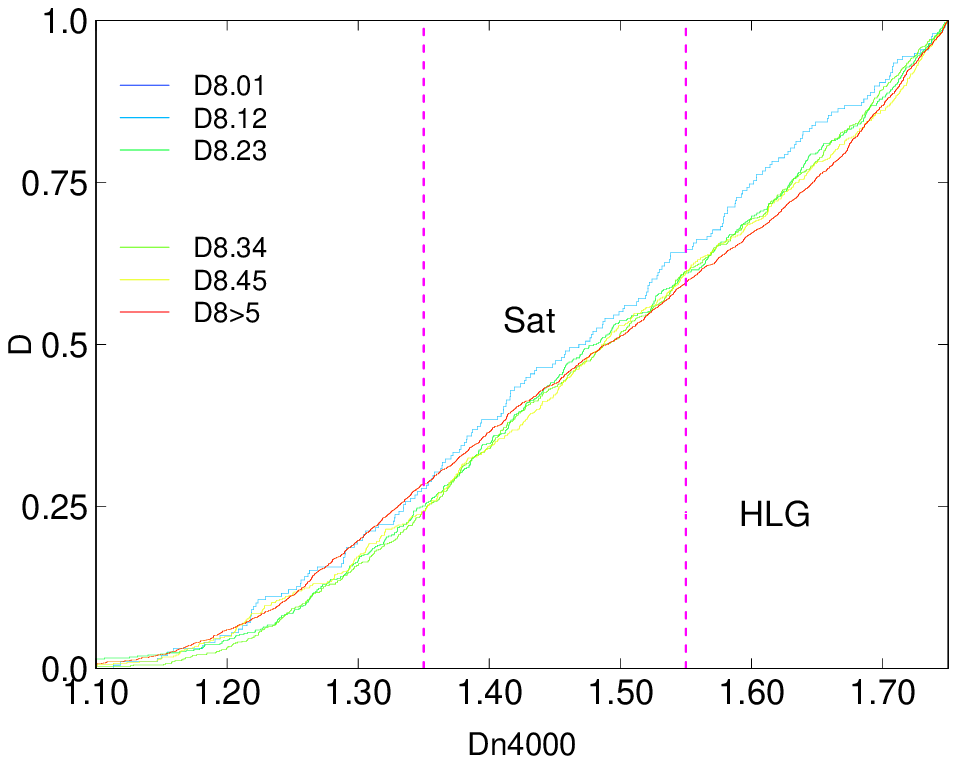}}
\caption{
YS galaxies:
distributions of stellar mass $M^{\mathrm{*}}$ (leftmost panels), 
stellar velocity dispersion $\sigma^{\mathrm{*}}$ (second panels),
concentration index $C$ (third panels), $\log \mathrm{SFR}$
(fourth panels), and $D_n(4000)$ index (rightmost panels)
for single galaxies (upper row), satellite galaxies and BGGs of 
LLGs (thin lines; middle row),
and satellite galaxies and BGGs of 
HLGs
(thick lines; lower row) in various global density $D8$ intervals, as shows in the legends.
In the rightmost panels, vertical lines indicate $D_n(4000)$ index
values $D_n(4000) = 1.35$ and $D_n(4000) = 1.55$ (this value separates star forming and 
(recently) quenched galaxies).
}
\label{fig:g0175d8int}
\end{figure*}

\subsection{The morphological properties of VO and YS galaxies in various global density environments}
\label{sect:galprop129}  

In our further analysis, we compare distributions of the stellar masses $M^{\mathrm{*}}$, 
stellar velocity dispersions $\sigma^{\mathrm{*}}$, concentration index $C$, and $\mathrm{SFR}$ 
of galaxies in various global density $D8$ intervals.
Although galaxies were selected according to their $D_n(4000)$ index, we also compare
the distributions of this index.
We further divide galaxies into single, satellite galaxies in groups,
or BGGs. 

We begin this analysis with {\it VO galaxies},
and compare the stellar masses, $M^{\mathrm{*}}$,
stellar velocity dispersion,  $\sigma^{\mathrm{*}}$,
and concentration index $C$ of VO galaxies
in various global density $D8$ regions.
We applied the Kolmogorov-Smirnov (KS) test to estimate the statistical significance 
of the differences between the galaxy populations.
We consider that the differences between distributions are significant and  
{\it \textup{highly}} significant 
when the $p$-value (the estimated probability of rejecting the hypothesis
that the distributions are statistically similar)  $p \leq 0.05$ and  $p \leq 0.01$, respectively. 

In Fig.~\ref{fig:g175d8int} we show the distribution of stellar masses of
galaxies with $D_n(4000) \geq 1.75$ in various global density $D8$ regions.
This figure shows that the largest difference is between the stellar masses of
galaxies in the lowest global density region, $D8 < 1$, and all other global density
regions $D8 > 1$. This difference is seen for all populations (single galaxies, 
satellite galaxies, and BGGs). 
This difference leads to a quantitative definition of the watershed region,
that is, a region with global luminosity--density   $D8 \leq 1$.
The KS test results in Tables~\ref{tab:satpop} - \ref{tab:singlepop} show that the
differences in stellar mass between the galaxies in the lowest density regions
and in all other global density regions are highly significant. Also, one can see that the
stellar masses of satellite galaxies are the lowest, BGGs have the highest
stellar masses, and 
the stellar masses of single galaxies have values between those of satellites
and BGGs. We note that, on average, more than half of single and satellite galaxies,
and more than $80$~\% of BGGs of LLGs,
have stellar masses higher than the stellar mass of the Milky Way galaxy 
as given in \citet{2015ApJ...806...96L}. All BGGs of HLGs have higher stellar masses
than the Milky Way.

We used the KS test to compare stellar masses of galaxies in other global density regions
($1 < D8 < 2$ and so on). This comparison showed that, as also suggested by Fig.~\ref{fig:g175d8int}, stellar masses of galaxies from the same population (single galaxies,
LLG satellite galaxies or BGGs, and HLG satellites or BGGs) 
are statistically similar with very high significance ($p > 0.05$, typically $p > 0.5$)
with a few exceptions. BGGs of HLGs have higher stellar masses in global
high-density environments. This is expected, as these groups are richer and of higher 
luminosity than HLGs in lower global density environments (Fig.~\ref{fig:grlumrich},
right panel).
This comparison shows  that although in the global density interval $1 < D8 < 2$
most galaxies are single or members of poor groups, as is also the case in the watershed region,
these galaxies have already higher stellar masses and other properties similar to those
of galaxies in higher global density regions where LLGs and HLGs dominate.

Similar, albeit weaker, trends are seen in the distribution of
the stellar velocity dispersion of galaxies  $\sigma^{\mathrm{*}}$, in the distribution of
the concentration index $C$, and in the distribution of SFR.
We note that approximately 85\%--90\% of VO galaxies have $C < 0.38,$ suggesting that these are mostly
early-type galaxies. This is expected, and was shown in \citet{2020A&A...641A.172E} 
using the probability of being an early-type galaxy. KS tests show that the differences in
stellar velocity dispersion and the concentration index of galaxies 
in the lowest global density regions and elsewhere are statistically highly significant,
and typically not significant between galaxies in other global density regions.
There is a weak trend that BGGs  in HLGs in low global density regions
have lower stellar velocity dispersions than in the highest global density region.
These are the same galaxies that have higher stellar masses than BGGs of HLGs
in the lower global density regions. 

We also looked at the distributions of the $D_n(4000)$ index of VO galaxies
(Fig.~\ref{fig:g175d8int}). 
For single galaxies, the distribution of the $D_n(4000)$ index is the same 
in all global density environments.
However, the $D_n(4000)$ index of galaxies in groups and especially for
the brightest galaxies of groups  (Fig.~\ref{fig:g175d8int}) shows that 
this index is the lowest in the lowest global density environment
($D8 < 1$). For satellites in HLGs, the trends with global environment
are weaker. The BGGs have the lowest  $D_n(4000)$ index values in the 
lowest global density environment, and the highest values of $D_n(4000)$ index 
in the highest global density environment. This 
suggests that they have
older stellar populations. However, the differences in the  
distribution of the $D_n(4000)$ index between satellites and BGGs are much larger than the
differences in $D_n(4000)$ between the same galaxy type in different global environments.



The average differences in the
$D_n(4000)$ index values between the satellite VO galaxies in the lowest global density
environment and in higher global density environments shown above may lead to the 
differences in the ages of stellar populations 
of up to approximately $2$~Gyr \citep{2003MNRAS.341...33K}. 
In addition, BGGs of LLGs in the lowest global density environment
have stopped their star formation later than the BGGs in high-global-density environment. For HLGs, the
trends with global density may also be affected by the increase in group richness with 
global density.
For single galaxies, there is no clear trend in the age of stellar populations with global
density environment.

Next we present
a similar analysis for {\it YS galaxies} with $D_n(4000) < 1.75$ (Fig.~\ref{fig:g0175d8int}
and Tables~\ref{tab:satpop0175} - \ref{tab:singlepop0175}). 
\citet{2020A&A...641A.172E} and \citet{2021A&A...649A..51E} divided galaxies with
$D_n(4000) < 1.75$ into several subclasses: blue star forming galaxies
with $D_n(4000) < 1.55$, and recently quenched and red star forming galaxies with
$1.35 < D_n(4000) < 1.75$ \citep[see Fig. 3 in][]{2021A&A...649A..51E}.
Overall, from Fig.~\ref{fig:g0175d8int} it appears that the trends of galaxy properties
with global density for YS galaxies are weaker than those for VO galaxies, although
the KS test shows that YS galaxies have lower stellar masses in the lowest global density
environments with very high statistical significance.  
Most YS galaxies have lower stellar masses than our Milky Way galaxy.

The distributions of the $D_n(4000)$ index for YS satellite galaxies, BGGs, and
single galaxies in Fig.~\ref{fig:g0175d8int} show that these galaxies have 
lower $D_n(4000)$ index values in the lowest global density environment.
YS LLG satellite galaxies have slightly lower stellar masses 
and lower $D_n(4000)$ index values in $1 < D8 < 2$ global
density regions than in higher global density regions. These differences are
highly significant.

On average, 
the median values of the $D_n(4000)$ index for galaxies  in LLGs (both for satellites and BGGs) and for single galaxies in all environments 
are $D_n(4000) \approx 1.40$ (for BGGs, $D_n(4000) \approx 1.45$) 
which according to \citet{2003MNRAS.341...33K} corresponds to an age of the stellar
populations of $\approx 0.95$~Gyr. The stellar age differences between satellites
and BGGs are small. Satellite galaxies in HLGs have slightly higher
values of $D_n(4000)$ index, suggesting  somewhat higher ages of their stellar populations.
Among satellites of HLGs, there are more galaxies with 
$D_n(4000) > 1.35$, and therefore more galaxies
 that are already quenched or are now in transition
 (red, star forming, and recently quenched galaxies). These galaxies can be found
preferentially in the infall zones of groups and clusters 
\citep{2020A&A...641A.172E, 2021A&A...649A..51E}. This may be an indication that
HLGs are dynamically more active than LLGs. 
The fraction of blue star forming galaxies with $D_n(4000) > 1.35$
is the highest among satellite galaxies of LLGs in
the lowest global density environment, namely approximately
45\%, where  even $\approx 35$\% of BGGs have $D_n(4000) > 1.35$.
Approximately 30\%--35\% of BGGs and single galaxies have  $D_n(4000) > 1.55$     
which suggest that they may already be quenched.         

The distribution of
stellar velocity dispersion of YS galaxies  $\sigma^{\mathrm{*}}$ shows that, on average,  $\sigma^{\mathrm{*}}$ values are lower than those for VO galaxies,
as expected in the case of  star-forming galaxies.
The concentration index $C$ values for YS galaxies are higher than those of VO galaxies,
and the percentage of late-type galaxies with $C > 0.38$ is larger than $25$~\% among them. 

The trends that we see between  global density 
and stellar velocity dispersion, concentration index, 
and SFRs of YS galaxies are very weak. 
This tells us that star-forming
galaxies with young stellar populations are very similar in all environments
from the lowest density watersheds to superclusters. This is confirmed by the results of the KS test.

We note that the relatively high percentage of YS galaxies among BGGs of
LLGs  should be taken with caution.
We analysed
the galaxy content of groups with between 5 and 9 member galaxies 
as such groups are more reliable than very poor groups.
We found that typically, even if the most luminous galaxy in such a group is
star forming, among the group member galaxies there is at least one quenched
galaxy. 
It is possible that the brightest galaxy in this case is not the central galaxy,
and groups are still forming, as was concluded for richer groups and clusters in \citet{2012A&A...540A.123E}. 
Furthermore, we only found three groups in the lowest global density
environment, and four groups in the highest global density environment, in which all the galaxies
have young stellar populations with $D_n(4000) < 1.35$.
Three such groups are filament members (two in high-density environment, and one
in low-density environment).

In summary, Figs.~\ref{fig:g175d8int} and \ref{fig:g0175d8int} present the
distributions of morphological properties  of galaxies for
different global density $D8$ ranges 
in terms of group membership:
single galaxies, satellite galaxies, and BGGs of LLGs and of HLGs.  
These figures demonstrate the similarity between galaxy properties  over a wide range of group 
luminosities (and therefore masses).
In particular, a remarkable similarity is observed between
the distributions of morphological properties of single galaxies and
satellites of HLGs.  
Figures~\ref{fig:g175d8int} and \ref{fig:g0175d8int} also
show another important aspect of morphological properties of galaxies:
great differences are seen between the distributions of morphological
properties of satellites and BGGs, both in LLGs and HLGs.  
These figures show that group membership type is the
major parameter determining the morphological properties of galaxies.

\subsection{Local environments of single galaxies and groups with
VO and YS BGGs}
\label{sect:locenv}

\begin{figure}
\centering
\resizebox{0.44\textwidth}{!}{\includegraphics[angle=0]{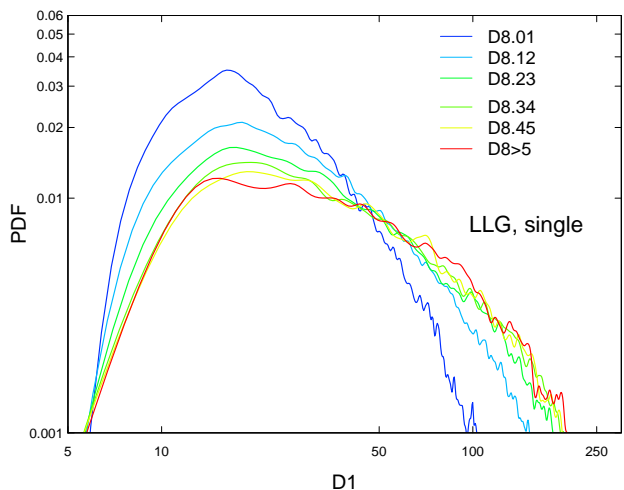}}
\caption{
Distributions of local luminosity densities $D1$ for single galaxies
and for LLG galaxies 
in various intervals of global luminosity--density $D8$.
}
\label{fig:d1d8llgsingle}
\end{figure}

Table~\ref{tab:ngalgr} shows that 27\%--36\% of  single galaxies
and  42\%--59\% of BGGs of LLGs are VO galaxies. 
The percentage of VO galaxies among these types of galaxies increases with global luminosity density. 
Next, we compare local environmental densities at the location of single galaxies
and galaxies in LLGs, as well as 
at the location of VO and YS single galaxies
and BGGs of LLGs.

Local luminosity--density around galaxies $D1$ is defined as the luminosity--density around 
galaxies calculated with a smoothing length of 1~\Mpc.
The distribution of local luminosity-densities $D1$
at the location of single galaxies and members of LLGs 
in various global luminosity--density $D8$ intervals 
is 
presented in Fig.~\ref{fig:d1d8llgsingle}.
We do not show local densities around HLG member galaxies as these are absent
in the lowest global density regions. 
Interestingly, these distributions show that in the global density interval $1 < D8 < 2$
local luminosity-densities $D1$ are higher than in watershed regions,
but lower than in higher global density regions, where 
local densities at the location of single galaxies and LLGs are 
similar on average.
Above, we show that the morphological properties of galaxies in
global density regions $D8 > 1$ are already similar.
We may  be seeing evidence of an interplay between local and global densities in shaping 
group formation and the morphological properties of galaxies.

\begin{figure}
\centering
\resizebox{0.22\textwidth}{!}{\includegraphics[angle=0]{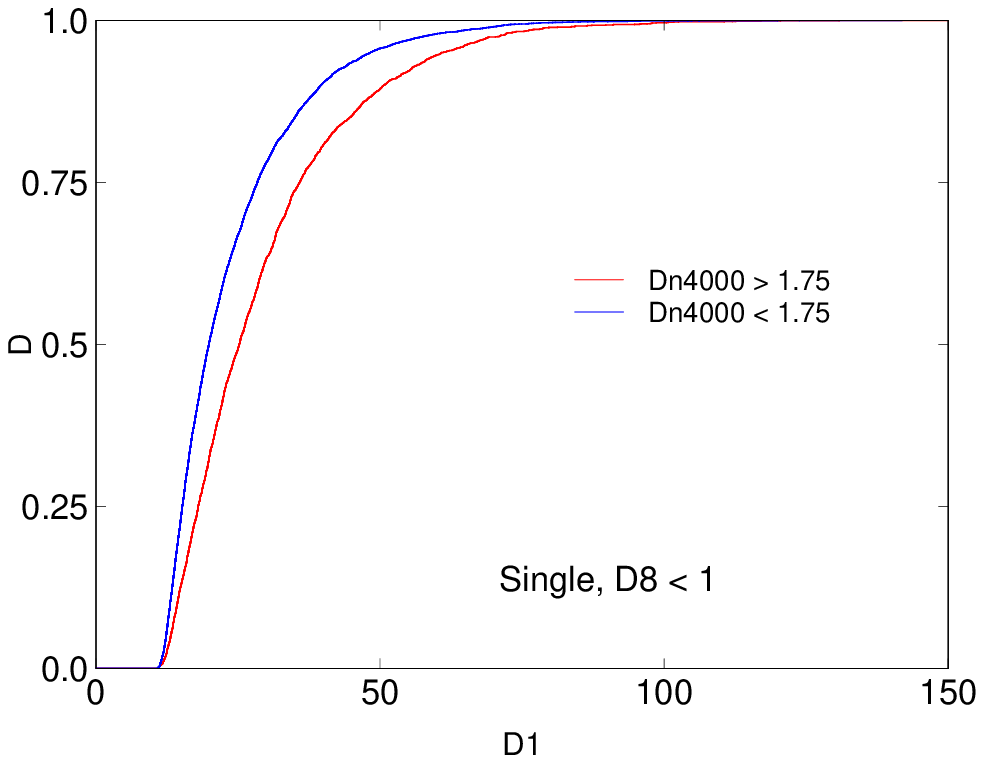}}
\resizebox{0.22\textwidth}{!}{\includegraphics[angle=0]{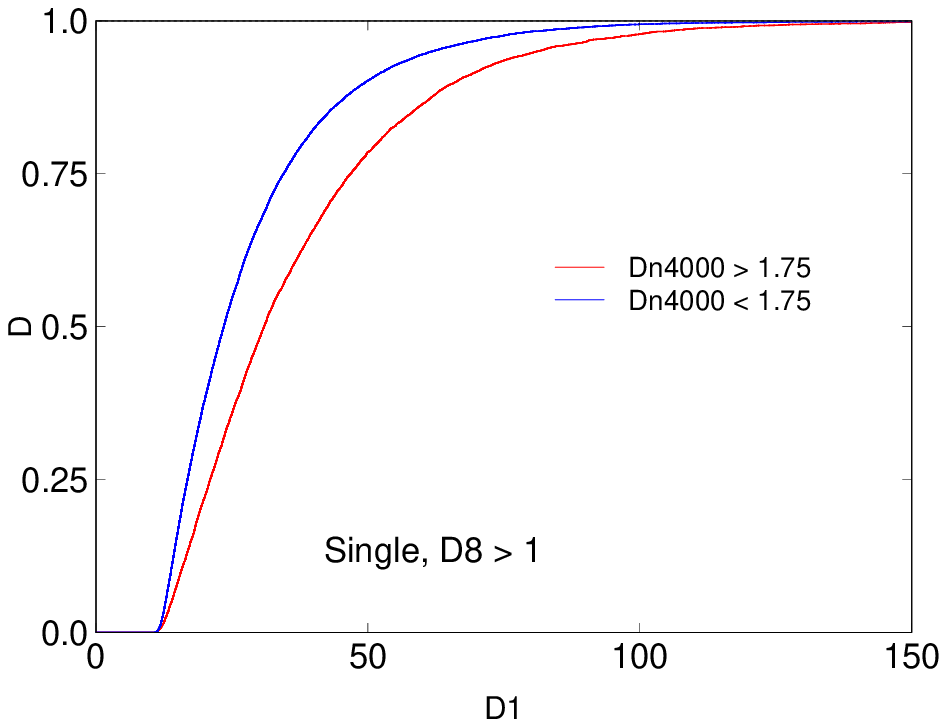}}

\resizebox{0.22\textwidth}{!}{\includegraphics[angle=0]{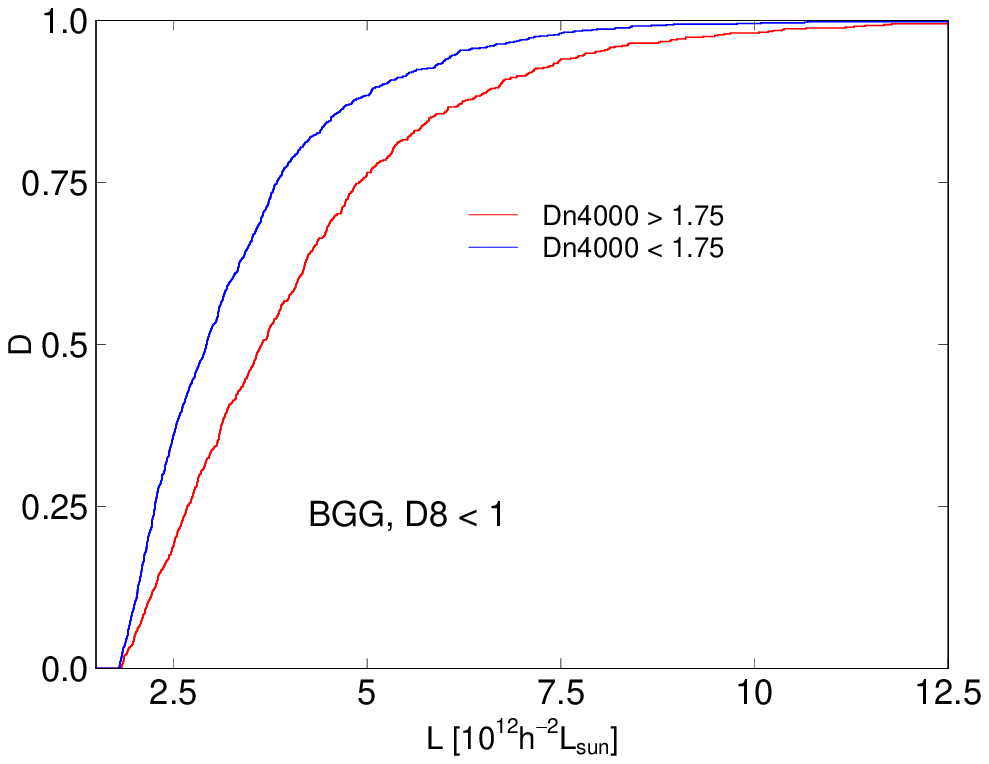}}
\resizebox{0.22\textwidth}{!}{\includegraphics[angle=0]{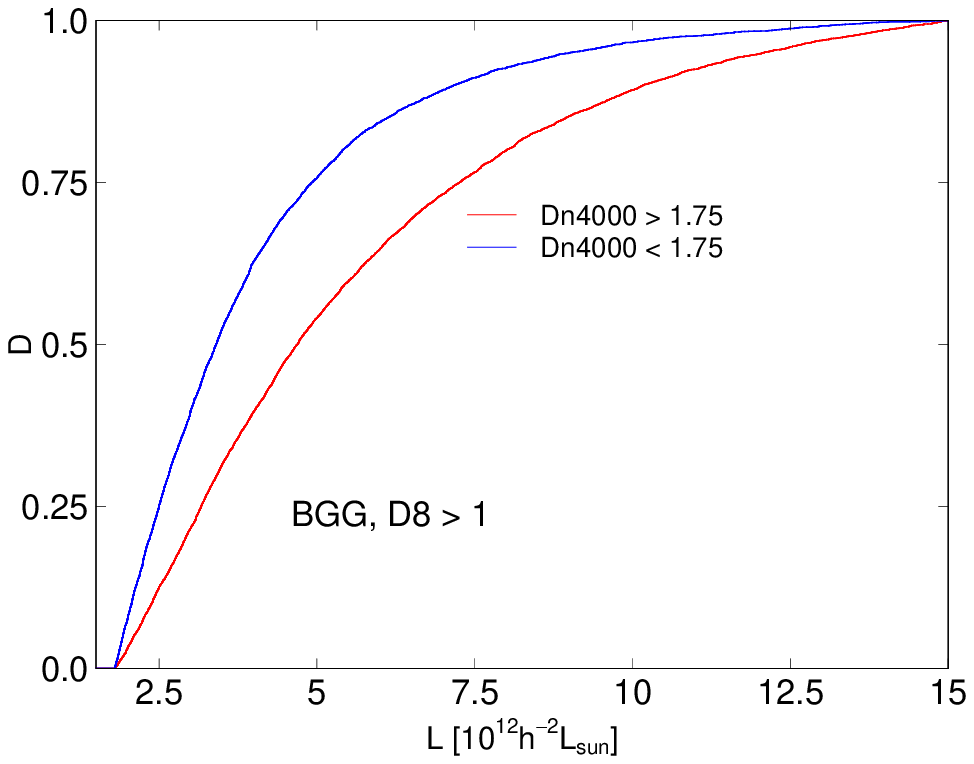}}
\caption{
Distributions of local luminosity densities $D1$ for single galaxies
(upper row) and group luminosities $L_{gr}$ for BGGs of LLGs (lower row)
in global low-luminosity regions ($D8 \leq 1$, left panels) and 
global high-luminosity regions ($D8 > 1$, right panels).
Red lines: Single galaxies and BGGs with $D_n(4000) > 1.75$ (VO galaxies).
 Blue lines: $D_n(4000) < 1.75$  (YS galaxies).
}
\label{fig:lumdenoy}
\end{figure}

In Fig.~\ref{fig:lumdenoy}, we present local environmental densities 
at the location of VO and YS single galaxies and BGGs of LLGs. 
In this figure, for BGGs we use  group luminosities $L_{gr}$ as a proxy for their local environment,
and for single galaxies we apply the values of the luminosity--density field
with smoothing length $1$~\Mpc, $D1$. 
Distributions of $L_{gr}$ and 
 $D1$ values separately for global luminosity densities $D8 \leq 1$ and $D8 > 1$
are given in Fig.~\ref{fig:lumdenoy}.

Figure~\ref{fig:lumdenoy} shows that in all global environments, in the case of both
single galaxies and BGGs, VO galaxies are located in 
higher local density environments than YS galaxies. 
Among single galaxies, in watershed regions ($D8 \leq 1$)
VO galaxies have median values
$D1_{med} = 25$, while galaxies with young stellar populations have 
$D1_{med} = 20$.
In regions of higher global density, ($D8 > 1$)
single VO galaxies have median values
$D1_{med} = 31$, and YS galaxies  have 
$D1_{med} = 24$. 

VO BGGs in watershed regions lie in groups with median
luminosity with $L_{gr}^{med} = 3.6 \times10^{10} h^{-2} L_{\sun}$,
while YS BGGs  have 
$L_{gr}^{med} = 2.9 \times10^{10} h^{-2} L_{\sun}$.
In higher global density regions, the median value of host
group luminosity for VO BGGs 
$L_{gr}^{med} = 4.7 \times10^{10} h^{-2} L_{\sun}$
and for YS BGGs 
$L_{gr}^{med} = 3.4 \times10^{10} h^{-2} L_{\sun}$.
The KS test shows that the differences between local densities (group luminosities for BGGs) 
are statistically very highly significant, with KS test $p$ value $p < 0.001$. 
Therefore, in global high-density environments, higher local densities lead to 
a higher percentage of quenched galaxies even in groups of the same luminosity.

\section{Analysis of the results and discussion}
\label{sect:discussion} 

\subsection{Does global density matter? }
\label{sect:gloden} 

In this study, our aim is to find out whether global density,
quantified with the luminosity--density field, has any influence on the
galaxy content of the cosmic
web.
Our results reveal that the global density  most strongly affects  the
group content of a given environment.
The lowest global density environments (watersheds between superclusters
with $D8 < 1$)
are populated by single galaxies, which form more than $70$~\% of galaxy
populations there, and by galaxies in poor, low-luminosity groups.
In contrast, the highest global density environments (superclusters)
are mostly populated 
by high-luminosity groups and clusters, although even in these environments 
single galaxies form approximately one-sixth of all galaxies.
This means that global and local environmental densities are correlated: 
local densities are higher, on average, at higher global densities.

The effect of global density on the star formation properties of galaxies is less strong.
Surprisingly, still almost one-third of all galaxies in the lowest global density
environments with $D8 \leq 1$ 
have $D_n(4000)$ index values $D_n(4000) > 1.75$.
In the highest global density environments, more than half of all galaxies 
have such high values of $D_n(4000)$ index. 
Even among single galaxies and satellites of LLGs (for which the local densities are
the lowest), 
the percentage of galaxies with $D_n(4000) > 1.75$ is  $27 - 30$. 
These galaxies can be considered as quenched, with old stellar populations.
Only approximately $2$\% of galaxies with $D_n(4000) > 1.75$ may still form stars,
and their high $D_n(4000)$ index values may be related to their metallicity.
Therefore,
the percentage of VO galaxies at the lowest densities is relatively high.
The percentage of galaxies with 
$D_n(4000) > 2$ increases from approximately $2$~\% in watershed regions
to $7$~\% in superclusters. Up to almost $40$~\% of BGGs of HLGs have such high
$D_n(4000)$ indexes, 
which indicates that they have not changed significantly over the last $10$~Gyr, since redshifts of
$z \approx 1.5 - 2$. 
 
Also somewhat surprisingly, we found that the effect of global density is the weakest on other
galaxy properties, such as their stellar masses, concentration indexes, and
stellar velocity dispersions.  These parameters are almost
the same in all global density environments, except in the regions of  lowest density.
In watershed regions with threshold density  $D8 = 1$, stellar masses and stellar velocity dispersions of 
galaxies from all types of systems (single galaxies, satellites, and BGGs) 
are lower than in higher global density environments,  and their concentration
index is higher.
We point out that all this holds for single galaxies, satellites, and BGGs, 
and is therefore independent of local density. This small but 
statistically significant difference may be an indicator that, at some level, 
global density is also important in this case. Evidence in support of this hypothesis is provided by the finding
that 
although in the global density interval $1 < D8 < 2$
most galaxies are still singles or members of LLGs, as also in the watershed region,
these galaxies have morphological properties similar to those
of galaxies in higher global density regions where LLGs and HLGs dominate.

In all global density environments, the properties of single galaxies
are between those of satellite galaxies in LLGs and HLGs.
Interestingly, satellite galaxies in HLGs are close in their properties to the
BGGs of LLGs.
BGGs of HLGs are different from
BGGs in LLGs, especially in the highest global density environments
(superclusters). There are almost no star-forming galaxies among the BGGs of HLGs, and they
have higher stellar masses and stellar velocity dispersions,
and lower concentration indexes than  their counterparts in LLGs.  
Next, we discuss what these results tell us about the evolution of galaxies and groups.

\subsection{Coevolution of galaxies and groups in the cosmic web}
\label{sect:waterq} 

\paragraph{Star formation quenching in single galaxies and LLG members: cosmic web detachment.}
\label{sect:sfcwd} 

In the lowest global density (watershed) regions, single VO galaxies 
and VO members of LLGs are somewhat different from
those in other global density regions: their median stellar mass is lower, and among them there
is a lower percentage of galaxies with extremely old stellar populations. 
This can be referred to as `environmental downsizing'. 
These are still
quenched VO galaxies, but their growth is not like that seen in denser environments.
As we show, even in the global density interval $1 < D8 < 2$, where single 
galaxies and members of LLG  still dominate, the morphological properties of VO galaxies
are similar to those in higher global density environments.

Due to very low environment density and lower stellar mass, one may conclude that
for these galaxies, mergers were suppressed but the galaxies are still quenched. Therefore, we might wonder  what caused star formation quenching in these environments. The answer may
be cessation of fresh gas supply due to the detachment of primordial filaments
as a galaxy enters a group 
(cosmic web detachment (CWD)),
which was theoretically predicted from simulations by \citet{2019OJAp....2E...7A} 
\citep[see also][]{2016A&A...588A..79L}.
Also, gas may be removed in galaxies due to ram pressure
as they fly across filaments of the cosmic web.
However, the  efficiency of this process is not yet clear. 
In addition, in the case of single VO galaxies the problem remains that 
for these processes there needs to be a significant amount of presently undetected LLGs or filaments.

Local densities at the locations of VO galaxies are higher than 
at the locations of YS galaxies.
This may be related to the distribution of filaments in various global environments.
Similar trends within filaments were noted by \citet{2021ApJ...906...68L}. 
From observations, we do not have information on primordial filaments,
but present-day galaxy filaments are closer together in superclusters than 
in the low-density environment around them \citep{2020A&A...641A.172E}.
Therefore, the probability that a galaxy will encounter a filament is higher in high-global-density environments,
and this makes CWD more effective in high-density environments.
This in turn leads to a
higher percentage of VO galaxies in the high-global-density environments, 
even among single galaxies in superclusters 
in comparison with watershed regions.

\paragraph{Preprocessing of galaxies: single galaxies and satellites versus 
BGGs, and LLGs versus HLGs.}
\label{sect:prep}

\begin{figure*}
\centering
\resizebox{0.44\textwidth}{!}{\includegraphics[angle=0]{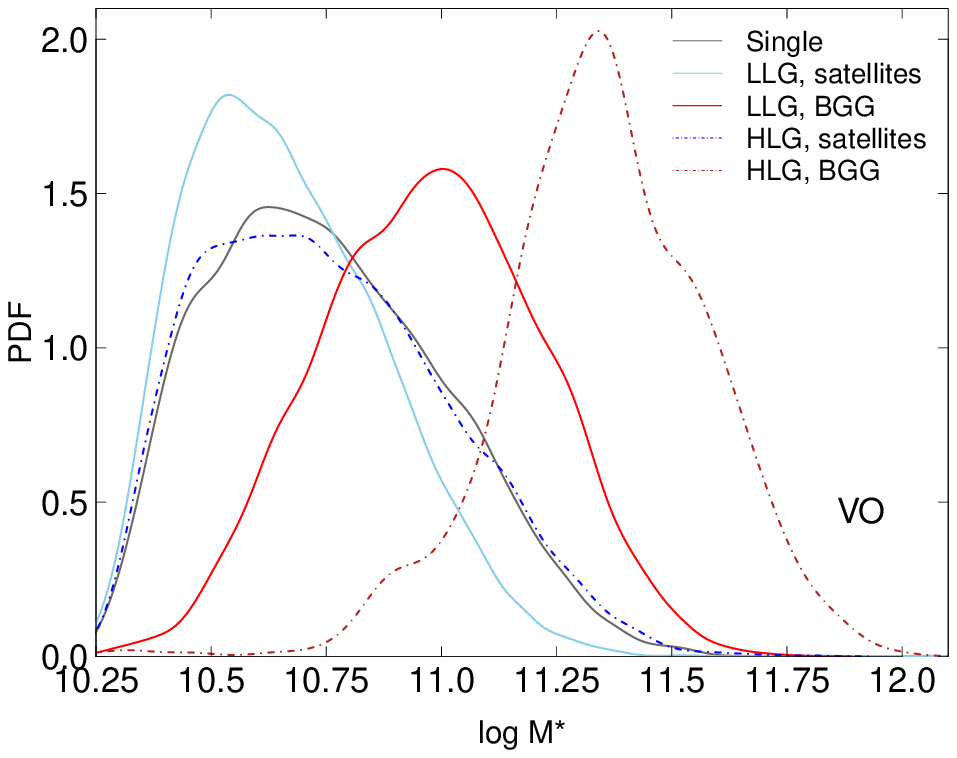}}
\resizebox{0.44\textwidth}{!}{\includegraphics[angle=0]{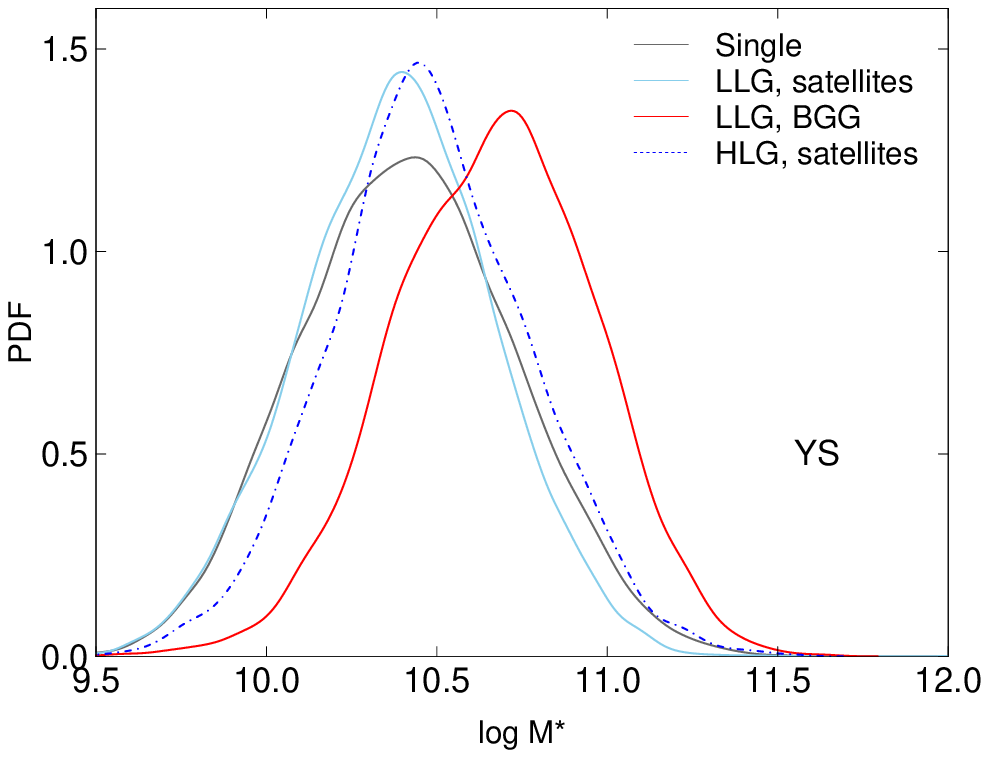}}
\caption{
Distributions of stellar mass $M^\star$ for single galaxies
(grey lines), satellites galaxies in LLGs and HLGs (light and dark
blue lines, respectively), and BGGs of LLGs and HLGs (light red and dark red 
lines, respectively) for VO galaxies (left panel) and YS galaxies (right panel)
in global luminosity--density regions $D8 > 1$.
Solid lines show single galaxies and LLG members, and dot-dashed lines show HLG members.
}
\label{fig:msvoys}
\end{figure*}

The lowest stellar masses of galaxies in this study are $M^{\mathrm{*}} = 1.8\times 10^{10}\,\mathrm{M}_{\sun}$. 
Due to this cutoff, we may not know all possible galaxy filaments. 
Detachment of the primordial filaments of VO galaxies in cosmic watersheds 
may be due to interactions with these undetected present-day filaments. 
But this is not the only  reason. 

The high fraction of VO galaxies among single galaxies and in poor groups is
clear evidence of the preprocessing of galaxies before falling into clusters,
as shown  for high-redshift clusters by \citet{2021gcf2.confE..20W}, who found that
at $z \sim 1$ most  massive galaxies were self-quenched or preprocessed in groups
before falling into clusters.
Isolated elliptical galaxies have also been found 
in filaments around Coma cluster in the Coma supercluster \citep{2016A&A...588A..79L}.
In the AMIGA sample of isolated galaxies, elliptical and lenticular galaxies
make up $14$~\% \citep{2006A&A...449..937S}.
Among the galaxies analysed in the present study, approximately $50$\% of those in LLGs and approximately $40$\% of the galaxies in HLGs and single
galaxies 
are members of filaments. 
We find that approximately $25$\% of filament member galaxies
from all populations (single galaxies, LLG members, and HLG members)
have $D_n(4000) \geq 1.75$. 
Therefore, preprocessing occurs also in filaments \citep{1992MNRAS.258..571E, 2014MNRAS.438.3465T,
2018MNRAS.474..547K, 2020A&A...639A..71K, 
2022ApJS..259...43C, 2021AJ....161..255E, 2022A&A...657A...9C, 2022arXiv220109540J}.
A dedicated and detailed analysis of galaxy quenching in filaments in various
global density environments is merited.

Next, we discuss other mechanisms in addition to the CWD that affect galaxy properties
and may lead to star-formation quenching in galaxies.
Various data point to accretion events and a wide range of merger scenarios 
in the history of the MW-like 
galaxies $11 - 8$~Gyr ago \citep{Haywood:2016un, Bell:2017vy, Di-Matteo:2019wt,
2022arXiv220604707Q}.
Accretion and merger events at least $4$~Gyr ago may also have occurred in 
isolated elliptical galaxies \citep{2016A&A...588A..79L, 2022ApJ...927..124M}, and
these could disturb gas inflow from primordial
filaments. This variety is interesting in the context of our study,
which reveals similarities in the morphological properties of single galaxies 
and satellites of LLGs in very different global environments.
Also, could the very different percentages of VO galaxies among single galaxies in
low- and high-global-density environment be related to the variety of accretion
histories of galaxies, with quenching being more effective in high-global-density
environment? 

The efficiency of the star-formation quenching of satellite galaxies as they fall into
a potential well of a main galaxy (in the case of single galaxies) or
of a group (for satellite galaxies in groups) depends on their
orbital parameters \citep{2020A&A...638A.133L, 2022MNRAS.514.6157R}.
Typically, it takes more than one peripassage to quench a satellite
galaxy. 
The star formation in galaxies deceases
under environmental processes described as slow-then-rapid quenching
with characteristic timescales of $4 - 5$~Gyr \citep{2014ApJ...796...65M, 2014MNRAS.442L.105M,
2019A&A...621A.131M, 2021A&A...647A..32K, 2022arXiv220512169K}.
Slow quenching by strangulation and overconsumption starts when a galaxy enters a group and 
gas inflow stops. When a galaxy reaches the central parts of a group or cluster, star formation is stopped by ram pressure  which strips gas from galaxies.
These timescales agree well with our finding that galaxies in the central parts of groups
have  $D_n(4000) > 1.75$, that is, they were quenched  several gigayears ago.
As the galaxies orbit
the group or cluster with many pericenter passages,
their shapes change from disky to triaxial,
and their rotation is suppressed and replaced by random
motions while they become increasingly red \citep{2020A&A...638A.133L}.
This may take at least $4$~Gyr from the first pericenter passage
\citep{2020A&A...638A.133L}, which agrees with our findings about the ages of stellar
populations of VO galaxies in groups and clusters.
This may lead to the differences in concentration index that we find for the galaxies included in our study.
Only rarely is the impact directed towards the BGG in a group, which could lead to the merger of
both galaxies and changes in the morphological properties of the BGGs.
   
To better understand the influence of group membership on the
distributions of morphological properties of galaxies, we show in Fig.~\ref{fig:msvoys} the
distribution of stellar masses $M^\star$ for galaxies in global
density regions $D8 > 1$ in different group membership classes,
separately for VO and YS galaxies. 
Watershed galaxies with clearly lower stellar masses than galaxies from the
same class in other global luminosity--density regions are not taken into account in this
figure. Figure~\ref{fig:msvoys} shows that the stellar mass $M^\star$
distributions of  satellites of LLGs and HLGs are very similar.  The $M^\star$
distributions for single galaxies are also close, but for VO galaxies they are slightly
shifted towards smaller stellar masses.  
These similarities suggest that 
some HLG satellites may have been single galaxies in the past.
For example, \citet{2009MNRAS.400..937M} found that approximately $40$\% of galaxies 
falling into groups and clusters are single galaxies and not members of groups. 
This interpretation agrees with the changes of the percentage of single galaxies 
and galaxies in LLGs and HLGs seen in Fig.~\ref{fig:frac}. While  the percentage of galaxies in LLGs remains almost unchanged over a wide
range of global 
luminosity densities, the percentage of
single galaxies decreases and the percentage of galaxies in HLGs increases.
The enhanced fraction of star-forming galaxies among single galaxies in comparison with HLG satellites
(Figs.~\ref{fig:dn4gr} and \ref{fig:dn4d8}) 
can be explained as evidence of star-formation triggering in galaxies while they fall into groups and clusters. 
The slight differences in distributions for single galaxies and satellites of LLGs
can be interpreted as evidence that the sample of satellite
galaxies of LLGs contains some galaxies that are statistically less massive
than our sample of single galaxies.
The distinction between satellites and single galaxies also depends on the parameters
used  to compile group catalogues and the magnitude limits 
of galaxies in the sample. The connection between satellites and single galaxies
is an interesting question by itself, and deserves a dedicated study.

Also, considering low local densities $D1$ at the location
of LLGs, even in global high-density environments shown in Fig.~\ref{fig:d1d8llgsingle}, 
it is  possible that
LLGs form at locations where the local density is sufficient for the formation of a
poor group,  but not yet sufficient for the formation of a rich group or 
cluster. 

The distributions for BGGs are shifted in respect to satellites towards larger
masses by a factor of about three, and the shapes of the distributions are also different.
These differences suggest that,  in terms of mass, the growth of BGGs is
governed by processes that are essentially different from those governing the growth of all other galaxies.

The $D_n(4000)$ index values of VO galaxies in $D8 < 1$ regions 
suggest that their stellar populations may be up to $2$~Gyr younger than in VO galaxies
from higher global density environments.
The percentage of galaxies with $D_n(4000) \geq 2.0$ among single galaxies is about 5\%--10\%.
In all global environments, approximately $10$~\% of poor group satellite galaxies
and $20$~\% of their BGGs have $D_n(4000) \geq 2.0$.
Among rich groups, about 15\%--20\% of satellite galaxies
and about $40$~\% of the BGGs have $D_n(4000) \geq 2.0$
These galaxies form a high-end tail in the stellar mass 
and $\sigma^{\mathrm{*}}$
distributions.
According to \citet{2003MNRAS.341...33K}, such high values of the $D_n(4000)$
index correspond to ages of the stellar population in these galaxies of at least
$10$~Gyr, that is, these galaxies stopped forming stars at redshifts $z \approx 1.5 - 2$
\citep[see also][]{2012ApJ...760...62B, 2021A&A...649A..42C}.
This is approximately the redshift of the morphology--density relation reversal:
at higher redshifts, high-density regions 
are populated by star-forming galaxies
that quenched faster than similar galaxies in lower density environments,
causing the reversal of the relation at lower redshifts at which high-density regions 
are populated by quenched galaxies 
\citep[][and references therein]{2017ApJ...844L..23C, 2019MNRAS.489..339H}.
Even so, observations have found  quenched galaxies in protoclusters
even at redshifts as high as $z = 3.37$
\citep{2022ApJ...926...37M}.

\citet{2019ApJ...872...50L} used the MaNGA data to study the quenching of
galaxies in different local environments at submegaparsec
scale, which is defined by halo mass, that is, luminosity-based group mass in which a galaxy resides.
These authors showed  that the dominant mechanism is inside-out
quenching where quenching first takes place in  the central region and then
proceeds outwards. This effect is dominant for central galaxies and
high-mass satellites, but is weaker for low-mass satellites with stellar masses
$log M* < 10.5$ \citep[see also][]{2021A&A...647A..32K}. 
This agrees well with our results, which showed that
the percentage of galaxies with $\log M* < 10.5$ is lower than $20$~\% among
satellite galaxies with $D_n(4000) > 1.75$, and there are almost no BGGs with such low
stellar mass. Almost half of YS galaxies with $D_n(4000) < 1.75$ ($25$~\% of BGGs) have stellar
masses $\log M* < 10.5$ (Figs.~\ref{fig:g175d8int} and \ref{fig:g0175d8int}).

The connection between the evolution of groups and the galaxies within them is also seen
in the star-formation properties of the BGGs: while BGGs in low-luminosity
groups may still be star-forming, in high-luminosity groups almost all
BGGs are quenched. 
Even in low-luminosity groups with star-forming BGGs, typically 
at least one galaxy is quenched and may become a BGG during  subsequent evolution
of galaxies and the whole group.
We found very few groups (among groups with $5 < N_{gal} < 10)$ for which
all the galaxies within have $D_n(4000) < 1.35$. The star-formation properties of BGGs and
single galaxies also depend on the local environment: those in a lower local
density environment are more likely to be actively star forming. 
Differences in the morphological properties of the BGGs and satellites suggest that 
BGGs have been through a special assembly history with a series of mergers during their 
evolution \citep{2015MNRAS.447.1491L, 2020MNRAS.491.2617E}.

The similarity in  the properties of galaxies found in different environments revealed by the present study
has been detected before.
In a review paper about galaxy properties \citet{2009ARA&A..47..159B} noted that
when galaxy mass and star formation history are fixed, other properties of galaxies
are the same over a range of different environments, with these different environments being typically defined using the
$Nth$ nearest neighbour of a galaxy, or their membership to clusters or groups.
\citet{2010MNRAS.402L..59B} found that the properties of galaxies in groups 
from a wide range of masses on a two-colour diagram are very homogeneous.
Questions as to why there are different types of galaxies (elliptical and spirals),
and why their properties are almost independent of environment were recently discussed
in \citet{2021arXiv210602672P} and \citet{2022MNRAS.511.5093P}, 
in which a series of unanswered questions about galaxy formation and evolution were 
outlined.

Here we outlined  how the evolution of galaxies and groups leads 
to the higher fraction of quenched galaxies in high-density environments,
known as the large-scale morphology-density relation.  
In our study, we used global luminosity density field to study galaxy environments, 
and divided galaxies according to their group membership. 
Our analysis shows that once we fix whether a galaxy is single, a satellite galaxy in a group, or a BGG, 
galaxy properties are almost independent of the global environment.
We propose various  processes that could lead to very similar galaxies
in hugely varying environments where conditions are very different.

\section{Summary}
\label{sect:sum}

Here we present a study of group content and  galaxy population in various global density
environments. We summarise our results as follows:

\begin{itemize}
\item[1)]
Based on galaxy and group properties, we define the watershed regions
as regions with global luminosity--density   $D8 \leq 1$.
Watersheds occupy approximately $65$\% of the total SDSS volume.
\item[2)]
The strongest effects of global environment are seen in the richness of groups.
Watershed regions
are mostly populated by single galaxies (70\% of all galaxies)
and poor, low-luminosity galaxy groups with 
$L_{gr} \leq 15\times10^{10} h^{-2} L_{\sun}$.
Richer and more luminous groups and clusters reside in higher global density regions.
\item[3)]
Global environment affects the star-formation properties of galaxies  less strongly.
In the watershed regions, approximately
30\% of all galaxies (including single galaxies) have very old stellar populations with
$D_n(4000) > 1.75 $ (VO galaxies). 
In the highest global density environments (superclusters), VO galaxies form
55\% of all galaxies.
\item[4)] 
The weakest effects of global environment are seen in the morphological properties of galaxies.
Single galaxies and satellites in watersheds have lower stellar masses 
and lower stellar velocity distributions than their counterparts
in higher global environments. In higher global density environments with $D8 > 1,$
the stellar masses, stellar velocity dispersions and concentration indexes of galaxies from a given
class (i.e. single galaxies, 
satellites, and BGGs), or, in other words, over a wide range of group 
luminosities (and therefore masses),  are statistically  similar.
\item[5)] Morphological properties of galaxies are mainly  determined
  by whether a galaxy is a single, satellite, or BGG. 
The largest differences in galaxy properties in all environments are between
BGGs and all other galaxies (satellites of groups and single galaxies).
BGGs have higher stellar masses, larger stellar velocity dispersions, and
lower concentration indexes than other galaxies. 
\subitem{5.1)}
BGGs of poor groups may have old and young stellar populations, but
BGGs of rich groups are almost all VOs.
\subitem{5.2)}
Up to  40\% of BGGs have $D_n(4000) > 2$, which suggests that these galaxies stopped star formation
approximately 10~Gyr ago, at redshifts $z \approx 1.5 - 2$.
\end{itemize}

The similarity of  galaxy properties  
in hugely varying environments where conditions are very different
suggests that the present-day galaxy properties are largely shaped by their 
birthplace in the cosmic web (initial conditions for galaxy formation) 
---which determines whether a galaxy remains single or becomes a member of a poor group or 
a rich cluster---
and internal processes that lead to star formation quenching
in galaxies.

To better understand the formation and evolution of galaxies in various global and local environments,
a next step is to analyse the morphological properties of galaxies in more detail
than in the present study. 
For this purpose, one could use data on galaxies in our
local cosmic neighbourhood, such as those provided by \citet{2022arXiv220604707Q},
and future multi-wavelength surveys with detailed information
on galaxies, such as the J-PAS survey, or the   HI Westerbork Coma Survey
\citep{2022A&A...659A..94M}, which showed that cluster galaxies are HI deficient in comparison with galaxies
found around the cluster.
Were galaxies born similar, turning into ellipticals or spirals  during evolution
in hugely different environments, or are the seeds of primeval galaxies
already different? 
Of special interest for answering such questions is the star formation quenching in single galaxies
in the lowest global density environments far from other galaxy systems
where the influence of external factors to galaxy evolution is the smallest.

\begin{acknowledgements}
We thank the referee for detailed comments and suggestions which helped us to
improve the paper. 
We thank Mirt Gramann for useful discussions.
We are pleased to thank the SDSS Team for the publicly available data
releases.  Funding for the Sloan Digital Sky Survey (SDSS) and SDSS-II has been
provided by the Alfred P. Sloan Foundation, the Participating Institutions,
the National Science Foundation, the U.S.  Department of Energy, the
National Aeronautics and Space Administration, the Japanese Monbukagakusho,
and the Max Planck Society, and the Higher Education Funding Council for
England.  The SDSS website is \texttt{http://www.sdss.org/}.
The SDSS is managed by the Astrophysical Research Consortium (ARC) for the
Participating Institutions.  The Participating Institutions are the American
Museum of Natural History, Astrophysical Institute Potsdam, University of
Basel, University of Cambridge, Case Western Reserve University, The
University of Chicago, Drexel University, Fermilab, the Institute for
Advanced Study, the Japan Participation Group, The Johns Hopkins University,
the Joint Institute for Nuclear Astrophysics, the Kavli Institute for
Particle Astrophysics and Cosmology, the Korean Scientist Group, the Chinese
Academy of Sciences (LAMOST), Los Alamos National Laboratory, the
Max-Planck-Institute for Astronomy (MPIA), the Max-Planck-Institute for
Astrophysics (MPA), New Mexico State University, Ohio State University,
University of Pittsburgh, University of Portsmouth, Princeton University,
the United States Naval Observatory, and the University of Washington.

The present study was supported by the ETAG projects 
PRG1006, PSG700, and by the European Structural Funds
grant for the Centre of Excellence "The Dark Side of the Universe" (TK133).
This work has also been supported by
ICRAnet through a professorship for Jaan Einasto.
We applied in this study R statistical environment 
\citep{ig96}.

\end{acknowledgements}

\bibliographystyle{aa}
\bibliography{ld}

\begin{appendix}

\section{The properties of galaxies in poor and rich groups and 
single galaxies in low- and high-global-density environments}
\label{sect:galprop10}  

Tables~\ref{tab:singlepop} - \ref{tab:bggpop0175}  present 
median values of the morphological properties of VO and YS galaxies
for single galaxies, satellites, and BGGs of LLGs  in various global density environments.
We also give the KS test $p$-values, which show the statistical significance
of the differences between distributions for watershed regions ($D8 < 1$) and other
global luminosity--density  regions.

\begin{table*}[ht]
{\footnotesize
  \centering
  \caption{Median values of the  galaxy properties,  and the KS test $p$-values
for VO single galaxies in various global density environments}
\begin{tabular}{lrrrrrrrrrrr} 
\hline\hline  
(1)&(2)&(3)&(4)&(5)&(6)&(7)&(8)&(9)&(10)&(11)&(12)\\      
\hline 
 $D8$ & $N_{\mathrm{gal}}$ & $M^{\mathrm{*}}_{\mathrm{med}}$ & $p$ & $\sigma^{\mathrm{*}}_{\mathrm{med}}$ & $p$& 
  $C_{\mathrm{med}}$ & $p$ & $D_n(4000)_{\mathrm{med}}$ & $p$ & $\mathrm{SFR}_{\mathrm{med}}$ & $p$  \\
\hline
 $0 - 1$ & 2797 & 10.69 &          &  157  &           & 0.35 &          &  1.88  &       & -1.30 &     \\
 $1 - 2$ & 3757 & 10.73 & $< 0.001$&  161  & $< 0.001$ & 0.34 & $< 0.001$&  1.88  & 0.121 & -1.26 & $< 0.001$  \\
 $2 - 3$ & 2100 & 10.73 & $< 0.001$&  163  & $< 0.001$ & 0.34 & $< 0.001$&  1.89  & 0.001 & -1.28 & 0.005 \\
 $3 - 4$ & 1086 & 10.73 & $< 0.001$&  161  & 0.018     & 0.34 & $< 0.001$&  1.89  & 0.230 & -1.28 & 0.124 \\
 $4 - 5$ &  570 & 10.74 & $< 0.001$&  164  & 0.008     & 0.34 & $< 0.001$&  1.90  & 0.017 & -1.26 & 0.030 \\
 $ > 5 $ &  870 & 10.72 & $< 0.001$&  161  & 0.009     & 0.34 &  0.003   &  1.89  & 0.122 & -1.29 & 0.300 \\
\hline
\label{tab:singlepop}  
\end{tabular}\\
\tablefoot{                                                                                 
Columns are as follows:
(1): Global luminosity--density $D8$ range;
(2): Number of galaxies in corresponding global luminosity--density range;
(3--4): Median value of the stellar mass, $M^{\mathrm{*}}_{\mathrm{med}}$,  and 
$p$-value of the KS test between a given population in
the lowest global density environment ($D8 \leq 1$) and in the given global density interval;
(5--6): Median value of the stellar velocity dispersion $\sigma^{\mathrm{*}}$ and 
$p$-value of the KS test as in column (3); 
(7--8): Median value of the concentration index $C$ and 
$p$-value of the KS test as in column (3); 
(9--10): Median value of the $D_n(4000)$ index, $D_n(4000)_{\mathrm{med}}$, and 
$p$-value of the KS test as in column (3);
(11--12): Median value of the star formation rate, $\mathrm{SFR}_{\mathrm{med}}$, and 
$p$-value of the KS test as in column (3). 
}
}
\end{table*}


\begin{table*}[ht]
{\footnotesize
  \centering
\caption{Median values of the  galaxy properties,  and the KS test $p$-values
for VO satellite galaxies in LLGs
in various global density environments}
\begin{tabular}{lrrrrrrrrrrr} 
\hline\hline  
(1)&(2)&(3)&(4)&(5)&(6)&(7)&(8)&(9)&(10)&(11)&(12)\\      
\hline 
 $D8$ & $N_{\mathrm{gal}}$ & $M^{\mathrm{*}}_{\mathrm{med}}$ & $p$ & $\sigma^{\mathrm{*}}_{\mathrm{med}}$ & $p$& 
  $C_{\mathrm{med}}$ & $p$ & $D_n(4000)_{\mathrm{med}}$ & $p$ & $\mathrm{SFR}_{\mathrm{med}}$ & $p$  \\
\hline
 $0 - 1$ & 595 & 10.61 &          & 151  &       & 0.35 &          &  1.88  &           & -1.39 &      \\
 $1 - 2$ &2069 & 10.64 & $< 0.001$& 155  & 0.040 & 0.35 & $< 0.001$&  1.89  & 0.108     & -1.33 & 0.02 \\
 $2 - 3$ &1732 & 10.65 & $< 0.001$& 156  & 0.001 & 0.34 & $< 0.001$&  1.89  & $< 0.001$ & -1.35 & 0.32 \\
 $3 - 4$ &1173 & 10.63 & $< 0.001$& 155  & 0.018 & 0.35 & $< 0.001$&  1.89  & 0.010     & -1.36 & 0.18 \\
 $4 - 5$ & 650 & 10.64 & $< 0.001$& 159  & 0.001 & 0.35 & $< 0.001$&  1.89  & 0.005     & -1.35 & 0.20 \\
 $ > 5 $ &1076 & 10.64 & $< 0.001$& 155  & 0.004 & 0.35 &  0.003   &  1.89  & 0.007     & -1.34 & 0.10 \\
\hline
\label{tab:satpop}  
\end{tabular}\\
\tablefoot{                                                                                 
Columns are as follows:
(1): Global luminosity--density $D8$ range;
(2): Number of galaxies in corresponding global luminosity--density range;
(3--4): Median value of the stellar mass, $M^{\mathrm{*}}_{\mathrm{med}}$,  and 
$p$-value of the KS test between a given population in
the lowest global density environment ($D8 \leq 1$) and in the given global density interval;
(5--6): Median value of the stellar velocity dispersion $\sigma^{\mathrm{*}}$ and 
$p$-value of the KS test as in column (3); 
(7--8): Median value of the concentration index $C$ and 
$p$-value of the KS test as in column (3); 
(9--10): Median value of the $D_n(4000)$ index, $D_n(4000)_{\mathrm{med}}$, and 
$p$-value of the KS test as in column (3);
(11--12): Median value of the star formation rate, $\mathrm{SFR}_{\mathrm{med}}$, and 
$p$-value of the KS test as in column (3). 
}}
\end{table*}


\begin{table*}[ht]
{\footnotesize
  \centering
\caption{Median values of the  galaxy properties,  and the KS test $p$-values
for VO BGGs in LLGs
in various global density environments}
\begin{tabular}{lrrrrrrrrrrr} 
\hline\hline  
(1)&(2)&(3)&(4)&(5)&(6)&(7)&(8)&(9)&(10)&(11)&(12)\\      
\hline 
 $D8$ & $N_{\mathrm{gal}}$ & $M^{\mathrm{*}}_{\mathrm{med}}$ & $p$ & $\sigma^{\mathrm{*}}_{\mathrm{med}}$ & $p$& 
  $C_{\mathrm{med}}$ & $p$ & $D_n(4000)_{\mathrm{med}}$ & $p$ & $\mathrm{SFR}_{\mathrm{med}}$ & $p$  \\
\hline
 $0 - 1$ & 770 & 10.90 &          & 182  &         &  0.33 &          &  1.90  &           & -1.21 &           \\
 $1 - 2$ &2042 & 10.97 & $< 0.001$& 190  & 0.098   &  0.33 & $< 0.001$&  1.92  & 0.053     & -1.16 & $< 0.001$ \\
 $2 - 3$ &1465 & 10.99 & $< 0.001$& 195  & 0.025   &  0.33 & $< 0.001$&  1.93  & $< 0.001$ & -1.16 & 0.002     \\
 $3 - 4$ & 963 & 10.98 & $< 0.001$& 198  &$< 0.001$&  0.32 & $< 0.001$&  1.92  & 0.001     & -1.17 & 0.013     \\
 $4 - 5$ & 498 & 10.98 & $< 0.001$& 194  & 0.074   &  0.32 & $< 0.001$&  1.93  & $< 0.001$ & -1.16 & 0.002     \\
 $ > 5 $ & 797 & 10.98 & $< 0.001$& 193  & 0.20    &  0.33 &  0.003   &  1.93  & $< 0.001$ & -1.16 & 0.006     \\
\hline
\label{tab:bggpop}  
\end{tabular}\\
\tablefoot{                                                                                 
Columns are as follows:
(1): Global luminosity--density $D8$ range;
(2): Number of galaxies in corresponding global luminosity--density range;
(3--4): Median value of the stellar mass, $M^{\mathrm{*}}_{\mathrm{med}}$,  and 
$p$-value of the KS test between a given population in
the lowest global density environment ($D8 \leq 1$) and in the given global density interval;
(5--6): Median value of the stellar velocity dispersion $\sigma^{\mathrm{*}}$ and 
$p$-value of the KS test as in column (3); 
(7--8): Median value of the concentration index $C$ and 
$p$-value of the KS test as in column (3); 
(9--10): Median value of the $D_n(4000)$ index, $D_n(4000)_{\mathrm{med}}$, and 
$p$-value of the KS test as in column (3);
(11--12): Median value of the star formation rate, $\mathrm{SFR}_{\mathrm{med}}$, and 
$p$-value of the KS test as in column (3). 
}}
\end{table*}

\begin{table*}[ht]
{\footnotesize
  \centering
\caption{Median values of the  galaxy properties,  and the KS test $p$-values
for YS single galaxies  with $D_n(4000) < 1.75 $ 
in various global density environments}
\begin{tabular}{lrrrrrrrrrrr} 
\hline\hline  
(1)&(2)&(3)&(4)&(5)&(6)&(7)&(8)&(9)&(10)&(11)&(12)\\      
\hline 
 $D8$ & $N_{\mathrm{gal}}$ & $M^{\mathrm{*}}_{\mathrm{med}}$ & $p$ & $\sigma^{\mathrm{*}}_{\mathrm{med}}$ & $p$& 
  $C_{\mathrm{med}}$ & $p$ & $D_n(4000)_{\mathrm{med}}$ & $p$ & $\mathrm{SFR}_{\mathrm{med}}$ & $p$  \\
\hline
 $0 - 1$ & 7425 & 10.38 &          &  84  &           & 0.42 &          &  1.39  &             & 0.21 &          \\
 $1 - 2$ & 8222 & 10.41 & $< 0.001$&  88  & $ 0.01  $ & 0.42 & $ 0.03  $&  1.40  & $< 0.001$   & 0.20 & 0.720     \\
 $2 - 3$ & 4098 & 10.42 & $< 0.001$&  89  & $< 0.001$ & 0.42 & $ 0.05  $&  1.41  & $< 0.001$   & 0.19 & 0.157     \\
 $3 - 4$ & 2054 & 10.42 & $< 0.001$&  91  & $< 0.001$ & 0.42 & $ 0.31$  &  1.41  & $< 0.001$   & 0.19 & 0.103         \\
 $4 - 5$ &  997 & 10.41 & $  0.005$&  89  & 0.013     & 0.42 & $ 0.04$  &  1.41  & $< 0.001$   & 0.18 & 0.071        \\
 $ > 5 $ & 1459 & 10.41 & $< 0.001$&  89  & 0.001     & 0.42 &  0.15    &  1.41  & $< 0.001$   & 0.19 & 0.002        \\
\hline
\label{tab:singlepop0175}  
\end{tabular}\\
\tablefoot{                                                                                 
Columns are as follows:
(1): Global luminosity--density $D8$ range;
(2): Number of galaxies in corresponding global luminosity--density range;
(3--4): Median value of the stellar mass, $M^{\mathrm{*}}_{\mathrm{med}}$,  and 
$p$-value of the KS test between a given population in
the lowest global density environment ($D8 \leq 1$) and in the given global density interval;
(5--6): Median value of the stellar velocity dispersion $\sigma^{\mathrm{*}}$ and 
$p$-value of the KS test as in column (3); 
(7--8): Median value of the concentration index $C$ and 
$p$-value of the KS test as in column (3); 
(9--10): Median value of the $D_n(4000)$ index, $D_n(4000)_{\mathrm{med}}$, and 
$p$-value of the KS test as in column (3);
(11--12): Median value of the star formation rate, $\mathrm{SFR}_{\mathrm{med}}$, and 
$p$-value of the KS test as in column (3). 
}}
\end{table*}

\begin{table*}[ht]
{\footnotesize
  \centering
\caption{Median values of the  galaxy properties,  and the KS test $p$-values
for YS satellite galaxies in LLGs
in various global density environments}
\begin{tabular}{lrrrrrrrrrrr} 
\hline\hline  
(1)&(2)&(3)&(4)&(5)&(6)&(7)&(8)&(9)&(10)&(11)&(12)\\      
\hline 
 $D8$ & $N_{\mathrm{gal}}$ & $M^{\mathrm{*}}_{\mathrm{med}}$ & $p$ & $\sigma^{\mathrm{*}}_{\mathrm{med}}$ & $p$& 
  $C_{\mathrm{med}}$ & $p$ & $D_n(4000)_{\mathrm{med}}$ & $p$ & $\mathrm{SFR}_{\mathrm{med}}$ & $p$  \\
\hline
 $0 - 1$ &1392 & 10.34 &          & 84  &            & 0.42 &          &  1.37  &            & 0.166 &          \\
 $1 - 2$ &3539 & 10.37 & $  0.002$& 88  & $ 0.01  $  & 0.42 & $ 0.37  $&  1.40  & $< 0.001$  & 0.150 & 0.526 \\
 $2 - 3$ &2611 & 10.40 & $< 0.001$& 89  & $< 0.001$  & 0.42 & $ 0.65 $ &  1.42  & $< 0.001$  & 0.125 & 0.014     \\
 $3 - 4$ &1797 & 10.41 & $< 0.001$& 91  & $< 0.001$  & 0.41 & $ 0.01  $&  1.44  & $< 0.001$  & 0.111 & 0.001     \\
 $4 - 5$ & 905 & 10.39 & $  0.002$& 89  & 0.013      & 0.41 & $0.01   $&  1.42  & $< 0.001$  & 0.105 & 0.010    \\
 $ > 5 $ &1460 & 10.49 & $< 0.001$& 89  & 0.001      & 0.42 & 0.64     &  1.42  & $< 0.001$  & 0.125 & 0.152    \\
\hline
\label{tab:satpop0175}  
\end{tabular}\\
\tablefoot{                                                                                 
Columns are as follows:
(1): Global luminosity--density $D8$ range;
(2): Number of galaxies in corresponding global luminosity--density range;
(3--4): Median value of the stellar mass, $M^{\mathrm{*}}_{\mathrm{med}}$,  and 
$p$-value of the KS test between a given population in
the lowest global density environment ($D8 \leq 1$) and in the given global density interval;
(5--6): Median value of the stellar velocity dispersion $\sigma^{\mathrm{*}}$ and 
$p$-value of the KS test as in column (3); 
(7--8): Median value of the concentration index $C$ and 
$p$-value of the KS test as in column (3); 
(9--10): Median value of the $D_n(4000)$ index, $D_n(4000)_{\mathrm{med}}$, and 
$p$-value of the KS test as in column (3);
(11--12): Median value of the star formation rate, ${\mathrm{SFR}}_{\mathrm{med}}$, and 
$p$-value of the KS test as in column (3). 
}}
\end{table*}

\begin{table*}[ht]
{\footnotesize
  \centering
\caption{Median values of the  galaxy properties,  and the KS test $p$-values
for YS BGGs in LLGs
in various global density environments}
\begin{tabular}{lrrrrrrrrrrr} 
\hline\hline  
(1)&(2)&(3)&(4)&(5)&(6)&(7)&(8)&(9)&(10)&(11)&(12)\\      
\hline 
 $D8$ & $N_{\mathrm{gal}}$ & $M^{\mathrm{*}}_{\mathrm{med}}$ & $p$ & $\sigma^{\mathrm{*}}_{\mathrm{med}}$ & $p$& 
  $C_{\mathrm{med}}$ & $p$ & $D_n(4000)_{\mathrm{med}}$ & $p$ & ${\mathrm{SFR}}_{\mathrm{med}}$ & $p$  \\
\hline
 $0 - 1$ &1043 & 10.60 &          & 102  &         &  0.42 &          &  1.42  &            & 0.352 &          \\
 $1 - 2$ &1806 & 10.67 & $< 0.001$& 109  &$< 0.001$&  0.41 & $< 0.001$&  1.45  & 0.013      & 0.359 & 0.180 \\
 $2 - 3$ &1132 & 10.70 & $< 0.001$& 111  &$< 0.001$&  0.41 & $< 0.001$&  1.47  & $< 0.001$  & 0.343 & 0.655     \\
 $3 - 4$ & 672 & 10.69 & $< 0.001$& 112  &$< 0.001$&  0.40 & $< 0.001$&  1.48  & $< 0.001$  & 0.290 & 0.010     \\
 $4 - 5$ & 373 & 10.67 & $< 0.001$& 111  & 0.003   &  0.41 & $ 0-93$  &  1.49  & $< 0.001$  & 0.303 & 0.15    \\
 $ > 5 $ & 559 & 10.66 & $  0.004$& 106  & 0.061   &  0.41 &  0.010   &  1.47  & $< 0.001$  & 0.299 & 0.010    \\
\hline
\label{tab:bggpop0175}  
\end{tabular}\\
\tablefoot{                                                                                 
Columns are as follows:
(1): Global luminosity--density $D8$ range;
(2): Number of galaxies in corresponding global luminosity--density range;
(3--4): Median value of the stellar mass, $M^{\mathrm{*}}_{\mathrm{med}}$,  and 
$p$-value of the KS test between a given population in
the lowest global density environment ($D8 \leq 1$) and in the given global density interval;
(5--6): Median value of the stellar velocity dispersion $\sigma^{\mathrm{*}}$ and 
$p$-value of the KS test as in column (3); 
(7--8): Median value of the concentration index $C$ and 
$p$-value of the KS test as in column (3); 
(9--10): Median value of the $D_n(4000)$ index, $D_n(4000)_{\mathrm{med}}$, and 
$p$-value of the KS test as in column (3);
(11--12): Median value of the star formation rate, $\mathrm{SFR}_{\mathrm{med}}$, and 
$p$-value of the KS test as in column (3). 
}
}
\end{table*}

\end{appendix}

\end{document}